# Infrared Spectroscopic Study of Vibrational Modes Across the Orthorhombic-Tetragonal Phase Transition in Methylammonium Lead Halide Single Crystals

*Götz Schuck,*[†] Daniel M. Többens,[†] Monika Koch-Müller,[‡] Ilias Efthimiopoulos,[‡] and Susan Schorr[†,§]*

[†]Helmholtz-Zentrum Berlin für Materialien und Energie, Hahn-Meitner-Platz 1, 14109 Berlin, Germany.

[‡]Deutsches GeoForschungsZentrum - GFZ, Telegrafenberg, 14473 Potsdam, Germany.

[§]Free University Berlin, Department of Geosciences, Malteserstraße 74–100, 12249 Berlin, Germany.

ABSTRACT

Single crystals of the methylammonium (MA) lead halides $MAPbI_3$, $MAPbBr_3$, and $MAPbCl_3$ have been investigated using infrared spectroscopy with the aim of analyzing structural and dynamical aspects of processes that enable the ordering of the MA molecule in the orthorhombic crystal structure of these hybrid perovskites. Our temperature-dependent studies were focused on the analysis of the CH/NH rocking, C-N stretching, and CH/NH bending modes of the MA molecule in the 800-1750 $cm^{-1}$ frequency range. They deliver a direct comparison of the behavior of the three halides on crossing the orthorhombic-tetragonal phase transition in MA lead halide single crystals. Drastic changes of all vibrational modes close to the phase transition were clearly observed. Additional spectral features that were not discussed previously are pointed out. The transformation of the 2-dimensional orthorhombic hydrogen bond layers into a more 3-dimensional arrangement in the tetragonal phase seems to be an important feature providing deeper insight into the mechanisms that lead to a free-rotating MA molecule in the inorganic host structure. The change of the molecules site symmetry in the tetragonal crystal structure seems to be an important feature of the orthorhombic-tetragonal phase transition. For low temperatures it can be stated that the iodide is stronger influenced by hydrogen bonding than the bromide and the chloride.

INTRODUCTION

A major current focus in photovoltaic research is the use of methylammonium (MA) lead halide as the light-harvesting active layer in thin film solar cells because it is cheap to produce, simple to manufacture and highly efficient.[1]

MA lead halide is a hybrid organic-inorganic $ABX_3$ perovskite with A = MA, B = Pb and X = I, Br or Cl. The Hydrogen bonding plays an important role in the interaction of the organic MA molecule with the inorganic lead halide host structure. At lower temperatures, the cubic aristotype-structure ($Pm\bar{3}m$) is symmetry-lowered due to tilting and distortion of the $[BX_6]^-$ octahedra, resulting in tetragonal and orthorhombic structural modifications (Table 1). In the tetragonal modification (*I*4/*mcm*), the MA molecule appears in eight different but symmetry-related positions due to orientational disorder at each site.[2] This orientational disorder of the MA molecule vanishes in the ordered orthorhombic (*Pnma*) low temperature phase.[8-11,26] In addition, the MA molecule shows more pronounced interaction with the inorganic lead-halide network in this ordered orthorhombic phase, due to shorter hydrogen bonds between the nitrogen and the halide. These shorter hydrogen bonds are arranged in two-dimensional hydrogen bond layers as illustrated in Fig. 1 and S1.

**TABLE 1** Phase transitions and transition temperatures of $MAPbX_3$ (X = I, Br and Cl). Phase transition I → II: cubic → tetragonal; Phase transition II → III: tetragonal → tetragonal; Phase transition III → IV: tetragonal → orthorhombic.

| Phase transition | Transition temperatures [K] (*references) | | |
|---|---|---|---|
| | $MAPbI_3$ | $MAPbBr_3$ | $MAPbCl_3$ |
| I → II | 330 (*[3]) | 236 (*[3]) | 177 (*[3]) |
| II → III | | 154 (*[49]) | |
| III → IV | 161 (*[3]) | 149 (*[3]) | 172 (*[3]) |

Based on first principle calculations, several recent publications emphasized the importance of the hydrogen bonding in $MAPbX_3$ in order to understand the interconnection of MA with the lead halide structure.[4-7] Some progress has been made in understanding the interplay of the organic MA molecule with inorganic host structures of $MAPbX_3$ by analyzing the fundamental frequencies of the MA molecule in $MAPbX_3$ and related compounds with IR spectroscopic methods.[3,14-16,18-23,27,30,31] Onoda-Yamamuro et al. reported temperature-dependent (88-299 K) infrared spectra of $MAPbX_3$ and stated that some of the MA vibrational modes were strongly temperature dependent, however only limited spectroscopic details about the orthorhombic-tetragonal phase transition were reported.[3] In an infrared spectroscopic study limited to ambient temperature, a shift to lower frequencies was observed for almost all MA vibrational modes in the series Cl-Br-I and was attributed to Lorentz-Lorenz shift effect by Glaser and coworkers.[21] In a more recent publication, Ivanovska et al. reported temperature-dependent (80-360 K) infrared absorption spectra in a broad frequency range of 30-3400 $cm^{-1}$ (IR) on $MAPbI_3$ thin films, providing information about the temperature evolution of spectral features of the MA vibrational modes.[14] It should be emphasized that infrared absorption spectroscopy is the method of choice for studying these materials, since it is not subject to the degradation issues[45,46] that were observed in Raman spectroscopic experiments when a photon excitation energy is used that is not well below the optical transition.[14,17,19]

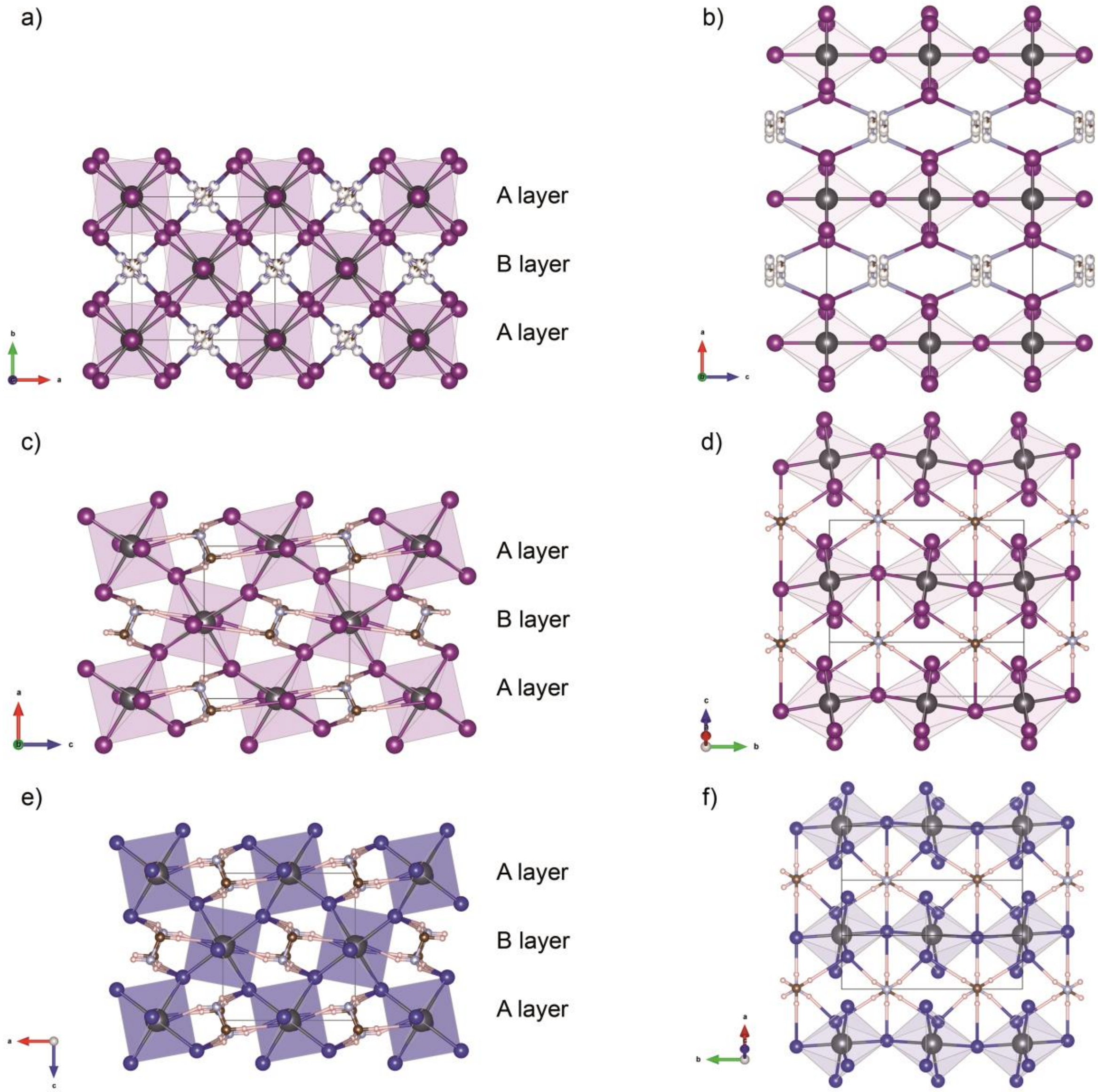

**FIGURE 1** Visualization of $MAPbI_3$ tetragonal (a, b), $MAPbI_3$ orthorhombic (c, d) and $MAPbBr_3$ orthorhombic (e, f) crystal structure. Two hydrogen bond layers, layer A and layer B, can be identified. Layer A is shown in b), d) and f). The only difference between layer A and layer B is the orientation of the MA molecule axis. All N-H...I bond lengths are the same for both layers A and B (Table S5). Hydrogen atoms are not shown in the tetragonal structure. The colors refer to the following elements: purple - iodine, dark blue - bromide, dark grey - lead, brown - carbon, light blue - nitrogen, light orange - hydrogen.

The room temperature spectra reported here are in good agreement with previous results.[3,14-16,18-20,22,23,51] Temperature dependent infrared absorption spectroscopy across the orthorhombic-tetragonal phase transition has been reported for $MAPbX_3$.[3,14,15] However our results show that there are additional spectral features present in the 1300-1750 $cm^{-1}$ frequency range, especially at low temperatures, that have not been analyzed so far. Here, we report on these additional spectral features and present a complete factor group analysis that considers the symmetry-reduced site symmetry $C_s$ of the MA molecule in the orthorhombic phase. Combined DFT calculations and factor group analysis provide vibrational modes' geometries and allowed unambiguous assignment of the MA vibrational modes in the low-temperature orthorhombic phase. We also provide detailed information about the temperature evolution (106-298 K) of spectral features of MA in the 800-1750 $cm^{-1}$ frequency range needed to understand the behavior of the MA molecules in the vicinity of the orthorhombic-tetragonal phase transition. Our investigations showed that especially the frequency range between 1300-1750 $cm^{-1}$ can only be interpreted if a) spectra with variable chemical composition are compared (as we show here), and b) all spectral features are taken into account, even though not all observed bands can be directly assigned to vibrational modes. Only then is it possible to follow the temperature-dependent spectral features for the investigated halides in a broad temperature range crossing the orthorhombic-tetragonal phase transition (Fig. 2).

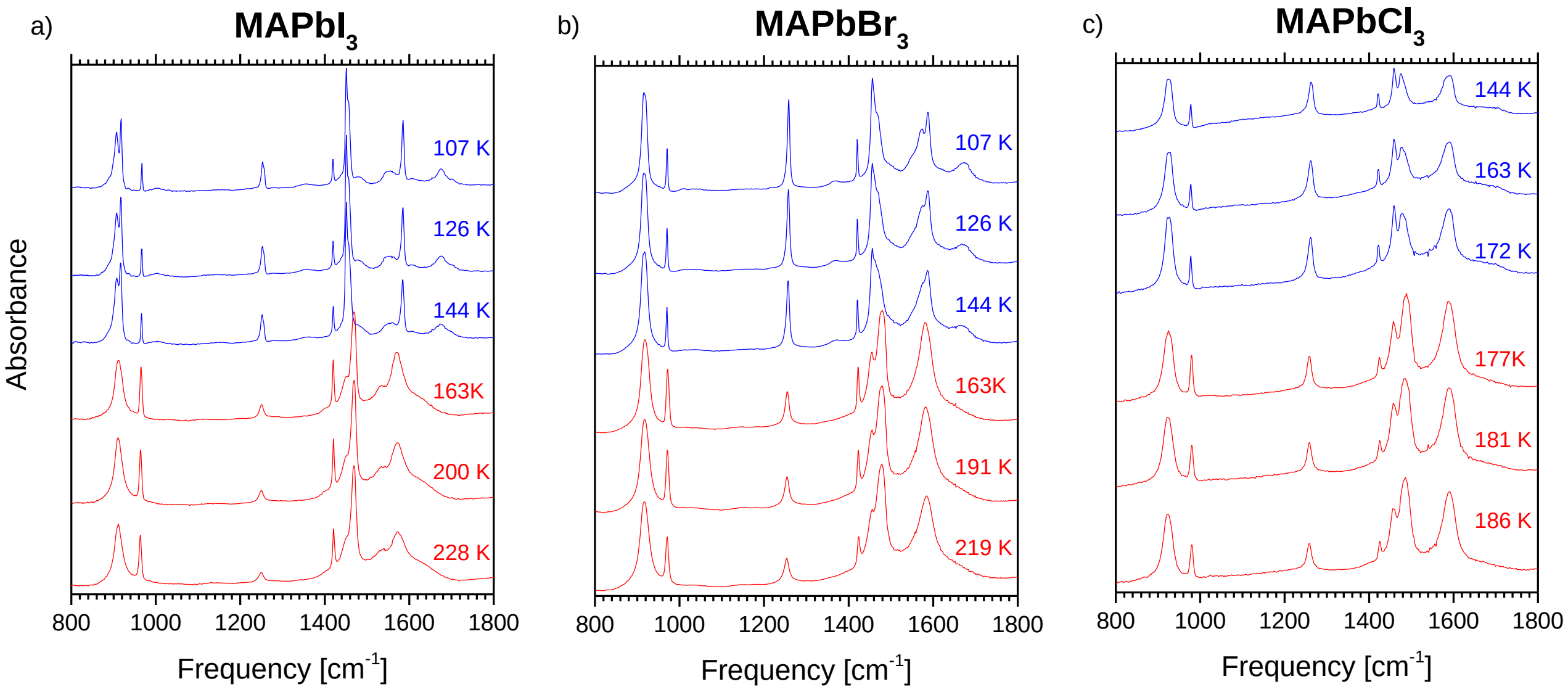

**FIGURE 2** Temperature-dependent infrared spectra of a) $MAPbI_3$, b) $MAPbBr_3$ and c) $MAPbCl_3$ between 800 $cm^{-1}$ and 1800 $cm^{-1}$

EXPERIMENTAL

## Crystal growth

$MAPbI_3$ single crystals were prepared according to the method reported by Baikie et al.[8]: yellow $PbI_2$ (Sigma-Aldrich, 99%) and methylammonium iodide (Sigma-Aldrich, 98%) was solved in a double-neck round-bottom flask with gamma-butyrolacton (GBL, Carl Roth, >99%). The solution was homogenized over night at 60°C under constant stirring. In the final step, the homogenized solution was put into a petri dish and the solvent was evaporated by heating it to 100 °C. $MAPbCl_3$ and $MAPbBr_3$ single crystals were prepared using the inverse temperature crystallization method.[24] Crystal growth was done in a glove box to avoid aging.

**X-ray diffraction**

Sample purity was proven by X-ray powder diffraction analysis. The measurement was performed by a Panalytical X'Pert Pro MRD powder diffractometer with sample spinner (Bragg-Brentano geometry, $\lambda$ = 1.5406 Å Cu K$\alpha$ with 40kV/30mA, step size 0.0032 and time per step 597.72 sec, 2$\theta$ from 11-81°). The analysis was done with HighScore Plus Version 3.05. No impurities were observed.

**FTIR microscope spectrometry**

Infrared spectra were recorded on a Bruker Vertex 80v FTIR spectrometer, equipped with a globar light source, a KBr beam-splitter and a Bruker Hyperion 2000 microscope using Cassegrainian objectives and an MCT Hg-Cd-Te detector. A detector nonlinear correction routine was applied using the OPUS software package (Bruker). A Linkam THMS600 stage was used for cooling under Argon atmosphere. The thermocouple was calibrated using melting points of different salts (melting points between 580 K and 1074 K) as described elsewhere.[50] The 0°C point (freezing of $H_2O$) was included in the calibration line. Based on the phase transition temperature of $MAPbI_3$ $T_t$ = 161 K (Tab. 1) the calibration of the cooling stage was corrected according to $T_{corr}$ [K] = 20.49965 + $T_{exp}$ * 0.9304 because the thermocouple was not calibrated in a suitable temperature range. Measurements were done in Argon atmosphere to avoid aging of the crystals (also: sample handling only in glove boxes and transportation only in mobile exsiccators). Between every temperature step there was a 5 minute temperature stabilization time so that the reproducibility of the temperature was around 0.1 K. To support the crystal fragments inside the Linkam THMS600 stage, a KBr disc was used that was also utilized to measure the reference spectra. The spectra were taken with an aperture size of 50 μm x 50 μm (for $MAPbBr_3$ and $MAPbI_3$) and 16 μm x 12 μm for $MAPbCl_3$.

Resolution of 2 $cm^{-1}$ was used. For each measurement, the spectrum was averaged over 256 scans. Temperature-dependent (107-349 K in 19 steps for $MAPbI_3$, 107-293 K in 9 steps for $MAPbBr_3$ and 126-293 K in 8 steps for $MAPbCl_3$) unpolarized transmission spectra on single crystal fragments with random orientation were measured between 400 $cm^{-1}$ and 4000 $cm^{-1}$. Dimensions of the measured crystal fragments was for $MAPbI_3$: 125 μm x 75 μm, thickness: 50 μm; for $MAPbCl_3$: 16 μm x 12 μm, thickness: 20 μm and for the $MAPBr_3$ crystal fragment: 100 μm x 75 μm, thickness: 22 μm.

COMPUTATIONAL DETAILS

**DFT calculations**

Calculations of IR frequencies and intensities were performed with the program CRYSTAL 14 from first principles using 3D-periodic density functional theory with Gaussian basis sets.[32] Vibrational frequencies were calculated at Γ-point from numerically computed second derivatives of the energy at a stationary point on the potential energy surface.[33] The basis sets of the atoms were selected as follows: for carbon, nitrogen, and hydrogen the Pople 6-311G** basis set was used.[34] For lead a pseudopotential ECP60MDF-6111(51d) was used, which has been successfully employed in perovskites before.[35] For these elements, which are present in all compounds, the exponents of the most diffuse shells have been set to common values taken from the optimization of $MAPbBr_3$. For chlorine a 86-311(1d),[36] for bromine a 976-311(51d),[37] and for iodine a 107663-111(631d) expansion were used.[37] Attempts using additional sp or d shells did not result in increased quality of the results, but in instability of the calculations. Introducing semiempirical dispersion correction terms also turned out to be not beneficial for the quality of the calculations.[38] For all basis sets the exponents of the most diffuse shells have been optimized to the following values: s(H) = 0.3172, 0.1, p(H) = 1.0874, sp(C) = 0.4547, 0.1209, d(C) = 0.8382, sp(N) = 0.6586, 0.2059, d(N) = 0.8616, sp(Cl) = 0.3162, 0.1238, d(Cl) = 0.3773, sp(Br) = 0.5837, 0.1831, d(Br) = 0.9570, sp(I) = 0.5212, 0.1621, d(I) = 0.4197, sp(Pb) = 1.2373, 0.4305, 0.1399, d(Pb) = 0.2242. For the level of numerical accuracy, a Pack-Monkhorst k net with 8x8x8 points in the Brillouin zone was used, a (99,1454) unpruned grid for the numerical integration of the DFT exchange-correlation contribution, overlap truncation limits for Coulomb and exchange sums of 10-8 (10-16 for n summation in the exchange series), and convergence thresholds of the SCF energy of 10−10 Eh for structure optimization and 10−11 Eh for frequency calculation. For the Hamiltonian the PBE0 functional was used.[39] Groupwise order of the mode frequencies and relative intensities allowed for a clear assignment of calculated to observed modes. From this a common correction factor of 0.9586 was obtained, which was applied to the calculated frequencies (tables S2-S4). It is known[40] that this systematic error in the calculated vibrational frequencies is the result of a too high Hartree-Fock contribution to the hybrid HF/DFT exchange-correlation term of 1/4 in the PBE0 Hamiltonian. The remaining deviations between observed and calculated IR frequencies were reasonable small, with standard deviation 10 $cm^{-1}$.

Following the symmetry analysis of space group *Pnma*, modes of symmetry $B_{1u}$ (22 modes), $B_{2u}$ (17), $B_{3u}$ (22) are generally IR-active; one each of these being acoustic modes. (Raman-active modes are $A_g$ (19), $B_{1g}$ (14), $B_{2g}$ (19), $B_{3g}$ (14), while modes with symmetry $A_u$ (17) are inactive. Compared with factor group analysis and other sources,[31] as well as the assignment herein, we found that within the symmetry analysis of CRYSTAL $B_{2u}$ and $B_{3u}$ as well as $B_{2g}$ and $B_{3g}$ are interchanged. DFT calculations for the tetragonal and cubic phase were not performed because of the disorder in these structures. The DFT algorithm works with complete atoms only, not splitted ones. Even extremely costly calculation would be of very limited validity given the selection rules in the presence of disorder in crystals.

Calculations for the low temperature phase of the chlorine compound were also carried out in the same unit cell as for the other two, although it is not the correct one of $MAPbCl_3$. However, this approach allows comparing the changes in the spectrum between the various compounds. The correct crystal structure,[11] with its twice as large unit cell, gives rise to twice as many vibrational modes, too, making assignment and meaningful interpretation more complex.

**Peak fitting**

To evaluate frequency, integrated absorbance, and half-widths at half maximum (HWHM) of the IR bands, Fityk 0.9.8 software[25] was used to fit the spectra with Lorentzian curves after a linear line background correction (only for $MAPbCl_3$ data in 1300-1750 $cm^{-1}$ frequency range constant background correction was used).

FACTOR GROUP ANALYSIS IN THE ORTHORHOMBIC PHASE

Starting from the ideal point group symmetry $C_{3v}$ for the isolated MA molecule (Fig. 3), it is possible to describe the 18 internal vibrations of MA (in general, a molecule with N atoms has 3N – 6 normal modes of vibration) by using the symmetry species $A_1$ (five modes), $A_2$ (one mode) and E (six doubly degenerated modes) which can be correlated to 12 fundamental frequencies ($\nu_1$ - $\nu_{12}$).[12] These fundamental frequencies $\nu_1$ - $\nu_{12}$ was assigned by Sandorfy and co-workers to MA stretching (two symmetric $A_1$ modes: $\nu_1$ for $NH_3$ and $\nu_2$ for $CH_3$ and two asymmetric double degenerated E modes: $\nu_7$ for $NH_3$ and $\nu_8$ for $CH_3$), MA bending (two symmetric $A_1$ modes: $\nu_3$ for $NH_3$ and $\nu_4$ for $CH_3$ and two asymmetric double degenerated E modes: $\nu_9$ for $NH_3$ and $\nu_{10}$ for $CH_3$), MA C-N stretching (one $A_1$ mode: $\nu_5$), two MA rocking modes (two double degenerated E modes: $\nu_{11}$ for $NH_3$ and $\nu_{12}$ for $CH_3$) and one MA torsional oscillation $A_2$ mode ($\nu_6$) at low frequencies.[13]

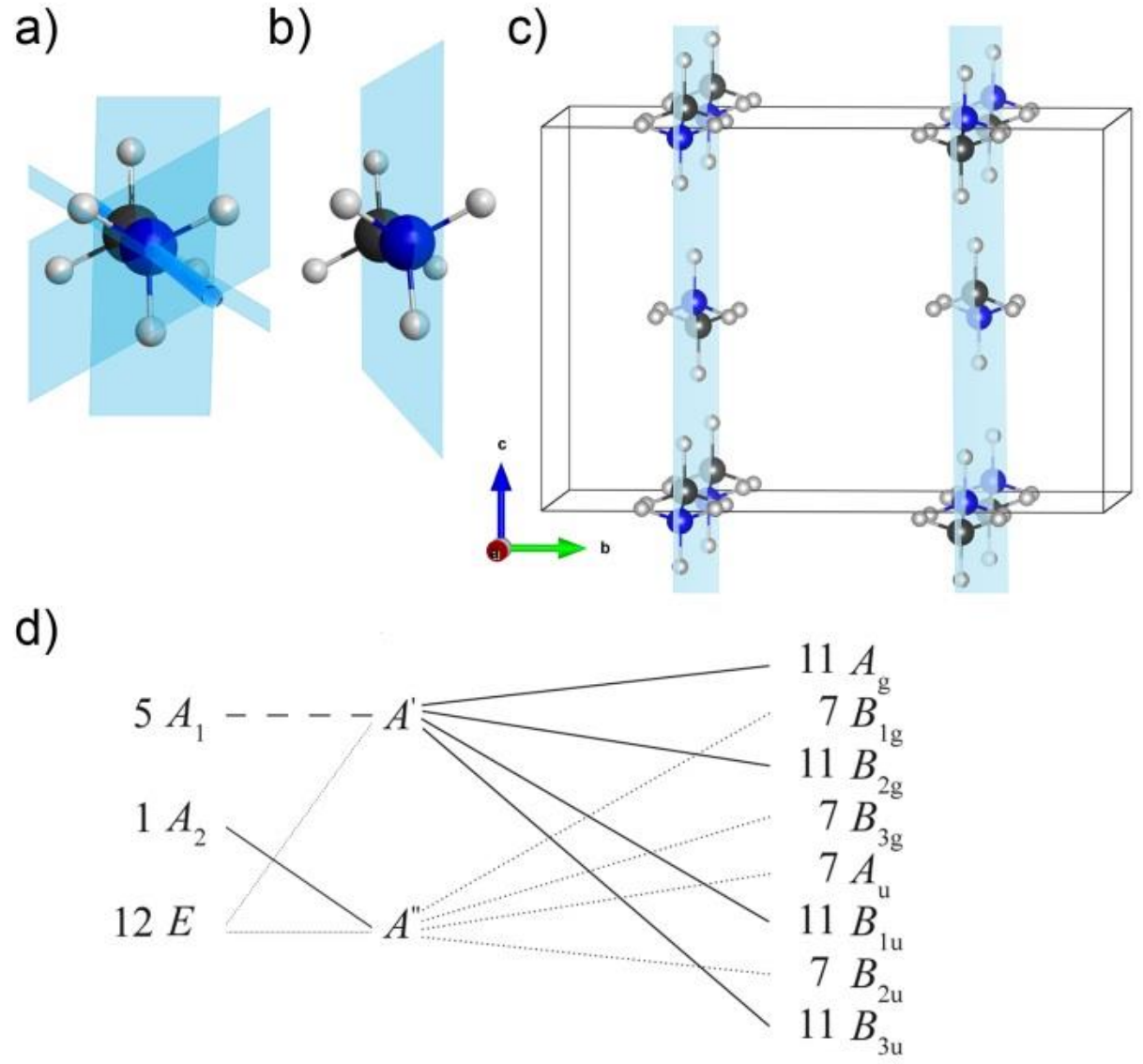

**FIGURE 3** Symmetry elements of the MA molecule a) ideal, $C_{3v}$ high symmetry, b) MA molecule with symmetry reduced $C_s$ symmetry, c) the four symmetry reduced MA molecules in the orthorhombic $MAPbI_3$ crystal structure, drawn in blue color are the *Pnma* mirror planes (0,1/4,0) and (0,3/4,0), d) correlation of the ideal, $C_{3v}$ high symmetry with the site symmetry $C_s$ and the factor group $D_{2h}$.

The 144 vibrational modes of the orthorhombic (*Pnma* space group with $D_{2h}$ factor group symmetry) phase of $MAPbX_3$ can be assigned through common databases,[31] whereas the differentiation between lattice vibrational modes that are caused by the $PbX_3$ cage and modes that are caused by internal vibrations of MA is only possible by means of factor group analysis.[41] The correlation technique[42] is used to predict number and type of intramolecular vibrations (Table S1 a-g). For the MA molecule, the site symmetry $C_s$ (m) had to be selected (using the $C_{3v}$ point group, as used in earlier attempts,[15] is not possible because this group is not part of the correlation tables for $D_{2h}$ factor group symmetry).

The crystallographic unit cell of $MAPbI_3$ and $MAPbBr_3$ in the orthorhombic phase contain four primitive Bravais cells so that the fundamental frequencies $5A_1 + 1A_2 + 6E$ of one MA molecule with $C_{3v}$ symmetry transform to intramolecular vibrations $5(A_g + B_{2g} + B_{1u} + B_{3u}) + 1(A_u + B_{1g} + B_{3g} + B_{2u}) + 6(A_g + B_{1g} + B_{2g} + B_{3g} + A_u + B_{1u} + B_{2u} + B_{3u})$ for four MA molecules with $C_s$ symmetry (Fig. 3). Because of cell doubling, one would have to consider twice as many vibrational modes for $MAPbCl_3$ (all irreducible representations in Table S1 a-f have to be doubled). For the chloride it follows that it is possible to distinguish the two types of MA molecules that are present in the orthorhombic low-temperature crystal structure. In the chloride, we have to consider the following intramolecular vibrations of the MA molecules: $2*5(A_g + B_{2g} + B_{1u} + B_{3u}) + 2*1(A_u + B_{1g} + B_{3g} + B_{2u}) + 2*6(A_g + B_{1g} + B_{2g} + B_{3g} + A_u + B_{1u} + B_{2u} + B_{3u})$. It follows that for the fundamental frequency $\nu_6$ (ideal infrared inactive MA molecule symmetry $A_2$), only one infrared-active mode $B_{2u}$ appears, but with a frequency outside our measured frequency range ([14]: IR-active mode: 350 $cm^{-1}$). The fundamental frequencies $\nu_1$ to $\nu_5$ (ideal infrared active MA molecule symmetry $A_1$) split into two infrared active modes $B_{1u}$ and $B_{3u}$. Three of these fundamental frequencies were investigated ($\nu_5$, $\nu_4$ and $\nu_3$). The fundamental frequencies $\nu_7$ to $\nu_{12}$ (ideal infrared active MA molecule symmetry E) split into three infrared active modes $B_{1u}$, $B_{2u}$ and $B_{3u}$. Four of these fundamental frequencies were investigated ($\nu_{12}$, $\nu_{11}$, $\nu_{10}$ and $\nu_9$).

RESULTS

The temperature-dependent analysis focused on the fundamental frequencies of CH / NH rocking ($\nu_{12}$ and $\nu_{11}$), C-N stretching mode ($\nu_5$) and CH / NH bending modes ($\nu_4$, $\nu_{10}$, $\nu_3$ and $\nu_9$) of the MA molecule in the 800-1750 $cm^{-1}$ frequency range (Fig. 2). Three frequency ranges with similar fitting models were used for data evaluation, in order to get comparable spectral information in the complete temperature range for all three halides (the same fitting procedures and band assignments were used for the chloride). In the frequency range 1300-1750 $cm^{-1}$ vibrational modes were observed that are most probably attributable to combinations of MA intramolecular vibrations with low frequency libration modes that have been identified around 100 $cm^{-1}$ in inelastic neutron scattering investigations[47,48] (these combinations have been observed but not assigned in previous publications[14,21,23] as shown in Fig. S2). The spectral features of the investigated vibrational modes are summarized for low temperatures in Tables S2-S4. Strong absorbance caused by relatively thick samples lead to slightly overexposed, somewhat broad vibrational modes in the 3100-3200 $cm^{-1}$ frequency range, especially for the bromide and the chloride spectra (Fig. S3). Comparison with literature data shows that this effect is less pronounced for the iodide (Fig. S4). Nevertheless, it was possible to analyze the CH and NH stretching vibrational modes $\nu_1$, $\nu_2$, $\nu_7$, and $\nu_8$ with a reduced completeness compared to the analysis of the vibrational modes in the 800-1750 $cm^{-1}$ frequency range. Therefore, only limited results for the spectral features of the CH and NH stretching vibrational modes are presented in Tables S2-S4**.**

**Part I: $\nu_{12}$ and $\nu_5$**

The first part, the 820-1000 $cm^{-1}$ frequency range, included the fundamental frequency $\nu_{12}$ (ideal MA molecule symmetry E) and the fundamental frequency $\nu_5$ (ideal MA molecule symmetry $A_1$). For spectra of the cubic/tetragonal phase, the two fundamental frequencies were modelled with two Lorentzians and a linear background term (Fig. 4 a, S5 a and b). Four Lorentzians plus a linear background term were utilized for the spectra of the orthorhombic phase (Fig. 4 b, S5 c and d). The additional Lorentzians in these spectra were needed because of the fundamental frequency $\nu_{12}$ degeneration into the IR active modes $B_{1u}$, $B_{2u}$, and $B_{3u}$. In the 173 K spectra of $MAPbCl_3$, only three Lorentzians were applied. For the fundamental frequency $\nu_5$, a narrowing towards low temperatures was observed, but this narrowing is not general for all $\nu_{12}$ Lorentzians (Fig. S6 a, and c). The low frequency Lorentzians cl1, i1, and br1 are broader than the other Lorentzians (cl2, cl3, i2, i3, br2, and br3) of the $\nu_{12}$ fundamental. In contrast to the broader $\nu_{12}$ Lorentzians (cl1, cl2, i1, i2, br1, and br3), fundamental frequency $\nu_5$ and the narrow $\nu_{12}$ Lorentzians (cl3, i3, and br3) showed an overall redshift with decreasing temperature for all halides (Fig. 5). However, near the phase transition, discontinuity diverse behavior in the $\nu_5$ frequency evolution could be observed. In the spectra of $MAPbBr_3$ and $MAPbCl_3$, $\nu_5$ jump-like blueshifts with decreasing temperature could be observed and in the spectra of $MAPbI_3$, $\nu_5$ jump-like redshifts (Fig. 5 b).

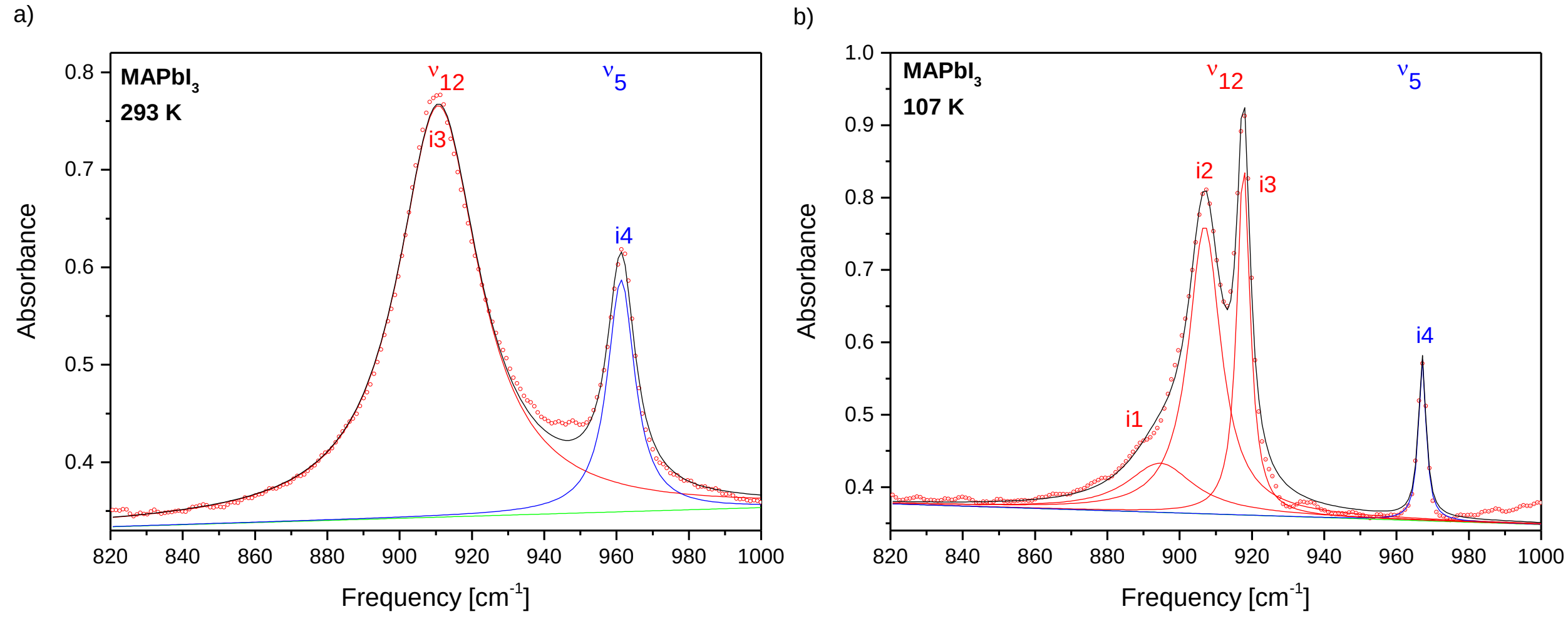

**FIGURE 4** Infrared spectra of $MAPbI_3$ between 820 $cm^{-1}$ and 1000 $cm^{-1}$ at a) 293 K (tetragonal/cubic) and at b) 107 K (orthorhombic phase); measurements of small single crystals using a FTIR-microscope spectrometer; fit of the measured spectra (red triangles) with a) two Lorentzian functions in the tetragonal/cubic phase and b) four Lorentzian functions in the orthorhombic phase, as well as with a linear background function (green); the solid black line corresponds to the summed-up Lorentzian functions plotted in colors. In red color: $CH_3$ rocking mode $\nu_{12}$; in blue color: C-N stretching mode $\nu_5$.

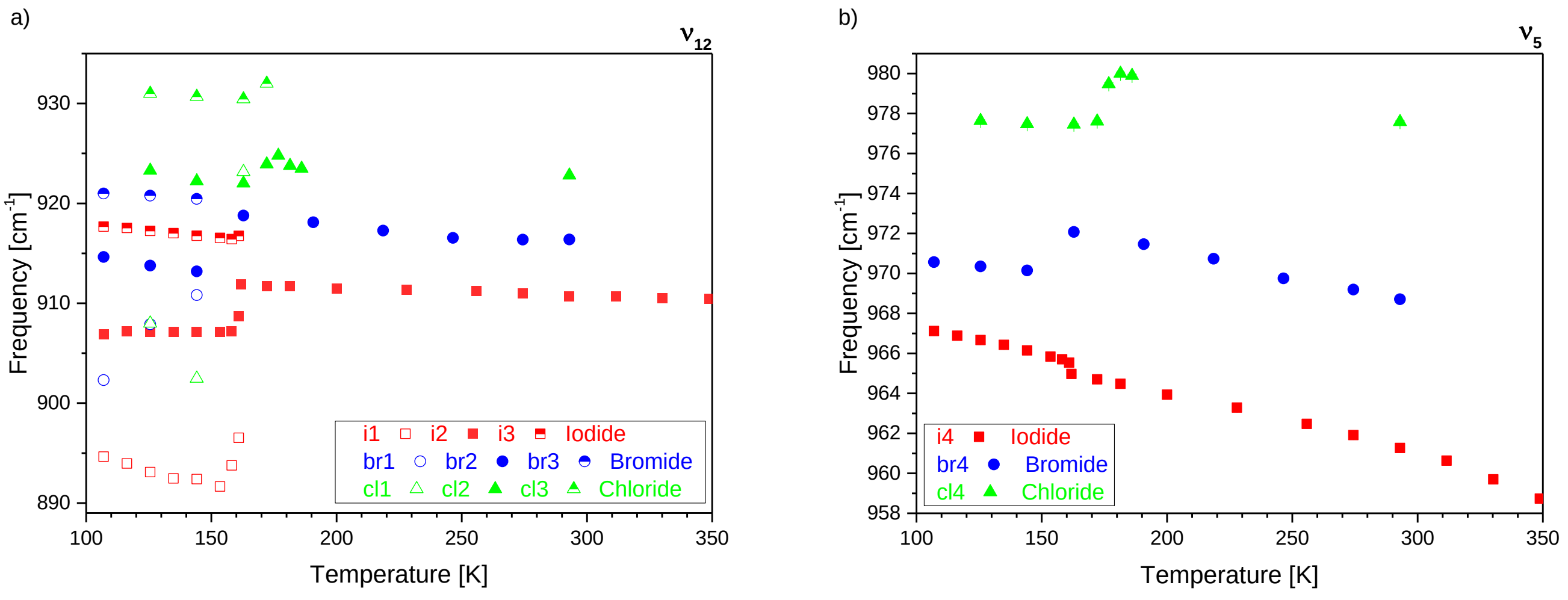

**FIGURE 5** Temperature-dependence of frequencies of a) $CH_3$ rocking mode $\nu_{12}$ and b) C-N stretching mode $\nu_5$. In red color: $MAPbI_3$, in blue color: $MAPbBr_3$ and in green color: $MAPbCl_3$.

## Part II: $\nu_{11}$

In the frequency range 1230-1280 cm$^{-1}$, the fundamental frequency $\nu_{11}$ (ideal MA molecule symmetry E) was observed. Spectra of the cubic/tetragonal phase were modelled with one Lorentzian plus a linear background term. Spectra of the orthorhombic phase were modelled with an additional Lorentzian for $MAPbI_3$ and two additional Lorentzians for $MAPbCl_3$ (beside 173 K data, where only one additional Lorentzian was used) and for $MAPbBr_3$ (Fig. 6 and S7). The frequency and HWHM seemed to behave fairly similar to the fundamental $\nu_{12}$ (Fig. S8 a-b), while it could be observed that the summed integrated absorbance of this $\nu_{11}$ fundamental mode increases remarkably on decreasing temperatures for all halides (Fig. S8 c).

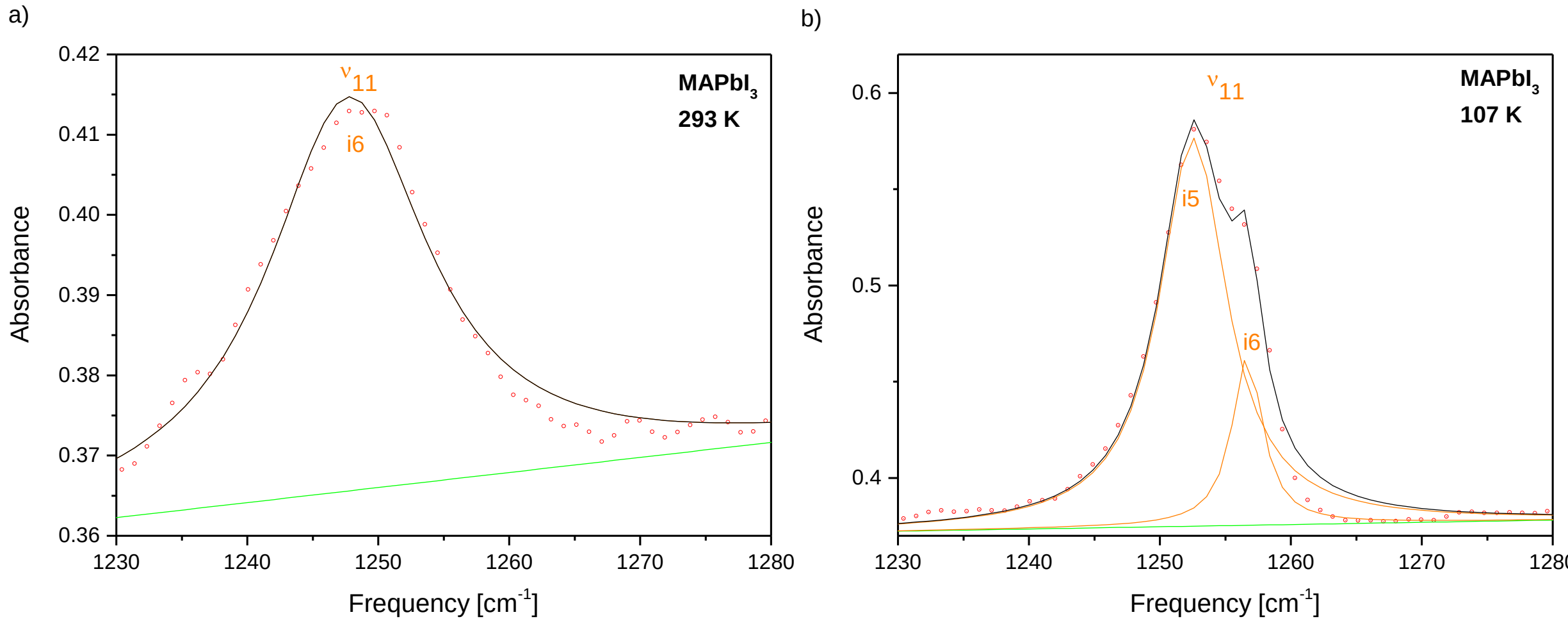

**FIGURE 6** Infrared spectra of $MAPbI_3$ between 1230 cm$^{-1}$ and 1280 cm$^{-1}$ at a) 293 K (tetragonal/cubic) and at b) 107 K (orthorhombic phase); measurements of small single crystals using a FTIR-microscope spectrometer; fit of the measured spectra (red triangles) with a) one Lorentzian function in the tetragonal/cubic phase and b) two Lorentzian functions in the orthorhombic phase, as well as with a linear background function (green); the solid black line corresponds to the summed-up Lorentzian functions plotted in colors. In orange color: $NH_3$ rocking mode $\nu_{11}$.

## Part III: $\nu_4$, $\nu_{10}$, $\nu_3$, and $\nu_9$

In the spectra of the tetragonal/cubic phase, the four fundamentals in the frequency range 1300-1750 cm$^{-1}$ were modelled with seven Lorentzians. A linear background term was used for $MAPbI_3$ and $MAPbBr_3$ spectra, whereas for the spectra of $MAPbCl_3$, a constant term for the background treatment was utilized (Fig. 7 a, and S9 a, c, e). For the spectra of the orthorhombic phase (Fig. 7 b, and S9 b, d, f), the modelling was more diverse: ten Lorentzians were used for $MAPbI_3$ (linear background), eleven for $MAPbBr_3$ (linear background) and twelve for $MAPbCl_3$ (constant background).

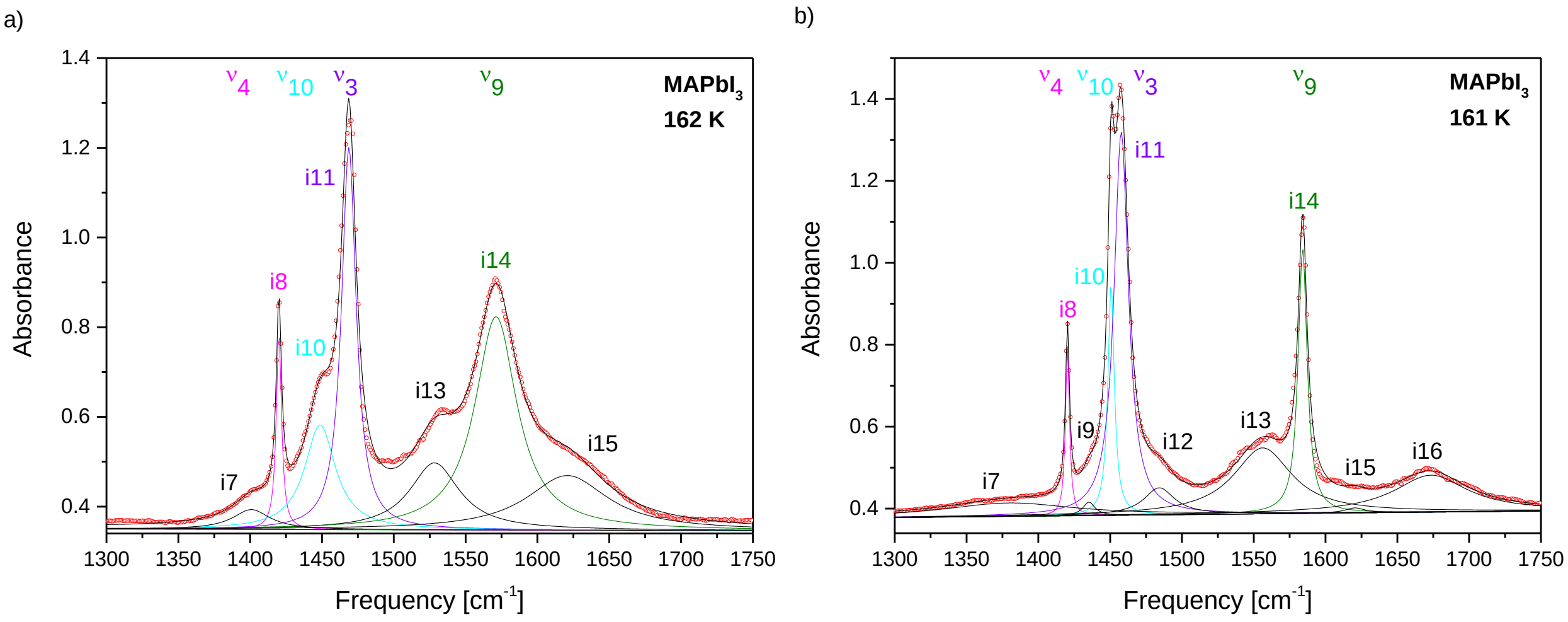

**FIGURE 7** Infrared spectra of a), b) $MAPbI_3$ between 1300 cm$^{-1}$ and 1750 cm$^{-1}$ at a) 162 K (tetragonal phase) and b) 161 K (orthorhombic phase); measurements of small single crystals using a FTIR-microscope spectrometer; fit of the measured spectra (red triangles) with b) ten Lorentzian functions in the orthorhombic phase and a) seven Lorentzian functions in the tetragonal phase, as well as with a linear background function (green); the solid black line corresponds to the summed-up Lorentzian functions plotted in colors. Magenta: symmetric $CH_3$ bending mode $\nu_4$; cyan: asymmetric $CH_3$ bending mode $\nu_{10}$; violet: symmetric $NH_3$ bending mode $\nu_3$ and in dark green color: asymmetric $NH_3$ bending mode $\nu_9$. In black color plotted Lorentzian functions correspond to modes that could not clearly assigned to MA vibrational modes. Note that the changes in the vibrational spectra observed here are completely reversible.

The assignment of the fundamental frequency $\nu_4$ (ideal MA molecule symmetry $A_1$) was straightforward, since no splitting or overlapping with other modes was observed ($MAPbI_3$: Lorentzian i8; $MAPbBr_3$: br9 and $MAPbCl_3$: cl9). The fundamental frequency $\nu_4$ showed an overall blueshift with decreasing temperature for all halides. However, a frequency discontinuity similar to $\nu_5$ could be observed near the phase transition (Fig. 8 a, and S10 a, b). The Lorentzians i7, br8, and cl8 between 1350-1400 $cm^{-1}$, that show more or less constant HWHM of about 30 $cm^{-1}$ and also remarkable changes in the integrated absorbance close to the phase transition, are tentatively assigned to combination modes.

The separation of the fundamental frequency $\nu_{10}$ (ideal MA molecule symmetry E) and $\nu_3$ (ideal MA molecule symmetry $A_1$) in the spectra of the orthorhombic phase was challenging, because these two fundamentals were quite close in frequency. Moreover, additional bands occur because of possible degeneration of the $\nu_{10}$ fundamental. For the 293 K spectra (Fig. S9 a, c, e), $\nu_{10}$ was assigned to a low frequency shoulder that could be observed for all halides (Lorentzian i10 at 1451 $cm^{-1}$ for $MAPbI_3$, br10 at 1455 $cm^{-1}$ for $MAPbBr_3$, and cl12 at 1455 $cm^{-1}$ for $MAPbCl_3$) and $\nu_3$ was assigned to the main peak at higher frequencies (Lorentzians i11 at 1469 $cm^{-1}$ for $MAPbI_3$, br12 at 1478 $cm^{-1}$ for $MAPbBr_3$, and cl15 at 1485 $cm^{-1}$ for $MAPbCl_3$). In recent publications[14-17,21] the assignment of the fundamental frequency $\nu_{10}$ did not show a consistent picture, mainly because the shoulder at around 1455 $cm^{-1}$ was ignored. Visual comparison of our data with published results clearly show the appearance of $\nu_{10}$ in all reported spectra at the frequencies proposed here (Fig. S2 a).

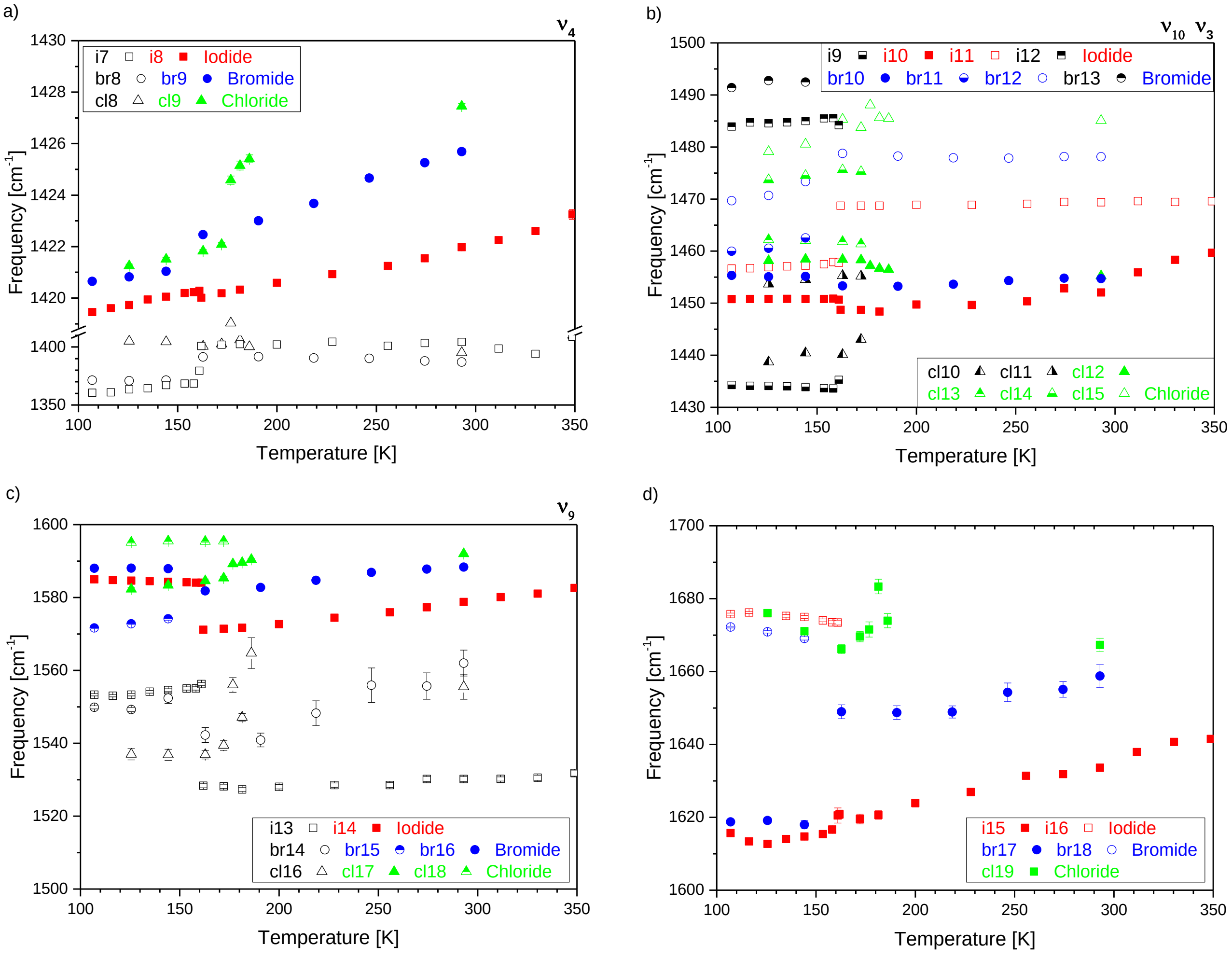

**FIGURE 8** Temperature-dependence of frequencies of symmetric $CH_3$ bending mode $\nu_4$ (a); asymmetric $CH_3$ bending mode $\nu_{10}$ and symmetric $NH_3$ bending mode $\nu_3$ (b); asymmetric $NH_3$ bending mode $\nu_9$ (c), and combinations (d). In red color: $MAPbI_3$, in blue color: $MAPbBr_3$ and in green color: $MAPbCl_3$. For most data points no error bars are provided for simplicity.

In the spectra of the orthorhombic phase, the deconvolution of $\nu_{10}$ and $\nu_3$ is more difficult. In the spectra of $MAPbCl_3$ both fundamentals showed splitting. The fundamental $\nu_{10}$ split into the Lorentzian cl12 and cl13 and the fundamental $\nu_3$ split into Lorentzian cl15 and cl14 (Fig. S9 f). Additionally, two low frequency components (cl10 and cl11) occurred that are tentatively assigned to combinational modes. Compared to the chloride, more drastic changes were observed for the $\nu_3$ fundamental in $MAPbBr_3$, whereby Lorentzian br12 redshifts of around 8 $cm^{-1}$ with decreasing temperatures (Fig. S9 d). The additional broad Lorentzian br13 cannot be clearly assigned to the $\nu_3$ fundamental and therefore also were assigned to a combination mode. The $\nu_{10}$ mode in $MAPbBr_3$ splits into the two components br10 and the additional Lorentzian br11. It was observed that br10 blueshifts into the direction of br12, so that there was a frequency difference of only about 10 $cm^{-1}$ between $\nu_{10}$ and $\nu_3$ in the low-temperature phase of the bromide. In the low-temperature spectra of $MAPbI_3$ (Fig. 7 b and S9 b), a frequency difference of 5.9 $cm^{-1}$ between the two modes $\nu_{10}$ and $\nu_3$ (Lorentzians i10 and i11) was observed (Fig. 8 b), accompanied with broad additional combinational modes on both sides of the $\nu_{10}$ / $\nu_3$ peak (i9 and i12). HWHM and integrated absorbance of Lorentzians that were assigned to the modes $\nu_{10}$ and $\nu_3$ decreased towards low temperatures for all halides (Fig. S10 c and d). The $\nu_3$ $B_{1u}/B_{3u}$ splitting was only observed for the chloride (5.8 $cm^{-1}$), whereby the DFT calculations suggest a splitting of 7.6 $cm^{-1}$ for the chloride, of 6.9 $cm^{-1}$ for the bromide, and 1.1 $cm^{-1}$ for the iodide (Tables S2-4). Théorêt and coworkers attributed the $B_{1u}/B_{3u}$ splitting of $\nu_3$ to correlation field effects.[27] However, based on the factor group analysis for the chloride, it is more plausible that the splitting is caused due to the cell doubling.

In the frequency range between $\nu_3$ and $\nu_9$, broad features (supplementary to i12 and br13) were observed for all temperatures in all halides that are tentatively assigned to combination modes ($MAPbI_3$: Lorentzian i13; $MAPbBr_3$: Lorentzian br14; $MAPbCl_3$: Lorentzian cl16). The Lorentzian i13 showed drastic changes around the orthorhombic-tetragonal phase transition in frequency and integrated absorbance, whereas the HWHM decreased in a more continuous manner towards low temperatures (Fig. 8 c and S11 a and b). The changes around the phase transition were less pronounced for the chloride and the bromide.

One Lorentzian was utilized for the fundamental frequency $\nu_9$ (ideal MA molecule symmetry E) for the room temperature spectra ($MAPbI_3$: i14; $MAPbBr_3$: Lorentzian br16; $MAPbCl_3$: cl17). For spectra in the orthorhombic phase of $MAPbBr_3$ and $MAPbCl_3$, an additional Lorentzian was assigned to the fundamental frequency $\nu_9$ due to splitting (br15, cl18). For the orthorhombic spectra of $MAPbI_3$, an assignment of additional spectral components due to splitting was less unambiguous, so that only i14 was assigned for the fundamental $\nu_9$ in the iodide. Fits for the orthorhombic 161 K and the tetragonal 162 K of the iodide are shown in figure 7 a and 7 b, to visualize more clearly that no splitting of i14 was observed. The same figure also shows the appearing of the Lorentzian i16 in the orthorhombic phase at 161 K, whereas the Lorentzian i15 that was present at temperatures above the orthorhombic-tetragonal phase transition in $MAPbI_3$ shows decreased integrated intensity in the low temperature phase (Fig. 8 d and S11 c and d). A quite similar behavior could be observed for $MAPbBr_3$ (Lorentzians br17 and br18), whereas for $MAPbCl_3$ data only one broad Lorentzian cl19 was present.

The temperature-dependent spectral features of all combination modes were investigated because they are a non-negligible integrated absorbance fraction in the 1300-1750 $cm^{-1}$ frequency range. All integrated areas of the Lorentzians of the combinational modes compromise about 50 % of the integrated absorbance for the analyzed frequency range (Fig. S12 a-c). When discussing the temperature-dependence of spectral features the combination modes have to be taken into account, because ignoring them would result in wrong spectral parameters for the MA fundamental frequencies. This is especially true when the temperature dependence of the spectral width of vibrational modes is analyzed and conclusions about the resulting activation energies are drawn.

**Part IV: $\nu_2$, $\nu_8$, $\nu_1$, and $\nu_7$**

Because of the broad features in the 2600-3600 $cm^{-1}$ frequency range that are caused by the intensive N-H stretching modes, an abridged description for the vibrational modes in this frequency range was used for modelling the fundamental frequencies $\nu_2$, $\nu_8$, $\nu_1$, and $\nu_7$. The approach was to utilize a simplified but consistent model for all three halides (Fig. S13). Only C-H and N-H stretching modes are included into the model, combinations are excluded in order to reduce the fitting parameter. The weak bands $\nu_2$ (ideal MA molecule symmetry $A_1$) and $\nu_8$ (ideal MA molecule symmetry E) of the C-H stretching modes were modelled first, in a narrow frequency range ($\nu_2$: 2940-2980 $cm^{-1}$ and $\nu_8$: 2980-3060 $cm^{-1}$), with one Lorentzian for $\nu_2$ and two Lorentzians for $\nu_8$ each with a linear background function. Because of the intense N-H stretching vibrational modes in the chloride and the bromide it was not possible to get useful results for $\nu_8$ in the cubic/tetragonal phase (Fig. S14 b). The refined values of $\nu_2$ and $\nu_8$ are included, as fixed values, in the fitting model for the broad vibrational modes of the $\nu_1$ (ideal MA molecule symmetry $A_1$) and $\nu_7$ (ideal MA molecule symmetry E) N-H stretching modes. The final model of the frequency range $2600\text{-}3600^{-1}$ (Fig. S13) consisted of two fitted Lorentzians for $\nu_1$ and $\nu_7$ and the fixed values of the Lorentzians for $\nu_2$ and $\nu_8$ that where utilized to model the C-H and N-H stretching modes. Because of the abridged character of the model for the description of the C-H and N-H stretching vibrational modes only the temperature-dependence of the frequencies is shown (Fig. S14).

DISCUSSION

The temperature dependence of the analyzed vibrational spectra exhibit clear discontinuous behavior at the orthorhombic-tetragonal phase transition for all compounds. Jumps in frequency, HWHM, and integrated absorbance were observed within a quite narrow temperature range in all three halides (Fig. 5, 7, 8, S6, S8, S10, S11 and S14). The band splitting in IR absorbance and the temperature dependence of the investigated vibrational modes were determined by peak fitting the measured spectra, enabling us to assign the vibrational modes, based on the factor group analysis and the DFT calculations.

Prior investigations on the discontinuous behavior were restricted to $MAPbI_3$ samples[14,17] or limited to Raman spectroscopy[17,19] so that the results here allow for the first time a comparative study of the temperature-dependence of MA lead halides from the IR spectroscopic and crystallographic perspective. The aim of this study is to understand how the apparent drastic changes of the investigated fundamental modes of MA (band splitting; jumps in frequency, HWHM and integrated absorbance) are interconnected with structural changes of the crystal structure during the orthorhombic-tetragonal phase transition.

The crystallographic data[8-11,26] for the orthorhombic *Pnma* space group symmetry show the presence of 2-dimensional hydrogen bond layers in the three MA lead halides (Fig. 1 and S1), which is confirmed by our DFT calculations. Comparing available N-H...X and H...X bond lengths of MA lead halides (Table S5) with literature data indicate that the hydrogen bonds in the halides are quite strong.[43,44] The linearity of the N-H...X bond is an additional indicator for the strength of the hydrogen bond character.[44] Analyzing these orthorhombic 2-dimensional layers of the MA lead iodide and bromide shows that the nitrogen atom of the MA molecule is connected to the halide atom through a short H...X bond that is positioned in a mirror plane (Table S5). It is evident that these mirror planes, (0,1/4,0) and (0,3/4,0), are essential symmetry elements because they structurally interconnect the MA molecule (with $C_s$ molecule symmetry) with the inorganic host structure in all orthorhombic halides (Fig. 3 and S15).

Analyzing the vibrational mode geometries of the DFT results, we identified in-plane ($B_{3u}$ and $B_{1u}$) and out-of-plane ($B_{2u}$) movements relative to the orthorhombic mirror planes in vibrational modes of the fundamental frequencies $\nu_{12}$, $\nu_{11}$, $\nu_{10}$, $\nu_9$, $\nu_8$, and $\nu_7$ (Fig. 9, S16, S17, and S18). In the DFT results of $MAPbI_3$, the out-of-plane ($B_{2u}$) modes consistently showed the lowest frequency for the investigated fundamental frequencies $\nu_{12}$, $\nu_{11}$, $\nu_{10}$, and $\nu_9$ (Table S3; for example $\nu_{12}$: out-of-plane ($B_{2u}$) DFT mode 77 at 912.5 $cm^{-1}$ and the both in-plane modes 82 ($B_{3u}$) and 84 ($B_{1u}$) at 918 $cm^{-1}$ and 920.7 $cm^{-1}$). The other two MA lead halides behaved differently because the out-of-plane mode showed mostly higher frequencies than the in-plane modes. Our DFT results and experimental data showed the absence of a consistent correlation between mode type (in-plane or out-of-plane) and HWHM (correlation times) for all halides. The $\nu_{12}$ $B_{2u}/B_{3u}$ splitting of 5.5 $cm^{-1}$ in $MAPbI_3$, predicted from our DFT calculations, suggests an observable splitting (based on the resolution of 2 $cm^{-1}$), and indeed it was possible to fit all three modes in the experimental spectra (i1, i2, and i3, Fig. 4 b). Only for $\nu_9$, all three halides showed the same frequency ordering of the three splitting modes: $B_{2u}$ -> $B_{3u}$ -> $B_{1u}$ but with quite different splitting (Br: 26.5 $cm^{-1}$, Cl: 24.7 $cm^{-1}$, I: 2.6 $cm^{-1}$). Ivanovska and coworkers[14] stated that the apparent overall blueshift with decreasing temperature in several vibrational modes indicates strong anharmonic contributions to the temperature-dependent frequency shifts of the investigated normal modes of MA (clearly observed in the fundamental frequencies $\nu_4$ and $\nu_3$). Our investigations show that these anharmonic contributions have to be considered for all three halides. The absolute values of classical amplitudes from DFT calculations clearly showed that the fundamental frequencies $\nu_4$, $\nu_{10}$, $\nu_3$, $\nu_9$, $\nu_2$, $\nu_8$, $\nu_1$, and $\nu_7$ were influenced either by C or CH atom movements ( $\nu_4$, $\nu_{10}$, $\nu_2$, and $\nu_8$) or by N or NH atom movements ($\nu_3$, $\nu_9$, $\nu_1$, and $\nu_7$) whereas the other analyzed modes ($\nu_{12}$, $\nu_5$, and $\nu_{11}$) were more or less equally influenced by C/CH and N/NH movements (Fig. S19). Most equally influenced by C/CH (50.6 %) and N/NH (49.4 %) vibrational movements is the $\nu_5$ mode, where a visualization of the vibrational movements clearly shows (Fig 10) that this vibrational mode is nearly identical to pure C-N stretching, showing that the reversed frequency behavior near the orthorhombic/tetragonal phase transition (blueshift of bromide/chloride and redshift of iodide) is most probably related to inversely arranged changes of the strength of the C-N bonding that is caused by the MA molecule reorientations. In vibrational modes that are not equally influenced by C/CH and N/NH vibrational movements a more diverse behavior of the spectral features can be observed. The $B_{2u}$ mode of the fundamental frequency $\nu_9$ ($MAPbI_3$ DFT mode 116) for example showed that 91.2% of the vibrational movement are caused by nitrogen and nitrogen bonding hydrogen atoms. At the same time, it was observed that N/NH dominated fundamental frequencies $\nu_3$ and $\nu_9$ show the largest discontinuous behaviour of the spectral features in the investigated vibrational modes. Upon crossing to the orthorhombic phase, the symmetrical $NH_3$ bending mode $\nu_3$ of $MAPbI_3$ shifted remarkably toward lower frequencies while the asymmetric one $\nu_9$ moved to higher frequencies (Fig. 8 b and c), similar to MA halides.[27] This behaviour was not observed in $MAPbCl_3$ and was less pronounced in $MAPbBr_3$. These $\nu_3$ and $\nu_9$ frequency changes near the phase transition in the iodide are most probably correlated with the change from three strong hydrogen bond interactions in the tetragonal phase to one strong hydrogen bond interaction in the orthorhombic phase.[7] We can also state that, generally, the hydrogen bonding is more prominent in the N-H...X bond than in the C-H...X bond. Therefore, it is evident that $\nu_3$ and $\nu_9$ are key indicators for the temperature evolution of the hydrogen bond crossing the orthorhombic-tetragonal phase transition. $MAPbI_3$ showed the largest changes in the spectral features of $\nu_3$ and $\nu_9$ near the phase transition (Fig. 8 b and c) and the most linear N-H...X bond in the low temperature orthorhombic phase, so that there is some evidence that the iodide is the halide that is influenced most by hydrogen bonding. This is supported by the results of the analysis of the N-H stretching vibrational mode $\nu_7$ frequency where a blueshift in the Cl-Br-I series was observed in our analysis (Fig. S14 d) as well as in previous investigations.[17,21] Compared to MP2 calculations of the free $CH_3NH_3^+$ cation ($\nu_7$: 3519 $cm^{-1}$)[21], vibrational modes of N-H stretching frequencies in $MAPbX_3$ are blueshifted more than 300 $cm^{-1}$ at room temperature. Low temperature results presented here indicate strengthening of N-H...X hydrogen bonding in the orthorhombic phase of $MAPbI_3$ compared to the tetragonal phase as the $\nu_7$ frequency is decreasing from 3193 $cm^{-1}$ at 293 K to 3179 $cm^{-1}$ at 107 K (Fig. S14 d).

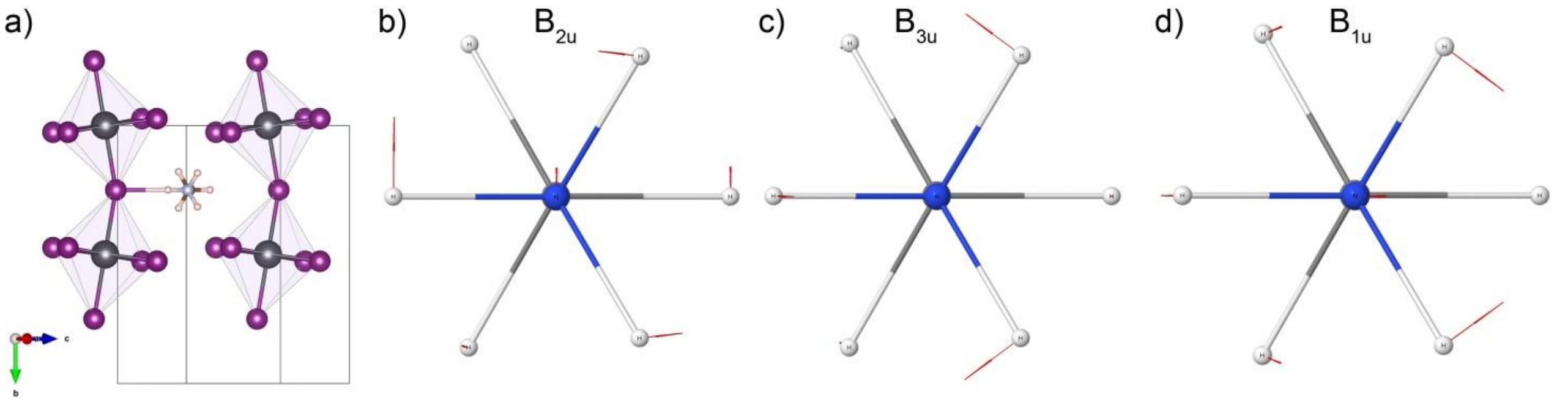

**FIGURE 9** a) Crystallographic orientation of the selected MA molecule, it can be distinguished between out-of-plane movements ($B_{2u}$, mode no. 116) and in-plane movements ($B_{3u}$, mode no. 117 and $B_{1u}$, mode no. 119) relative to the *Pnma* mirror plane (0,1/4,0). b), c) and d) Selected infrared active vibrational modes of $CH_3NH_3$ (MA) in $MAPbI_3$ calculated with DFT (DFT mode no. see also table S3).

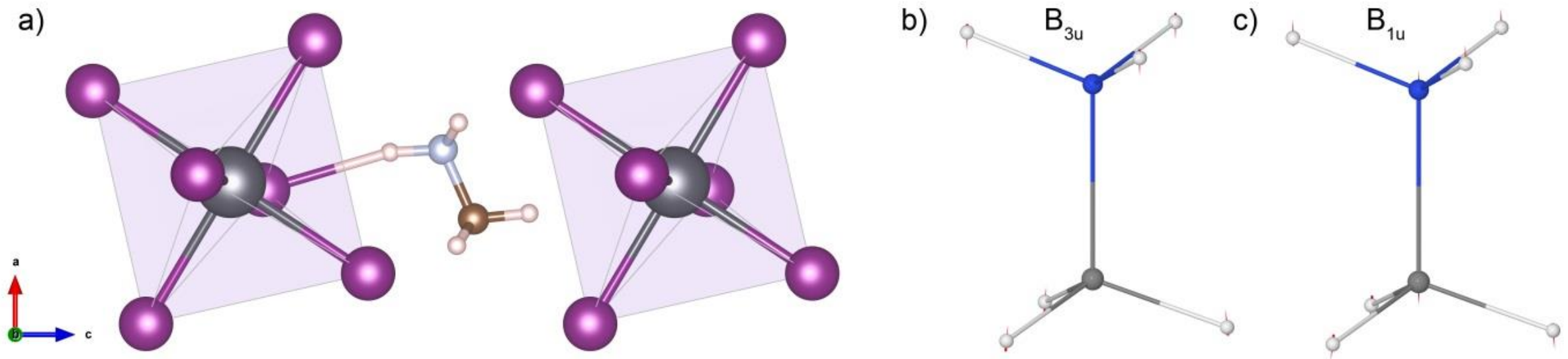

**FIGURE 10** a) Selected infrared active vibrational modes of $CH_3NH_3$ (MA) in $MAPbI_3$ calculated with DFT (DFT mode no. see also table S3). a) Crystallographic orientation of the selected MA molecule. The classical amplitudes (see also Fig. S19) and absolute directions (shown in red color) of the two IR active $v_5$ C-N stretching modes b) $B_{3u}$ (mode no. 86) and c) $B_{1u}$ (mode no. 87) showing mainly vibrational movements that are related to pure C-N stretching (Blue: N atom; grey: C atom; white: H atom).

The reorientation of the $[BX_6]^-$ octahedra in the tetragonal phase [2] and the appearance of a 3-dimensional hydrogen bond structure where the orientation of the shortest hydrogen bond is quite different to the orthorhombic phase, is the likely explanation for the drastic changes in the vibrational modes. In the tetragonal structure, several symmetry elements appear (including mirror planes) but not the mirror planes present in the ordered low temperature phase. Glaser et al. argue that the N-H...X hydrogen bonding does not differ significantly between the three halides at room temperature.[21] And indeed, from the phenomenological point of view the room temperature vibrational spectra are quite similar for the three halides, a frequency shift to lower frequencies in the series Cl-Br-I (Lorentz-Lorenz shift effect) [15] and quite similar peak widths for all halides were observed.

CONCLUSIONS

We have studied temperature-dependent infrared vibrational spectra of a series of MA lead halide single crystals. In the analyzed frequency and temperature range, all three halide compounds show a discontinuous behavior of the vibrational modes crossing the orthorhombic-tetragonal phase transition. Without fully understanding the vibrational modes in the investigated frequency range, it is questionable if further conclusions can be distilled from fragmentary spectral characteristics. We demonstrated that only in-depth analysis of the measured spectral features makes it possible to detangle the overlapping vibrational modes of MA in the orthorhombic and tetragonal phase. This allows an unambiguous assignment of the MA vibrational modes. These results were interconnected with the known crystal structural features (crystal structure refinements and DFT calculations) through the complete factor group analysis that was carried out here. To our knowledge, this is the first time that hydrogen-bond layers were discussed to elucidate the interplay of the organic MA molecule with the inorganic host structure. It was observed that the bromide and the chloride generally show several spectral similarities where the iodide seems to behave differently. There are several indications that the iodide is influenced more strongly by hydrogen-bonding than the bromide and the chloride at low temperatures. It is evident that the hydrogen-bonding scheme is highly interconnected with the MA vibrational modes investigated here.

ASSOCIATED CONTENT

**Supporting Information**. Factor group analysis, vibrational modes, lattice constants and N-H...X bonds, visualization of $MAPbCl_3$ orthorhombic crystal structure, symmetry reduced MA molecules in the orthorhombic $MAPbI_3$ crystal structure, infrared spectra, temperature-dependence of frequencies, HWHM and integrated absorbance, summed integrated absorbance, orthorhombic $MAPbI_3$ DFT results: absolute values of classical amplitudes, selected infrared active vibrational modes of MA in $MAPbI_3$ calculated with DFT.

The following files are available free of charge.

file type, PDF

AUTHOR INFORMATION

**Corresponding Author**

*Götz Schuck, e-mail: goetz.schuck@helmholtz-berlin.de

ACKNOWLEDGEMENTS

We acknowledge the use of the High-Performance-Computing-Cluster DIRAC at the HZB.

Electronic Supplementary Information for

# Infrared Spectroscopic Study of Vibrational Modes Across the Orthorhombic-Tetragonal Phase Transition in Methylammonium Lead Halide Single Crystals

*Götz Schuck,*† Daniel M. Többens, † Monika Koch-Müller, ‡ Ilias Efthimiopoulos, ‡ and Susan Schorr †,§*

†Helmholtz-Zentrum Berlin für Materialien und Energie, Hahn-Meitner-Platz 1, 14109 Berlin, Germany.

‡Deutsches GeoForschungsZentrum - GFZ, Telegrafenberg, 14473 Potsdam, Germany.

§Free University Berlin, Department of Geosciences, Malteserstraße 74–100, 12249 Berlin, Germany.

**Corresponding Author: *Götz Schuck, e-mail: goetz.schuck@helmholtz-berlin.de**

**Table S1**: Factor group analysis of $MAPbI_3$ in the orthorhombic phase using the correlation method.[42] Site symmetry of the MA molecule is $C_{s(zx)}$. The total irreducible representation of the crystal g) is the combined irreducible representation of each equivalent set of the atoms Pb, I(1) and I(2) a) - c) (lattice vibrations) plus the irreducible representation of the translations (lattice vibrations), rotations (libration) and intra-molecular vibrations of the MA molecule d) -f). $Z^B$: number of molecules per primitive Bravais cell, $t^\gamma$: number of translations of a site species $\gamma$, $a_\zeta$ is the number of lattice vibrations of the equivalent set of atoms in species $\zeta$ of the factor group, $\nu_{vib}$: degrees of vibration freedom of a single MA molecule.

a)

The Correlation for the Lead Atom with Site Symmetry $C_i$ (multiplicity = 4), $Z^B = 4$

| $f^\gamma$ | $t^\gamma$ | Site $C_i$ | Correlation → | Factor group $D_{2h}$ | $a_\zeta$ |
|---|---|---|---|---|---|
| 12 | 3 ($T_x$, $T_y$, $T_z$) | $A_u$ | | $A_u$ | 3 |
| | | | | $B_{1u}$ | 3 |
| | | | | $B_{2u}$ | 3 |
| | | | | $B_{3u}$ | 3 |

$\Gamma^{Pb} = 3A_u + 3B_{1u} + 3B_{2u} + 3B_{3u}$

b)

The Correlation for the Iodine Atom (1) with Site Symmetry $C_s$ (multiplicity = 4), $Z^B = 4$

| $f^\gamma$ | $t^\gamma$ | Site $C_{s(zx)}$ | Correlation → | Factor group $D_{2h}$ | $a_\zeta = a_{A'} + a_{A''}$ |
|---|---|---|---|---|---|
| 8 | 2 ($T_x$, $T_y$) | $A'$ | | $A_g$, $B_{2g}$, $B_{1u}$, $B_{3u}$ | |
| 4 | 1 ($T_z$) | $A''$ | | $B_{1g}$, $B_{3g}$, $A_u$, $B_{2u}$ | |
| | | | | $A_g$ | 2 = 2 + 0 |
| | | | | $B_{1g}$ | 1 = 0 + 1 |
| | | | | $B_{2g}$ | 2 = 2 + 0 |
| | | | | $B_{3g}$ | 1 = 0 + 1 |
| | | | | $A_u$ | 1 = 0 + 1 |
| | | | | $B_{1u}$ | 2 = 2 + 0 |
| | | | | $B_{2u}$ | 1 = 0 + 1 |
| | | | | $B_{3u}$ | 2 = 2 + 0 |

$\Gamma^{I(1)} = 2A_g + B_{1g} + 2B_{2g} + B_{3g} + A_u + 2B_{1u} + B_{2u} + 2B_{3u}$

c)

The Correlation for the Iodine Atom (2) with Site Symmetry $C_1$ (multiplicity = 8), $Z^B = 4$

| $f^\gamma$ | $t^\gamma$ | Site $C_1$ | Correlation → | Factor group $D_{2h}$ | $a_\zeta$ |
|---|---|---|---|---|---|
| 24 | 3 ($T_x$, $T_y$, $T_z$) | A | | $A_g$ | 3 |
| | | | | $B_{1g}$ | 3 |
| | | | | $B_{2g}$ | 3 |
| | | | | $B_{3g}$ | 3 |
| | | | | $A_u$ | 3 |
| | | | | $B_{1u}$ | 3 |
| | | | | $B_{2u}$ | 3 |
| | | | | $B_{3u}$ | 3 |

$\Gamma^{I(2)} = 3A_g + 3B_{1g} + 3B_{2g} + 3B_{3g} + 3A_u + 3B_{1u} + 3B_{2u} + 3B_{3u}$

d)

MA ion Translation, $\Gamma_{MA}^{trans}$, $Z^B = 4$

| $f^\gamma$ | $t^\gamma$ | Site $C_{s(zx)}$ | Correlation → | Factor group $D_{2h}$ | $a_\zeta = a_{A'} + a_{A''}$ |
|---|---|---|---|---|---|
| | | | | $A_g$ | 2 = 2 + 0 |
| 8 | 2 ($T_x$, $T_y$) | $A'$ | | $B_{1g}$ | 1 = 0 + 1 |
| | | | | $B_{2g}$ | 2 = 2 + 0 |
| | | | | $B_{3g}$ | 1 = 0 + 1 |
| | | | | $A_u$ | 1 = 0 + 1 |
| 4 | 1 ($T_z$) | $A''$ | | $B_{1u}$ | 2 = 2 + 0 |
| | | | | $B_{2u}$ | 1 = 0 + 1 |
| | | | | $B_{3u}$ | 2 = 2 + 0 |

$\Gamma_{MA}^{trans} = 2A_g + B_{1g} + 2B_{2g} + B_{3g} + A_u + 2B_{1u} + B_{2u} + 2B_{3u}$

e)

MA ion Libration, $\Gamma_{MA}^{lib}$, $Z^B = 4$

| $f^\gamma$ | $t^\gamma$ | Site $C_{s(zx)}$ | Correlation → | Factor group $D_{2h}$ | $a_\zeta = a_A + a_A$ |
|---|---|---|---|---|---|
| | | | | $A_g$ | 1 = 1 + 0 |
| 4 | 1 ($R_z$) | $A'$ | | $B_{1g}$ | 2 = 0 + 2 |
| | | | | $B_{2g}$ | 1 = 1 + 0 |
| | | | | $B_{3g}$ | 2 = 0 + 2 |
| | | | | $A_u$ | 2 = 0 + 2 |
| 8 | 2 ($R_x$, $R_y$) | $A''$ | | $B_{1u}$ | 1 = 1 + 0 |
| | | | | $B_{2u}$ | 2 = 0 + 2 |
| | | | | $B_{3u}$ | 1 = 1 + 0 |

$\Gamma_{MA}^{lib} = A_g + 2B_{1g} + B_{2g} + 2B_{3g} + 2A_u + B_{1u} + 2B_{2u} + B_{3u}$

f)

$\Gamma_{MA}^{intramol, vib}$, $Z^B = 4$

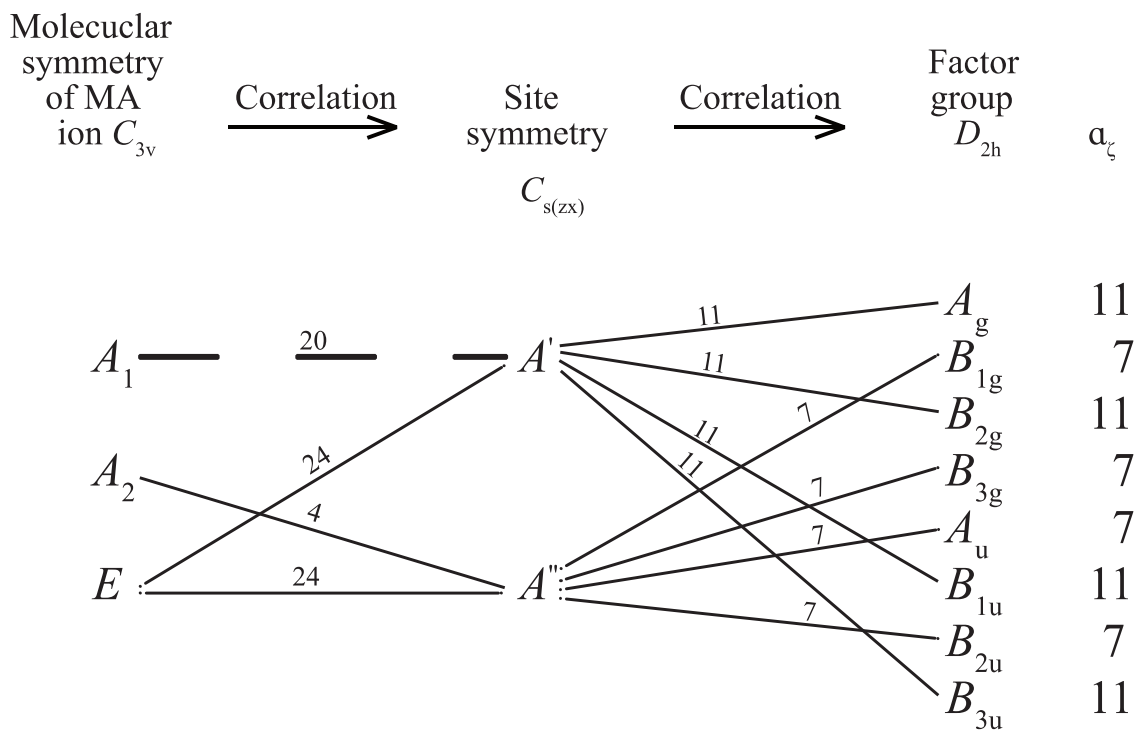

| $f^\gamma$ | $v_{vib}$ |
|---|---|
| 20 | 5 ($\nu_1$, $\nu_2$, $\nu_3$, $\nu_4$, $\nu_5$) |
| 4 | 1 ($\nu_6$) |
| 48 | 12 ($\nu_7$, $\nu_8$, $\nu_9$, $\nu_{10}$, $\nu_{11}$, $\nu_{12}$) |

$$\Gamma_{MA}^{intramol} = 11A_g + 7B_{1g} + 11B_{2g} + 7B_{3g} + 7A_u + 11B_{1u} + 7B_{2u} + 11B_{3u}$$

g)

| | | | | | | | | | | | | | | | | |
|---|---|---|---|---|---|---|---|---|---|---|---|---|---|---|---|---|
| $\Gamma^{Pb}$ | = | | | | | | | | | $3A_u$ | + | $3B_{1u}$ | + | $3B_{2u}$ | + | $3B_{3u}$ |
| $\Gamma^{I(1)}$ | = | $2A_g$ | + | $B_{1g}$ | + | $2B_{2g}$ | + | $B_{3g}$ | + | $A_u$ | + | $2B_{1u}$ | + | $B_{2u}$ | + | $2B_{3u}$ |
| $\Gamma^{I(2)}$ | = | $3A_g$ | + | $3B_{1g}$ | + | $3B_{2g}$ | + | $3B_{3g}$ | + | $3A_u$ | + | $3B_{1u}$ | + | $3B_{2u}$ | + | $3B_{3u}$ |
| $\Gamma_{MA}^{trans}$ | = | $2A_g$ | + | $B_{1g}$ | + | $2B_{2g}$ | + | $B_{3g}$ | + | $A_u$ | + | $2B_{1u}$ | + | $B_{2u}$ | + | $2B_{3u}$ |
| $\Gamma_{MA}^{lib}$ | = | $A_g$ | + | $2B_{1g}$ | + | $B_{2g}$ | + | $2B_{3g}$ | + | $2A_u$ | + | $B_{1u}$ | + | $2B_{2u}$ | + | $B_{3u}$ |
| $\Gamma_{MA}^{intramol}$ | = | $11A_g$ | + | $7B_{1g}$ | + | $11B_{2g}$ | + | $7B_{3g}$ | + | $7A_u$ | + | $11B_{1u}$ | + | $7B_{2u}$ | + | $11B_{3u}$ |
| $\Gamma_{MAPbI3}^{optic}$ | = | $19A_g$ | + | $14B_{1g}$ | + | $19B_{2g}$ | + | $14B_{3g}$ | + | $17A_u$ | + | $21B_{1u}$ | + | $16B_{2u}$ | + | $21B_{3u}$ |
| $\Gamma_{MAPbI3}^{acust}$ | = | | | | | | | | | | | $B_{1u}$ | + | $B_{2u}$ | + | $B_{3u}$ |

**Table S2**: Vibrational modes of $CH_3NH_3$ (MA) in **$MAPbBr_3$** calculated with DFT compared with fit results of experimental data at 107 K and 293 K

| | | DFT | | | | Lorentzian fit of experimental data | | | | | |
|---|---|---|---|---|---|---|---|---|---|---|---|
| | | | | | | 107 K | | | 293 K | | |
| Band assignment | Name | DFT mode No. | DFT cor. Frequency [$cm^{-1}$] | DFT IRREP | DFT Intensity [KM/MOL] | Frequency [$cm^{-1}$] | HWHM [$cm^{-1}$] | Area [$cm^{-1}$] | Frequency [$cm^{-1}$] | HWHM [$cm^{-1}$] | Area [$cm^{-1}$] |
| $\nu_{12}$ (E)<br>$CH_3$ rocking | br1<br>br2<br>br3 | 78<br>80<br>82 | 913.2<br>916.3<br>922.0 | B1u<br>B3u<br>B2u | 54.61<br>510.25<br>348.47 | 902(2)<br>914.63(11)<br>920.99(10) | 34(3)<br>5.32(14)<br>3.97(14) | 7.54<br>9.83<br>5.12 | -<br>916.39(6)<br>- | -<br>16.1(1)<br>- | -<br>30.00<br>- |
| $\nu_5$ ($A_1$)<br>C-N strectching | br4 | 86<br>87 | 983.6<br>983.6 | B3u<br>B1u | 12.16<br>53.43 | 970.56(3) | 1.68(5) | 1.82 | 968.71(9) | 6.2(2) | 4.31 |
| $\nu_{11}$ (E)<br>$NH_3$ rocking | br5<br>br6<br>br7 | 89<br>91<br>93 | 1260.8<br>1261.7<br>1264.8 | B1u<br>B3u<br>B2u | 17.80<br>138.73<br>30.19 | 1253(1)<br>1257.3(4)<br>1259.3(3) | 5.2(5)<br>2.7(5)<br>2.3(3) | 1.33<br>3.21<br>2.48 | -<br>1251.67(7)<br>- | -<br>10.0(2)<br>- | -<br>3.90<br>- |
| combination | br8 | | | | | 1371.4(7) | 23(2) | 2.82 | 1387(1) | 27(3) | 2.50 |
| $\nu_4$ ($A_1$)<br>sym. $CH_3$ bending | br9 | 98<br>99 | 1406.5<br>1406.8 | B3u<br>B1u | 26.33<br>41.32 | 1420.65(2) | 1.66(3) | 1.66 | 1425.7(1) | 5.2(3) | 1.97 |
| $\nu_{10}$ (E)<br>asym. $CH_3$ bending | br10<br>br11 | 101<br>103<br>105 | 1447.3<br>1451.2<br>1453.7 | B3u<br>B1u<br>B2u | 99.76<br>5.25<br>95.42 | 1455.34(9)<br>1460.0(3) | 3.47(15)<br>5.0(5) | 5.24<br>5.04 | 1454.7(5)<br>- | 21.3(9)<br>- | 17.55<br>- |
| $\nu_3$ ($A_1$)<br>sym. $NH_3$ bending | br12 | 110<br>111 | 1462.2<br>1469.1 | B3u<br>B1u | 84.91<br>293.84 | 1469.7(2) | 8.0(2) | 8.91 | 1478.1(1) | 14.3(2) | 31.91 |
| combination | br13<br>br14 | | | | | 1491.4(10)<br>1549.9(4) | 40.0(13)<br>13.1(8) | 14.81<br>4.38 | -<br>1562(4) | -<br>63(3) | -<br>48.81 |
| $\nu_9$ (E)<br>asym. $NH_3$ bending | br15<br>br16 | 113<br>117<br>119 | 1582.4<br>1604.4<br>1608.9 | B2u<br>B3u<br>B1u | 67.84<br>123.65<br>26.32 | 1571.6(1)<br>1588.01(4) | 13.1(8)<br>11.8(4) | 11.76<br>7.05 | -<br>1588.3(1) | -<br>19.5(7) | -<br>19.20 |
| combination | br17<br>br18 | | | | | 1618.8(7)<br>1672.2(2) | 25(2)<br>22.8(5) | 5.05<br>11.30 | 1659(3)<br>- | 50(6)<br>- | 10.06<br>- |
| $\nu_2$ ($A_1$)<br>sym. $CH_3$ stretching | br19 | 123<br>124 | 2976.7<br>2976.8 | B3u<br>B1u | 3.46<br>4.76 | 2964.62(4) | 1.61(8) | 0.52 | 2964.7(4) | 3(1) | 0.20 |
| $\nu_8$ (E)<br>asym. $CH_3$ stretching | br20<br>br21 | 128<br>129<br>131 | 3076.7<br>3080.0<br>3080.1 | B2u<br>B1u<br>B3u | 13.86<br>0.03<br>5.37 | 3029.3(2)<br>3043.5(2) | 4.0(4)<br>4.5(3) | 0.62<br>1.18 | - | - | - |
| $\nu_1$ ($A_1$)<br>sym. $NH_3$ stretching | br22 | 134<br>136 | 3167.7<br>3168.9 | B3u<br>B1u | 102.06<br>224.59 | 3136(1) | 55(1) | 163.85 | 3123(1) | 82(1) | 240.77 |
| $\nu_7$ (E)<br>asym. $NH_3$ stretching | br23 | 138<br>139<br>143 | 3218.9<br>3219.5<br>3224.2 | B1u<br>B3u<br>B2u | 585.11<br>4108.24<br>3262.32 | 3210.1(7) | 32(1) | 75.23 | 3221.2(7) | 40(1) | 90.45 |

Note: The calculated frequencies given in the column "DFT cor. Frequency" for Tables S2-S4 have been calculated from the raw results of the DFT calculation by applying a scissor factor of 0.9586(13). This scissor factor is the same for all data sets and was determined from a linear best-fit to all data after assignment of observed to calculated modes based on grouping and relative intensity. After this, calculated and observed frequencies are in agreement with a standard deviation of only ±11 $cm^{-1}$).

**Table S3**: Vibrational modes of $CH_3NH_3$ (MA) in **$MAPbI_3$** calculated with DFT compared with fit results of experimental data at 107 K and 293 K

| | | DFT | | | | Lorentzian fit of experimental data | | | | | |
|---|---|---|---|---|---|---|---|---|---|---|---|
| | | | | | | 107 K | | | 293 K | | |
| Band assignment | Name | DFT mode No. | DFT cor. Frequency [$cm^{-1}$] | DFT IRREP | DFT Intensity [KM/MOL] | Frequency [$cm^{-1}$] | HWHM [$cm^{-1}$] | Area [$cm^{-1}$] | Frequency [$cm^{-1}$] | HWHM [$cm^{-1}$] | Area [$cm^{-1}$] |
| $\nu_{12}$ (E)<br>$CH_3$ rocking | i1<br>i2<br>i3 | 77<br>82<br>84 | 912.5<br>918.0<br>920.7 | B2u<br>B3u<br>B1u | 520.10<br>69.94<br>446.30 | 895(2)<br>906.92(12)<br>917.68(4) | 11(2)<br>5.4(3)<br>2.20(7) | 2.41<br>6.79<br>3.34 | -<br>910.71(7)<br>- | -<br>13.7(1)<br>- | -<br>18.15<br>- |
| $\nu_5$ ($A_1$)<br>C-N streching | i4 | 86<br>87 | 985.1<br>985.7 | B3u<br>B1u | 54.25<br>6.15 | 967.12(7) | 1.28(9) | 0.90 | 961.27(7) | 4.6(1) | 3.43 |
| $\nu_{11}$ (E)<br>$NH_3$ rocking | i5<br>i6 | 91<br>93<br>95 | 1259.1<br>1260.4<br>1261.2 | B2u<br>B1u<br>B3u | 31.43<br>29.52<br>7.95 | 1252.5(1)<br>1256.7(1) | 3.1(1)<br>1.4(2) | 1.98<br>1.43 | 1247.7(1)<br>- | 7.4(2)<br>- | 1.15<br>- |
| combination | i7 | | | | | 1361(2) | 35(3) | 3.30 | 1404(3) | 36(3) | 4.18 |
| $\nu_4$ ($A_1$)<br>sym. $CH_3$ bending | i8 | 98<br>99 | 1399.2<br>1400.1 | B1u<br>B3u | 48.41<br>18.41 | 1419.45(6) | 1.94(9) | 1.16 | 1421.97(5) | 3.6(1) | 2.04 |
| combination | i9 | | | | | 1434.3(5) | 6(1) | 0.84 | - | - | - |
| $\nu_{10}$ (E)<br>asym. $CH_3$ bending | i10 | 104<br>106<br>107 | 1439.7<br>1440.8<br>1440.9 | B2u<br>B1u<br>B3u | 150.86<br>102.03<br>45.54 | 1450.79(2) | 2.36(4) | 5.71 | 1452.1(4) | 17.9(6) | 12.57 |
| $\nu_3$ ($A_1$)<br>sym. $NH_3$ bending | i11 | 110<br>111 | 1470.9<br>1472.0 | B1u<br>B3u | 63.53<br>288.19 | 1456.65(5) | 3.76(7) | 6.58 | 1469.38(3) | 7.42(7) | 17.11 |
| combination | i12<br>i13 | | | | | 1483.9(4)<br>1553.3(3) | 9.5(7)<br>18.6(6) | 1.81<br>7.00 | -<br>1530.2(7) | -<br>46(1) | -<br>25.11 |
| $\nu_9$ (E)<br>asym. $NH_3$ bending | i14 | 116<br>117<br>119 | 1598.5<br>1599.9<br>1601.1 | B2u<br>B3u<br>B1u | 5.47<br>53.37<br>24.82 | 1584.97(3) | 2.96(5) | 4.47 | 1578.8(2) | 20.6(6) | 12.81 |
| combination | i15<br>i16 | | | | | 1616(1)<br>1675.7(3) | 16(2)<br>18.1(6) | 1.69<br>6.83 | 1633.6(8)<br>- | 47(1)<br>- | 16.91<br>- |
| $\nu_2$ ($A_1$)<br>sym. $CH_3$ stretching | i17 | 121<br>124 | 2853.2<br>2853.5 | B1u<br>B3u | 19.17<br>4.75 | 2957.93(7) | 1.1(1) | 0.21 | 2957.7(2) | 2.7(3) | 0.22 |
| $\nu_8$ (E)<br>asym. $CH_3$ stretching | i18<br>i19 | 125<br>127<br>132 | 2948.8<br>2949.3<br>2952.6 | B1u<br>B3u<br>B2u | 5.55<br>1.95<br>0.16 | 3024.36(6)<br>3031.9(1) | 3.0(1)<br>4.3(2) | 2.02<br>2.13 | 3023.5(3) | 8.0(8) | 0.54 |
| $\nu_1$ ($A_1$)<br>sym. $NH_3$ stretching | i20 | 133<br>135 | 3035.7<br>3037.5 | B3u<br>B1u | 12.26<br>1346.10 | 3121.9(6) | 46.5(6) | 36.33 | 3126(1) | 58.9(8) | 169.73 |
| $\nu_7$ (E)<br>asym. $NH_3$ stretching | i21 | 139<br>140<br>141 | 3086.2<br>3086.5<br>3088.0 | B1u<br>B3u<br>B2u | 5444.71<br>1117.39<br>4968.09 | 3178.9(4) | 15.8(8) | 193.08 | 3193.6(6) | 35.5(9) | 89.71 |

**Table S4**: Vibrational modes of $CH_3NH_3$ (MA) in **$MAPbCl_3$** calculated with DFT[*1] compared with fit results of experimental data at 126 K and 293 K

| | | DFT | | | | Lorentzian fit of experimental data | | | | | |
|---|---|---|---|---|---|---|---|---|---|---|---|
| | | | | | | 126 K | | | 293 K | | |
| Band assignment | Name | DFT mode No. | DFT cor. Frequency [cm$^{-1}$] | DFT IRREP | DFT Intensity [KM/MOL] | Frequency [cm$^{-1}$] | HWHM [cm$^{-1}$] | Area [cm$^{-1}$] | Frequency [cm$^{-1}$] | HWHM [cm$^{-1}$] | Area [cm$^{-1}$] |
| $\nu_{12}$ (E)<br>$CH_3$ rocking | cl1<br>cl2<br>cl3 | 78<br>80<br>81 | 910.9<br>915.2<br>925.0 | B1u<br>B3u<br>B2u | 47.08<br>390.31<br>287.31 | 908(5)<br>923.3(2)<br>931.0(2) | 21(3)<br>7.6(5)<br>3.7(4) | 2.03<br>6.69<br>1.47 | -<br>922.81(4)<br>- | -<br>15.5(1)<br>- | -<br>20.68<br>- |
| $\nu_5$ ($A_1$)<br>C-N<br>strectching | cl4 | 86<br>87 | 985.8<br>986.0 | B3u<br>B1u | 7.54<br>45.13 | 977.64(5) | 2.12(8) | 1.02 | 977.58(6) | 5.6(8) | 2.62 |
| $\nu_{11}$ (E)<br>$NH_3$ rocking | cl5<br>cl6<br>cl7 | 89<br>91<br>94 | 1258.5<br>1259.8<br>1260.5 | B1u<br>B3u<br>B2u | 15.79<br>83.97<br>33.80 | 1251(2)<br>1262.3(2)<br>1267.1(2) | 14(2)<br>5.0(4)<br>3.0(3) | 1.45<br>2.94<br>0.82 | -<br>1255.27(5)<br>- | -10.4(1)<br>- | -<br>3.52<br>- |
| combination | cl8 | | | | | 1405(3) | 23(4) | 1.19 | 1395(1) | 26(2) | 1.82 |
| $\nu_4$ ($A_1$)<br>sym. $CH_3$<br>bending | cl9 | 98<br>100 | 1404.3<br>1405.4 | B3u<br>B1u | 13.33<br>34.14 | 1421.25(6) | 2.2(1) | 0.82 | 1427.5(1) | 6.0(2) | 1.73 |
| combination | cl10<br>cl11 | | | | | 1439(1)<br>1454(1) | 7(3)<br>5(2) | 0.40<br>0.60 | -<br>- | -<br>- | -<br>- |
| $\nu_{10}$ (E)<br>asym. $CH_3$<br>bending | cl12<br>cl13 | 101<br>102<br>107 | 1449.9<br>1450.2<br>1454.1 | B3u<br>B2u<br>B1u | 66.69<br>101.48<br>7.04 | 1458.2(1)<br>1462.2(3) | 2.3(3)<br>3.4(4) | 1.34<br>1.34 | 1455.3(2)<br>- | 18.8(4)<br>- | 12.27<br>- |
| $\nu_3$ ($A_1$)<br>sym. $NH_3$<br>bending | cl14<br>cl15 | 109<br>111 | 1470.5<br>1478.1 | B3u<br>B1u | 50.38<br>326.56 | 1473.7(2)<br>1479(1) | 3.5(7)<br>10.3(6) | 1.03<br>5.22 | -<br>1485.07(4) | -<br>14.9(1) | -<br>28.61 |
| combination | cl16 | | | | | 1537(2) | 36(4) | 5.86 | 1555(3) | 54(4) | 17.64 |
| $\nu_9$ (E)<br>asym. $NH_3$<br>bending | cl17<br>cl18 | 113<br>117<br>119 | 1587.8<br>1606.8<br>1612.5 | B2u<br>B3u<br>B1u | 58.93<br>100.04<br>16.89 | 1582.3(4)<br>1595.1(2) | 13.6(7)<br>5.9(4) | 7.57<br>2.26 | -<br>1592.0(1) | -<br>22.3(4) | -<br>25.74 |
| combination | cl19 | | | | | 1676(1) | 50(2) | 7.03 | 1667(2) | 53(3) | 6.44 |
| $\nu_2$ ($A_1$)<br>sym. $CH_3$<br>stretching | cl20 | 121<br>124 | 2971.0<br>2971.0 | B3u<br>B1u | 8.52<br>11.59 | 2970.1(3) | 3.1(6) | 0.10 | 2968.9(3) | 6(1) | 0.03 |
| $\nu_8$ (E)<br>asym. $CH_3$<br>stretching | cl21<br>cl22 | 127<br>129<br>130 | 3059.5<br>3094.3<br>3094.3 | B2u<br>B3u<br>B1u | 8.93<br>0.72<br>2.46 | 3035.4(4)<br>3053.6(5) | 5.3(6)<br>6(1) | 0.83<br>0.70 | - | - | - |
| $\nu_1$ ($A_1$)<br>sym. $NH_3$<br>stretching | cl23 | 135<br>136 | 3146.1<br>3146.3 | B1u<br>B3u | 529.04<br>63.35 | 3120(2) | 95(2) | 132.17 | 3121(1) | 94(1) | 239.51 |
| $\nu_7$ (E)<br>asym. $NH_3$<br>stretching | cl24 | 139<br>142<br>143 | 3191.2<br>3206.7<br>3207.6 | B3u<br>B2u<br>B1u | 3700.66<br>3815.82<br>352.58 | 3220(1) | 37(3) | 32.17 | 3219.7(8) | 41(2) | 80.84 |

[*1]Note: The DFT calculations for the chloride were carried out in the same unit cell as for the iodide and the bromide

**Table S5**: Lattice constants and N-H...X bonds of the orthorhombic crystal structures of methylammonium lead halide calculated with DTF compared with crystal structure refinement from literature data (comparison of tow hydrogen bonds: a) hydrogen bond "1" (X and H atom are positioned on 1 site symmetry); b) hydrogen bond "m" (X and H atom are positioned on m site symmetry); red color = long bond; blue color = short bond). Because of cell doubling we have to consider two MA molecules for the chloride (Fig. S1: MA molecule 2 is present in layer B whereas the other N...X and N-H...X hydrogen bonds are present in layer A).

| | $MAPbX_3$; X = **Cl** | | | $MAPbX_3$; X = **Br** | | $MAPbX_3$; X = **I** | |
|---|---|---|---|---|---|---|---|
| | this work | this work; doubled cell | Ref. [11]; doubled cell; 80 K | this work | Ref. [10]; 11 K; $CH_3ND_3PbBr_3$ | this work | Ref. [9]; 100 K |
| a [Å] | 7.6301 | 11.3587 | 11.1747(2) | 7.8993 | 7.9434(4) | 9.0901 | 8.8657(3) |
| b [Å] | 11.5786 | 11.5820 | 11.3552(1) | 11.9693 | 11.8499(5) | 12.7324 | 12.6293(4) |
| c [Å] | 8.5284 | 11.4874 | 11.2820(1) | 8.7830 | 8.5918(4) | 8.6787 | 8.5769(3) |
| $H_{(1)}...X_{(1)}$ [Å] | 2.2977 | 2.2692 | 2.292 | 2.4437 | 2.5250 | 2.6066 | 2.8075 |
| $H_{(1)}...X_{(1)}$ [Å] Molecule 2 | - | 2.3056 | 2.337 | - | - | - | - |
| $N...X_{(1)}$ [Å] | 3.2817 | 3.2697 | 3.273 | 3.3949 | 3.4573 | 3.5787 | 3.6804 |
| $N...X_{(1)}$ [Å] Molecule 2 | - | 3.2966 | 3.300 | - | - | - | - |
| $N-H...X_{(1)}$ [°] | 159.4 | 163.2 | 158.8 | 153.4 | 151.2 | 157.1 | 145.8 |
| $N-H...X_{(1)}$ [°] Molecule 2 | - | 161.1 | 155.1 | - | - | - | - |
| $H_{(m)}...X_{(m)}$ [Å] | 2.3906 | 2.3939 | 2.320 | 2.4413 | 2.4996 | 2.5855 | 2.6126 |
| $H_{(m)}...X_{(m)}$ [Å] Molecule 2 | - | 2.4122 | 2.373 | - | - | - | - |
| $N...X_{(m)}$ [Å] | 3.3532 | 3.3449 | 3.336 | 3.4223 | 3.4718 | 3.5970 | 3.6113 |
| $N...X_{(m)}$ [Å] Molecule 2 | - | 3.3611 | 3.346 | - | - | - | - |
| $N-H...X_{(m)}$ [°] | 155.6 | 153.5 | 150.8 | 159.2 | 156.9 | 166.6 | 174.1 |
| $N-H...X_{(m)}$ [°] Molecule 2 | - | 153.2 | 156.9 | - | - | - | - |

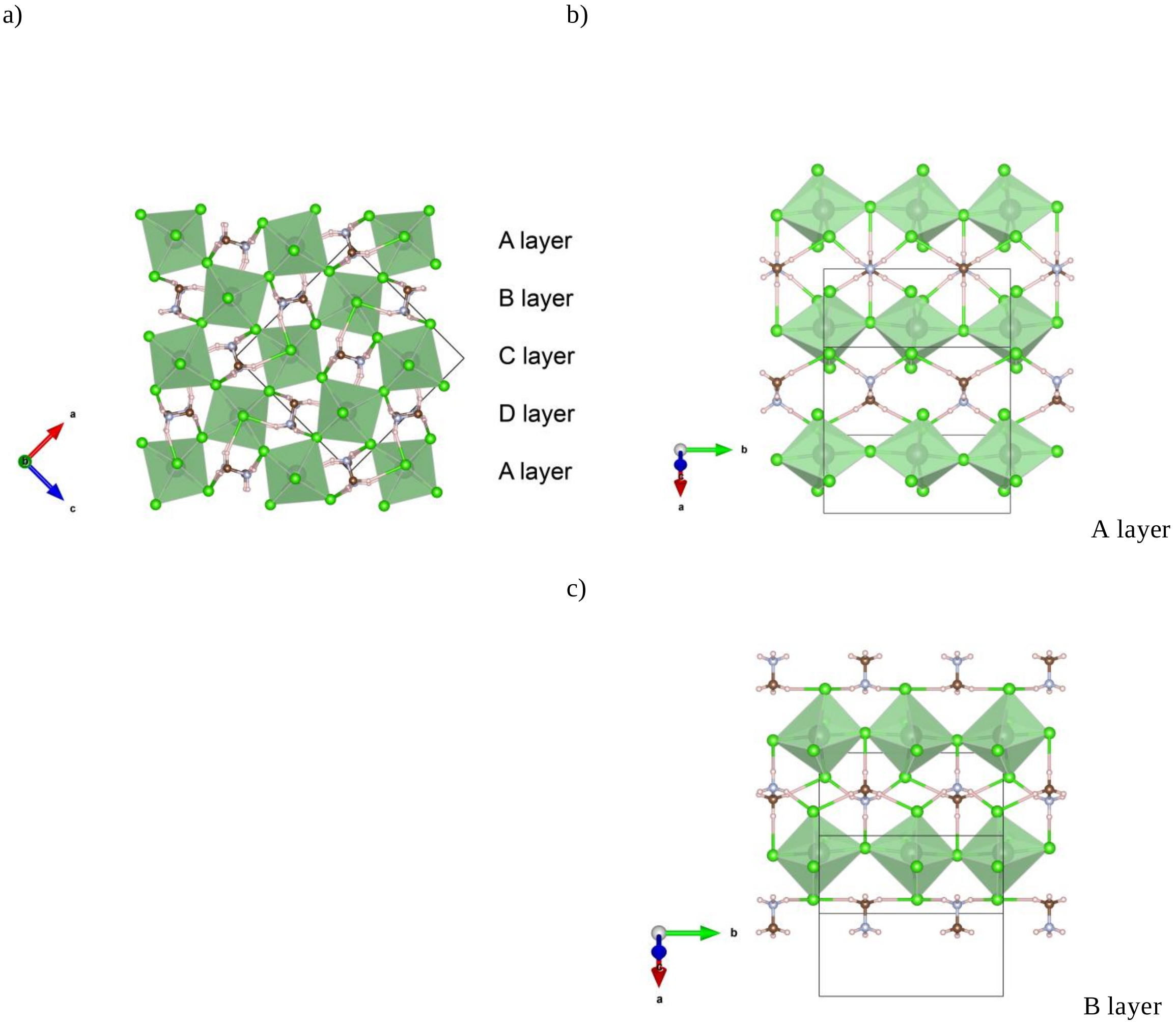

**Figure S1** Visualization of $MAPbCl_3$ orthorhombic crystal structure. Because of the doubled cell (compared to the iodide and bromide) four hydrogen bond layers, layer A, B, C and layer D, can be identified. Layer A is shown in b) and Layer B is shown in c). The only difference between layer A and layer C, and between layer B and D is the orientation of the MA molecule axis. The N-H...I bond lengths are different in the layers A and B (Table S5: MA molecule 1 is present in layer A and C, MA molecule 2 is present in layer B and D). The colors refer to the following elements: green - chloride, dark grey - lead, brown - carbon, light blue - nitrogen, light orange - hydrogen.

a)

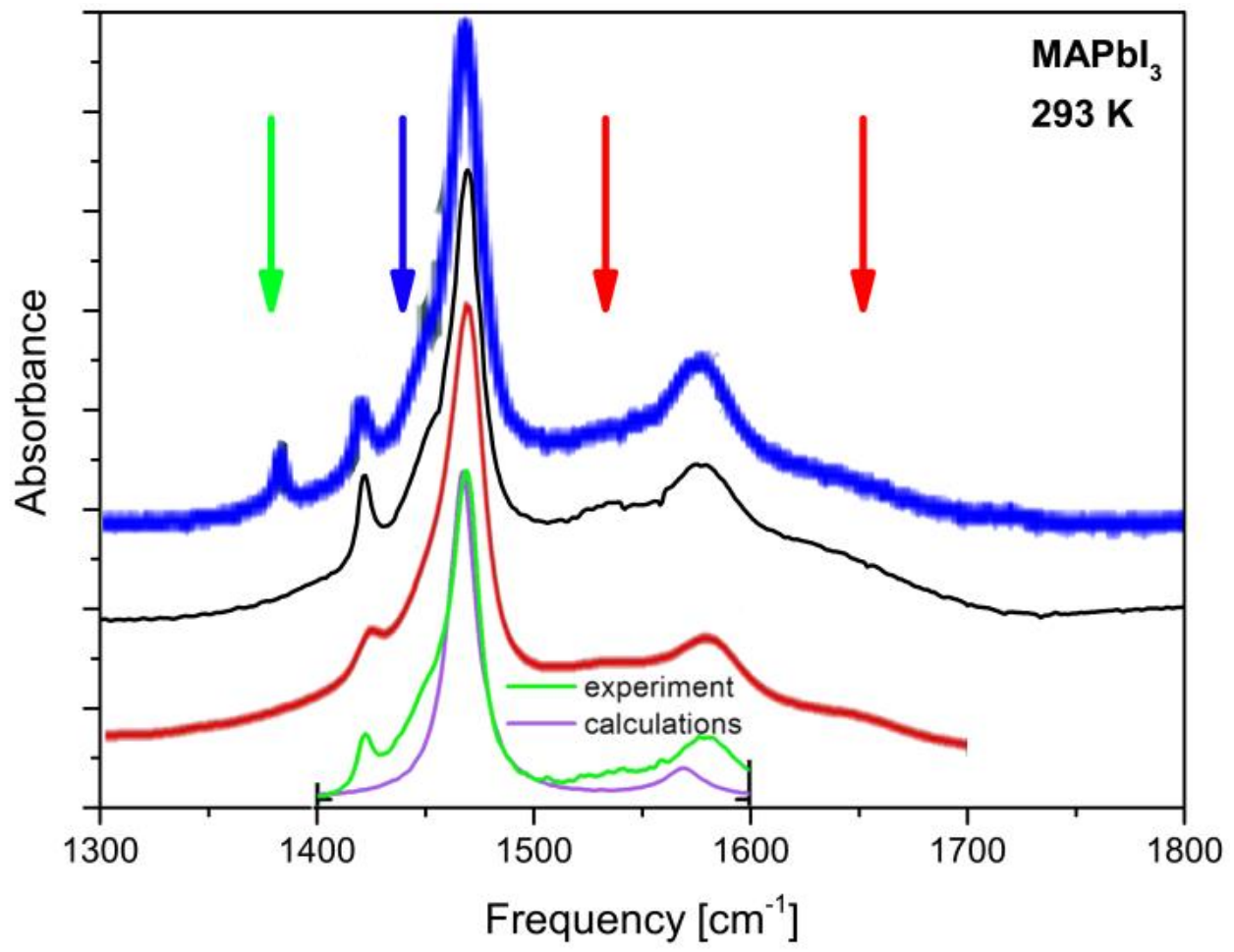

b)

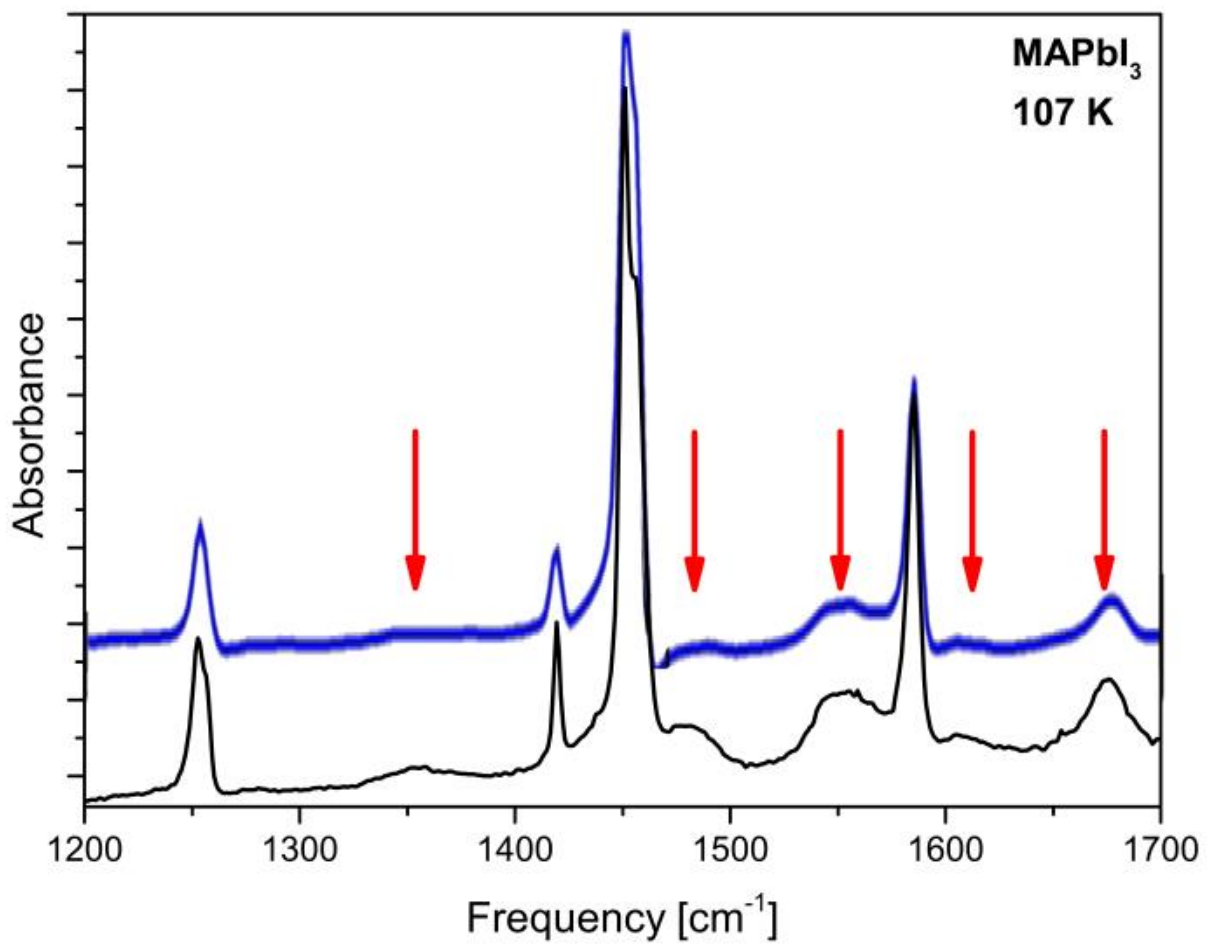

**Figure S2** Infrared spectra of $MAPbI_3$ a) room temperature; measurement of small single crystals using a FTIR-microscope spectrometer (black line); for comparison to our data: blue line: room temperature-relative transmission spectra of $MAPbI_3$ thin film from Glaser et al. (scaled data from Fig. 2a from reference [21]); red line: experimental IR spectra of $MAPbI_3$ thin film from Ivanovska et al. (scaled data from Fig. 2a from reference [14]); green and violet line: experimental IR spectra of $MAPbI_3$ thin film (green line) and calculated absorbance spectrum of the NH3 bending modes (violet line) from Bakulin et al. (scaled data from Fig. 1d from reference [23]). According to Sender [20] the band at 1385 $cm^{-1}$ (green arrow) was not observed in single crystal data. b) 107 K measurement of small single crystals using a FTIR-microscope spectrometer (black line); for comparison blue line: experimental IR spectra of $MAPbI_3$ thin film from Ivanovska et al. (scaled data from Fig. 2a from reference [14]). Red arrows: combination modes not assigned in previous works; blue arrow: $\nu_{10}$ discussed in text.

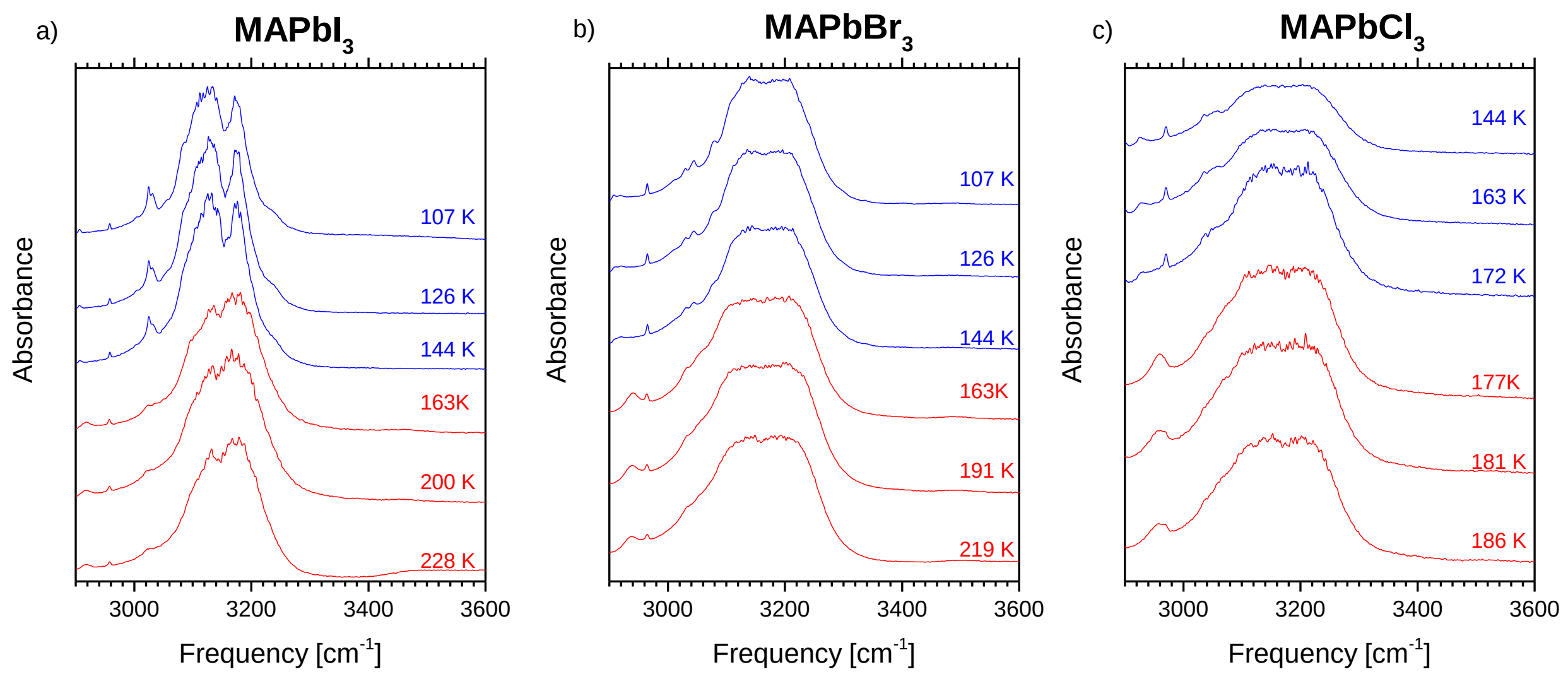

**FIGURE S3** Temperature-dependent infrared spectra of a) $MAPbI_3$, b) $MAPbBr_3$ and c) $MAPbCl_3$; frequency range: 2900-3600 $cm^{-1}$. The vibrational modes in the 3100-3200 $cm^{-1}$ frequency range are somewhat broad (especially for $MAPbBr_3$ and $MAPbCl_3$) and slightly overexposed, most probably due to relatively thick samples.

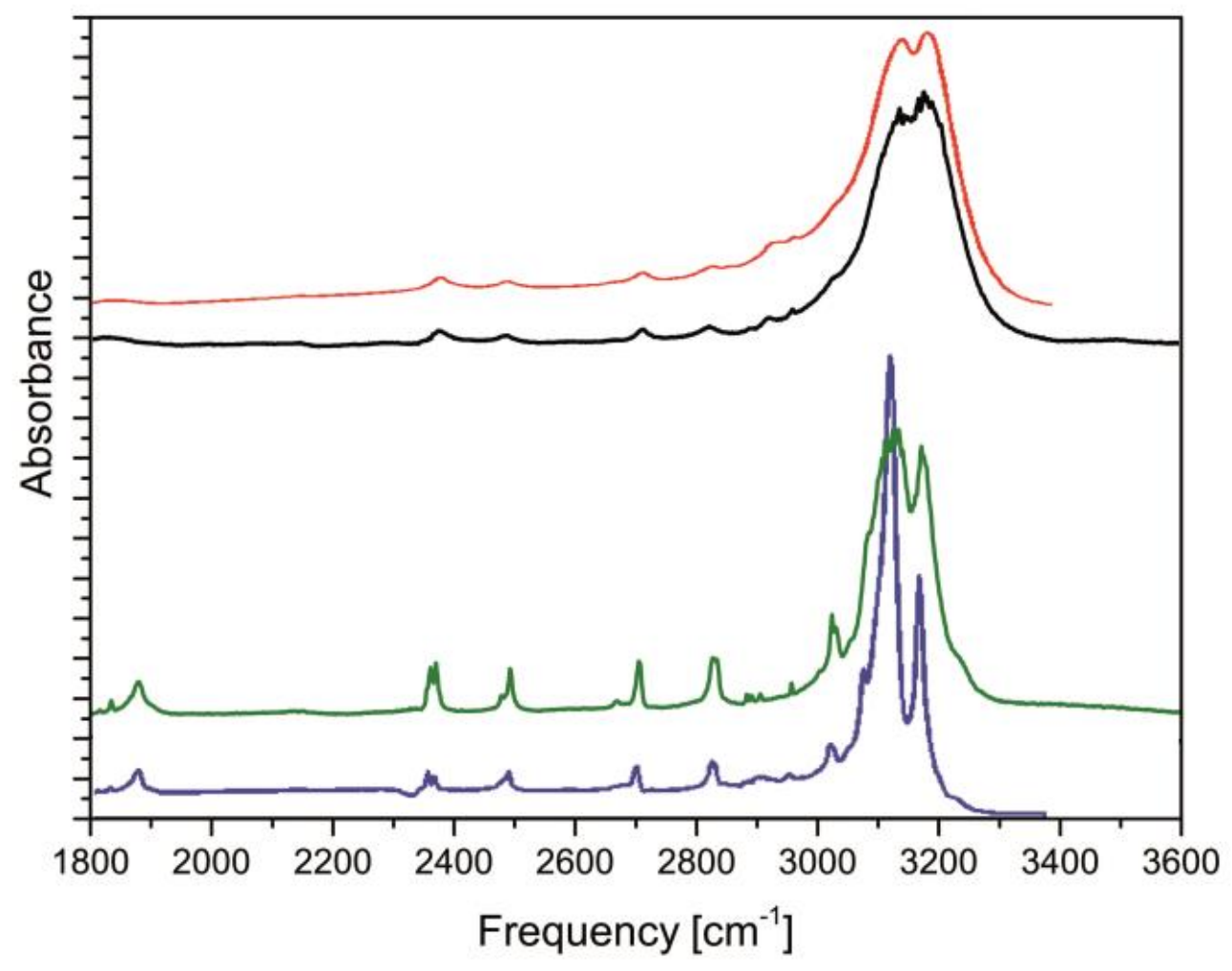

**FIGURE S4** Comparison of temperature-dependent infrared spectra of $MAPbI_3$ based on literature data. Red and blue: experimental IR spectra from Ivanovska et al. [14] (Figure taken from figure S1 in [14]) in comparison with our infrared spectra (black line: 293 K; green line: 107 K). The spectra are quite similar for the room temperature data as well as for the low temperatures. The 80 K (blue) spectre from Ivanovska et al. [14] is somewhat sharper in the 3000-3300 $cm^{-1}$ frequency range than our spectra at 107 K.

a)

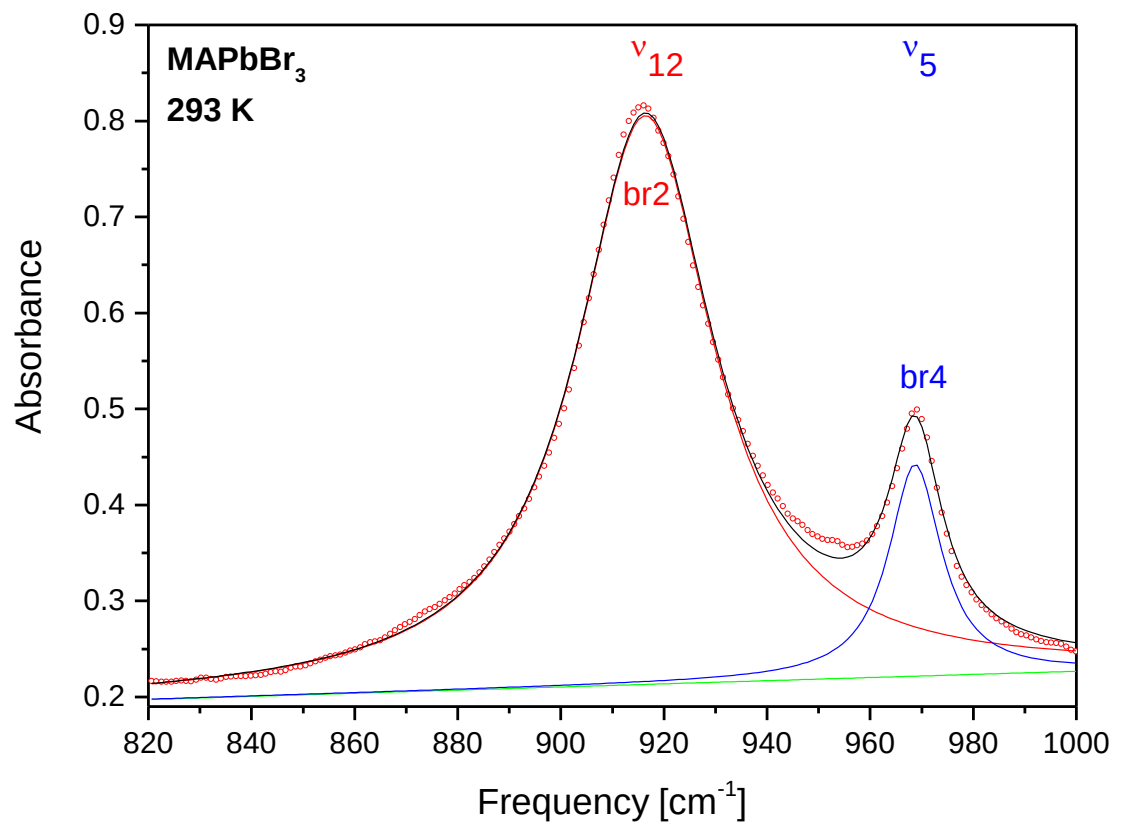

b)

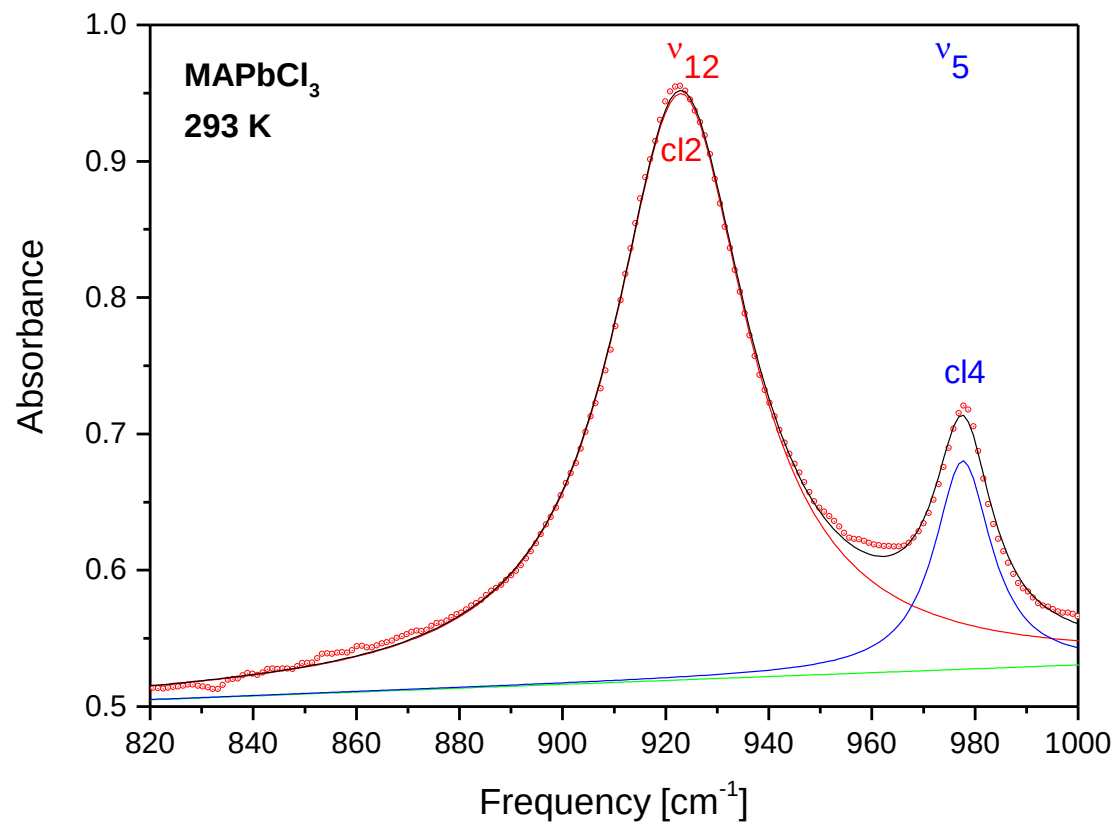

c)

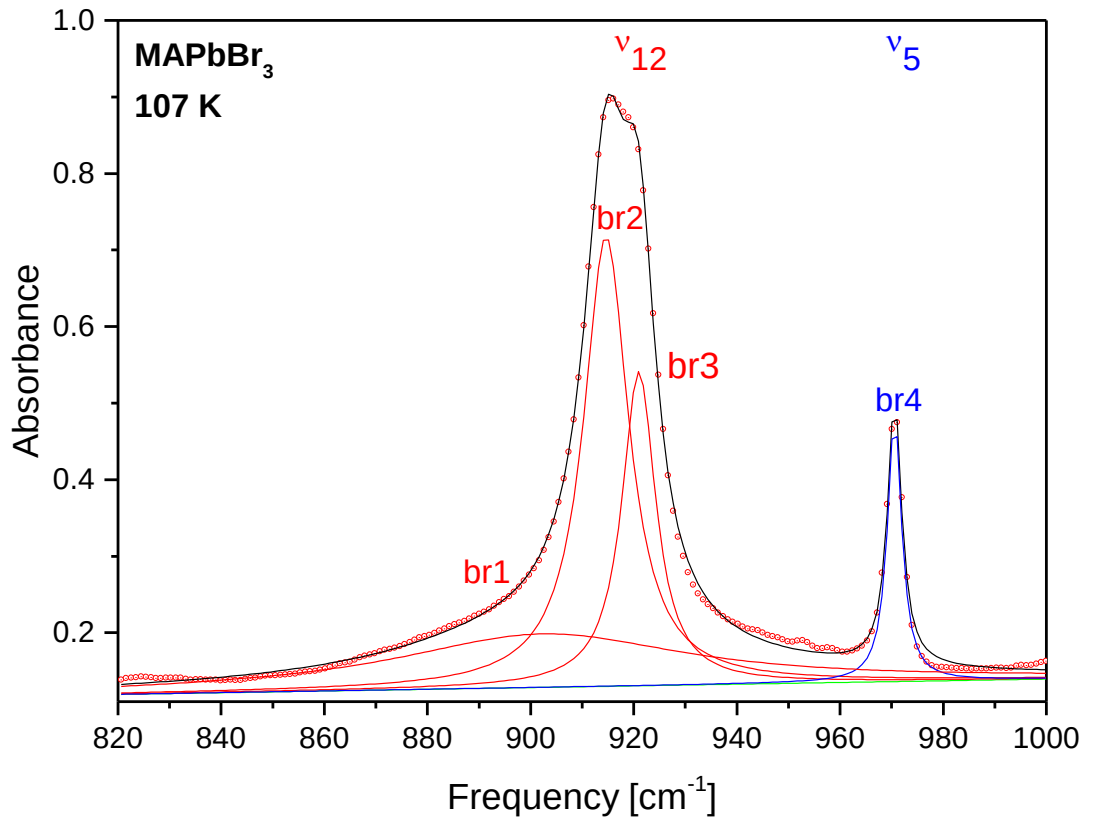

d)

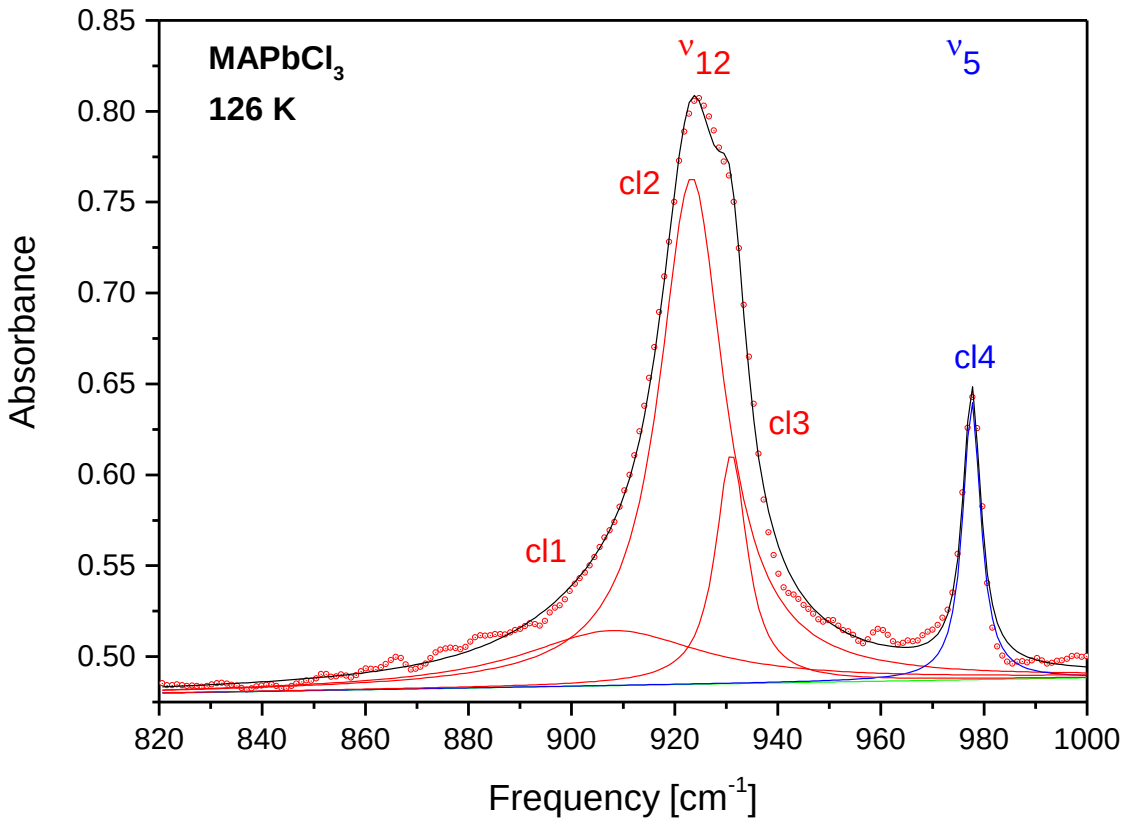

**Figure S5** Infrared spectra of a), c) $MAPbBr_3$ and b), d) $MAPbCl_3$ between 820 $cm^{-1}$ and 1000 $cm^{-1}$ at a), b) 293 K (tetragonal/cubic) and at c) 107 K and, d) 126 K (orthorhombic phase); measurements of small single crystals using a FTIR-microscope spectrometer; fit of the measured spectra (red circles) with a), b) two Lorentzian functions in the tetragonal/cubic phase and c), d) four Lorentzian functions in the orthorhombic phase, as well as with a linear background function (green); the solid black line corresponds to the summed-up Lorentzian functions plotted in colors. In red color: $CH_3$ rocking mode $\nu_{12}$; in blue color: C-N stretching mode $\nu_5$.

a)

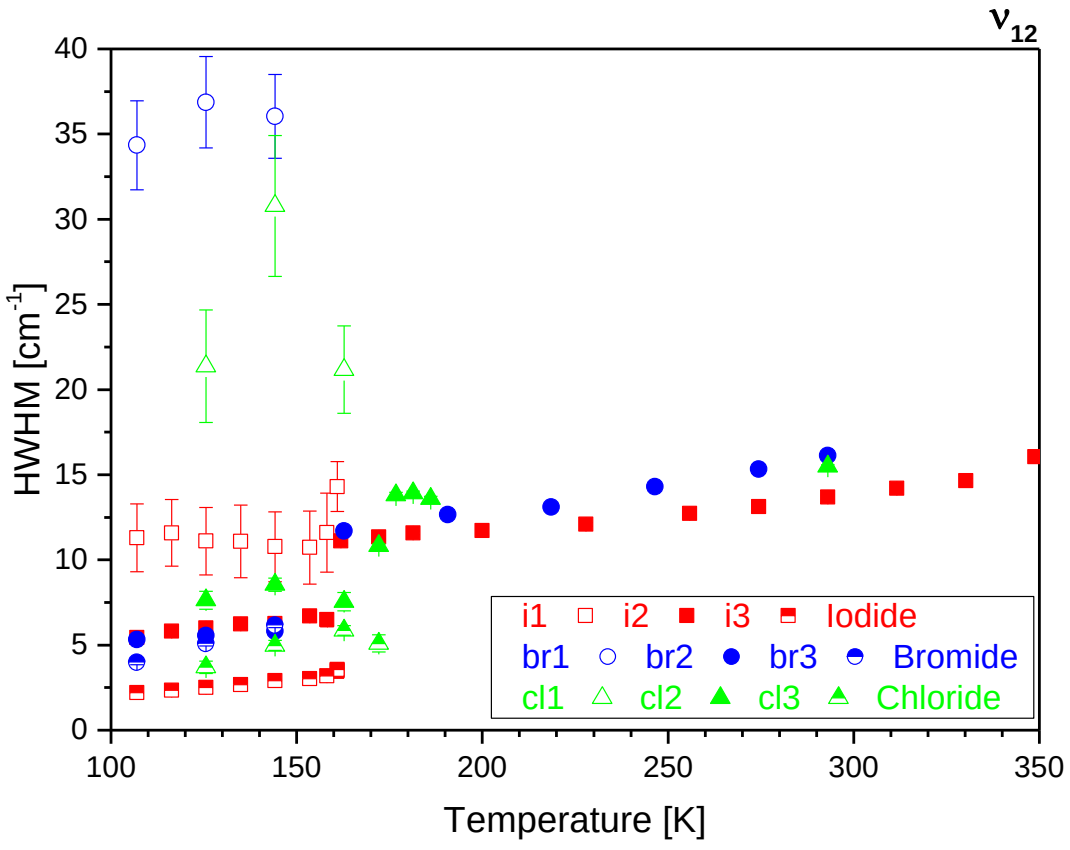

b)

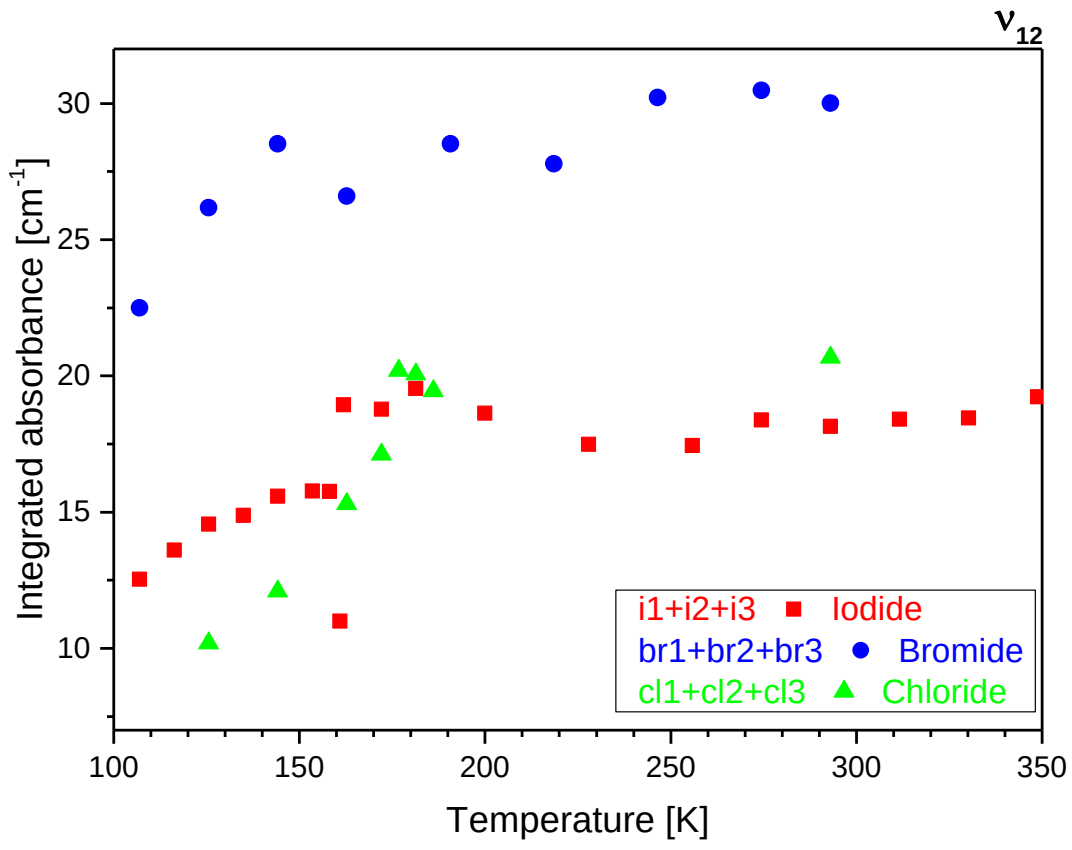

c)

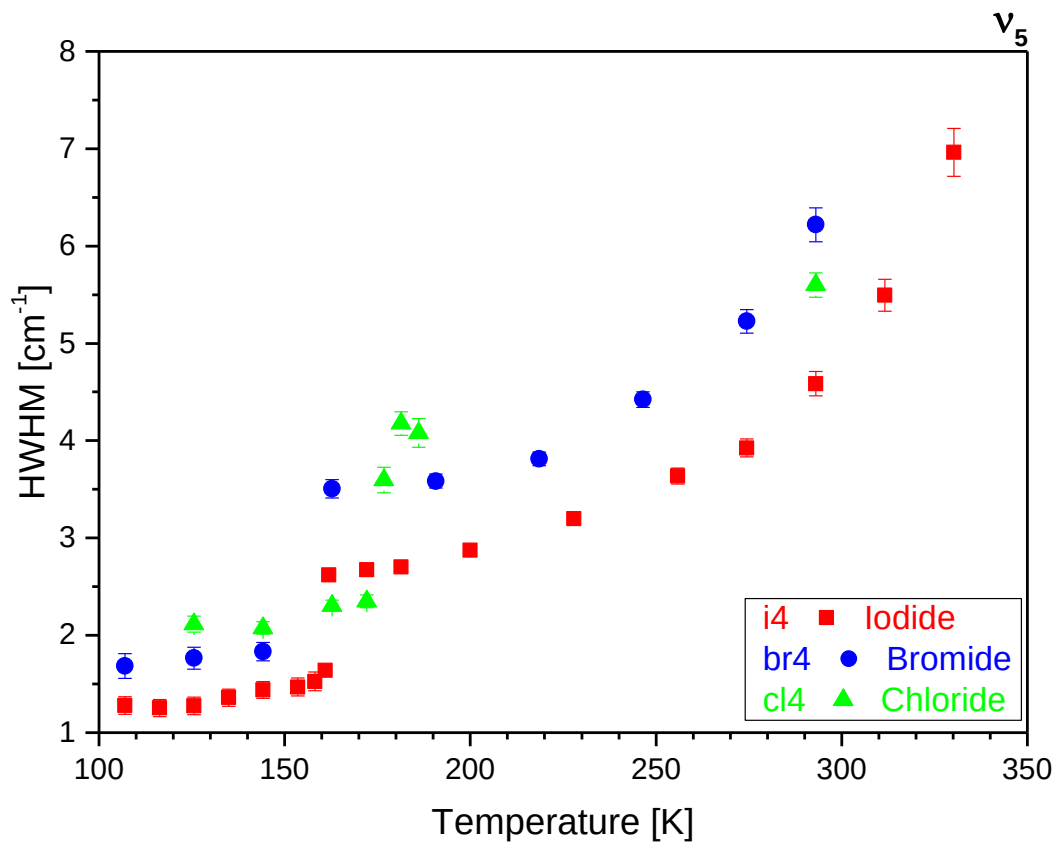

d)

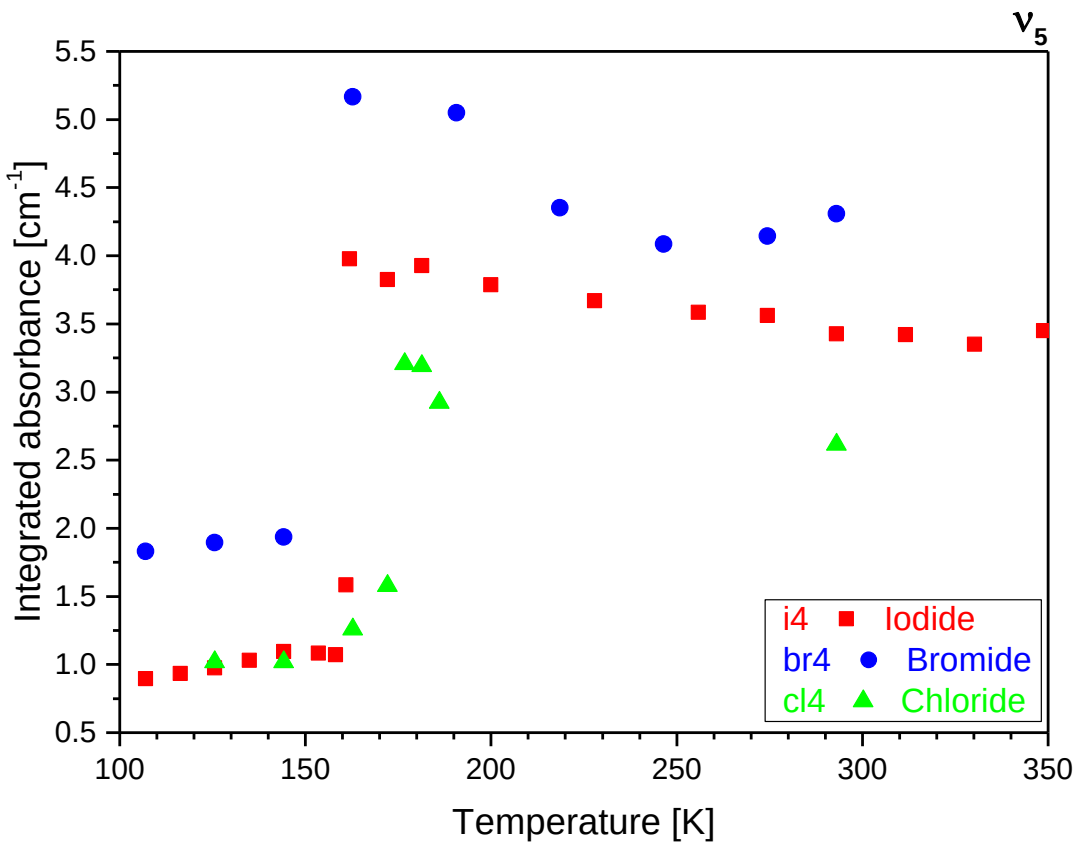

**Figure S6** Temperature-dependence of a), c) HWHM and b), d) integrated absorbance of a), b) $CH_3$ rocking mode $\nu_{12}$ and c), d) C-N stretching mode $\nu_5$. In red color: $MAPbI_3$, in blue color: $MAPbBr_3$ and in green color: $MAPbCl_3$. The open and half-filled symbols in a) correspond to the same frequencies in Fig. 5a). The integrated absorbance of degenerated modes in b) are summed up.

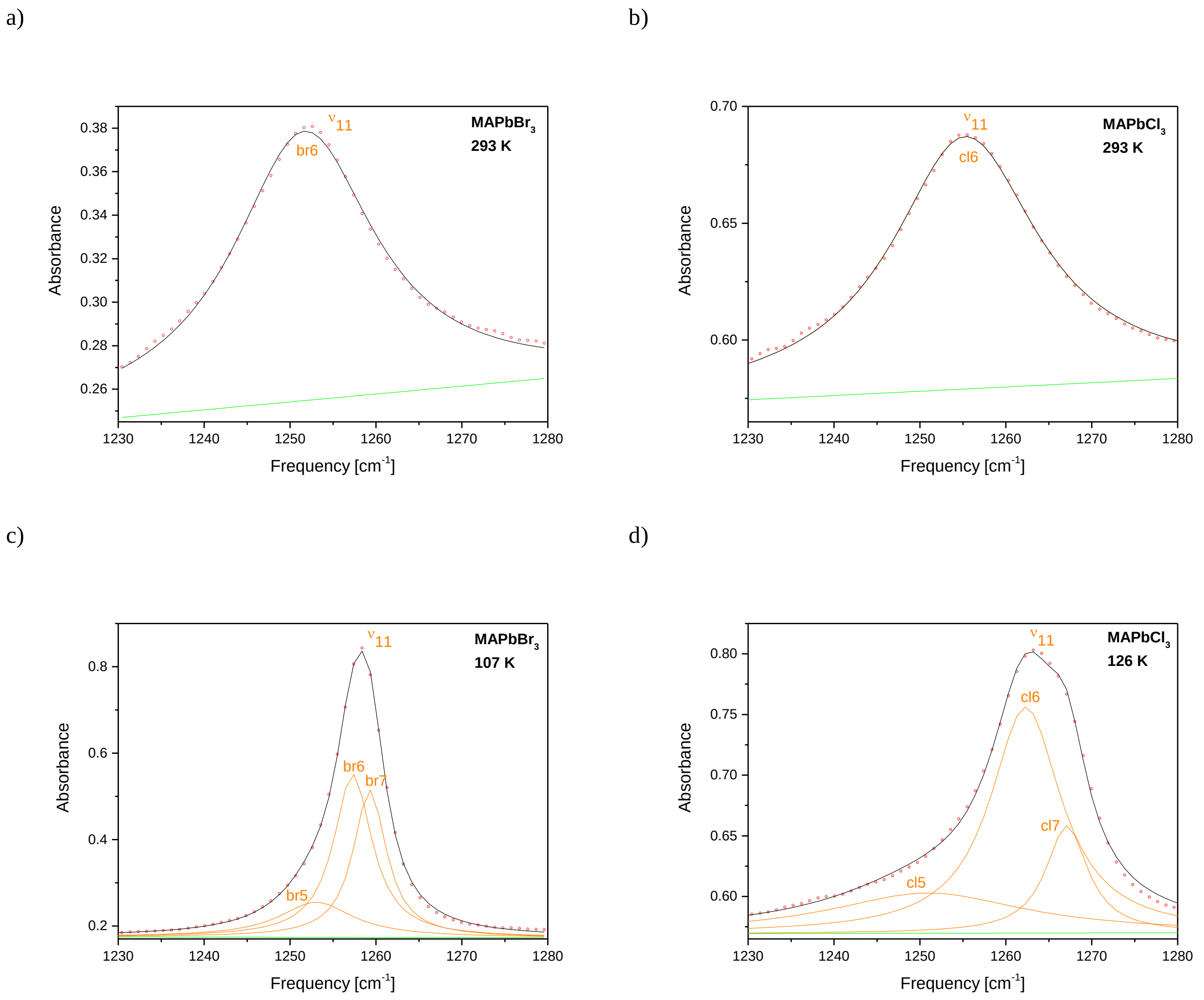

**Figure S7** Infrared spectra of a), c) $MAPbBr_3$ and b), d) $MAPbCl_3$ between 1230 $cm^{-1}$ and 1280 $cm^{-1}$ at a), b) 293 K (tetragonal/cubic) and at c) 107 K and d) 126 K (orthorhombic phase); measurements of small single crystals using a FTIR-microscope spectrometer; fit of the measured spectra (red circles) with a), b) one Lorentzian function in the tetragonal/cubic phase and c), d) three Lorentzian functions in the orthorhombic phase, as well as with a linear background function (green); the solid black line corresponds to the summed-up Lorentzian functions plotted in colors. In orange color: $NH_3$ rocking mode $\nu_{11}$.

a)

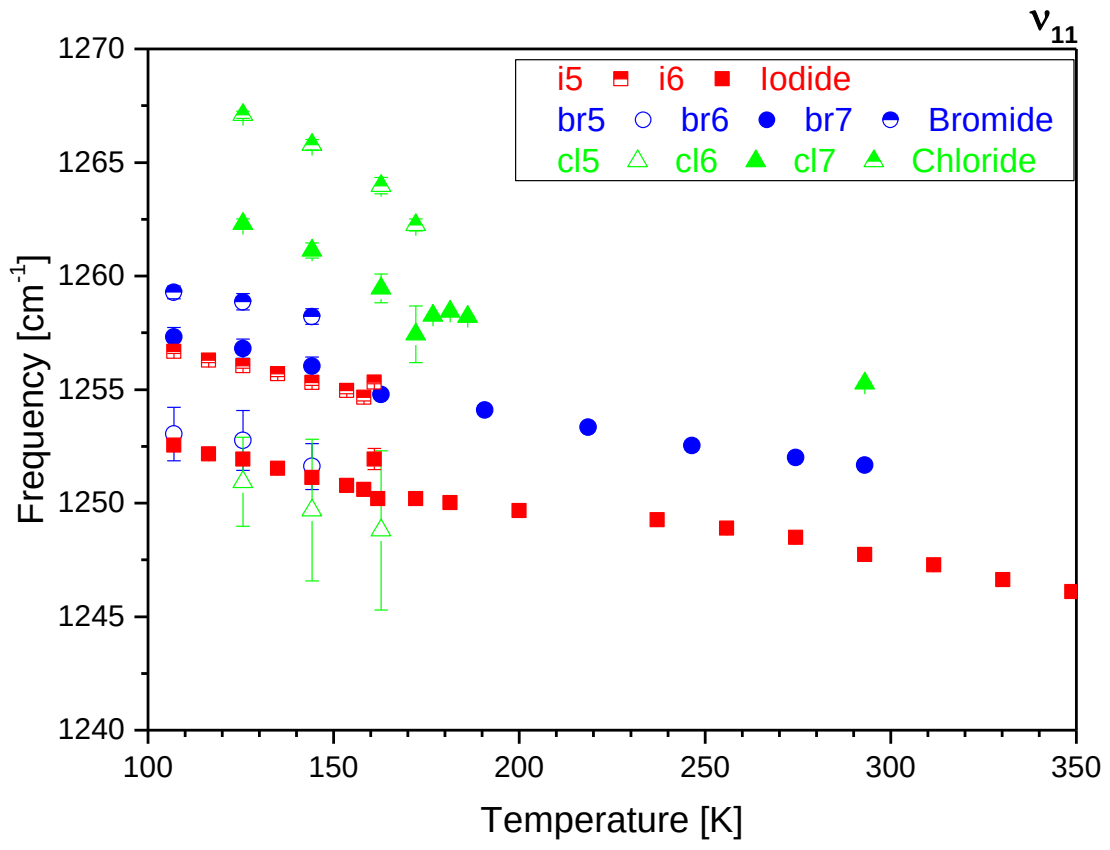

b)

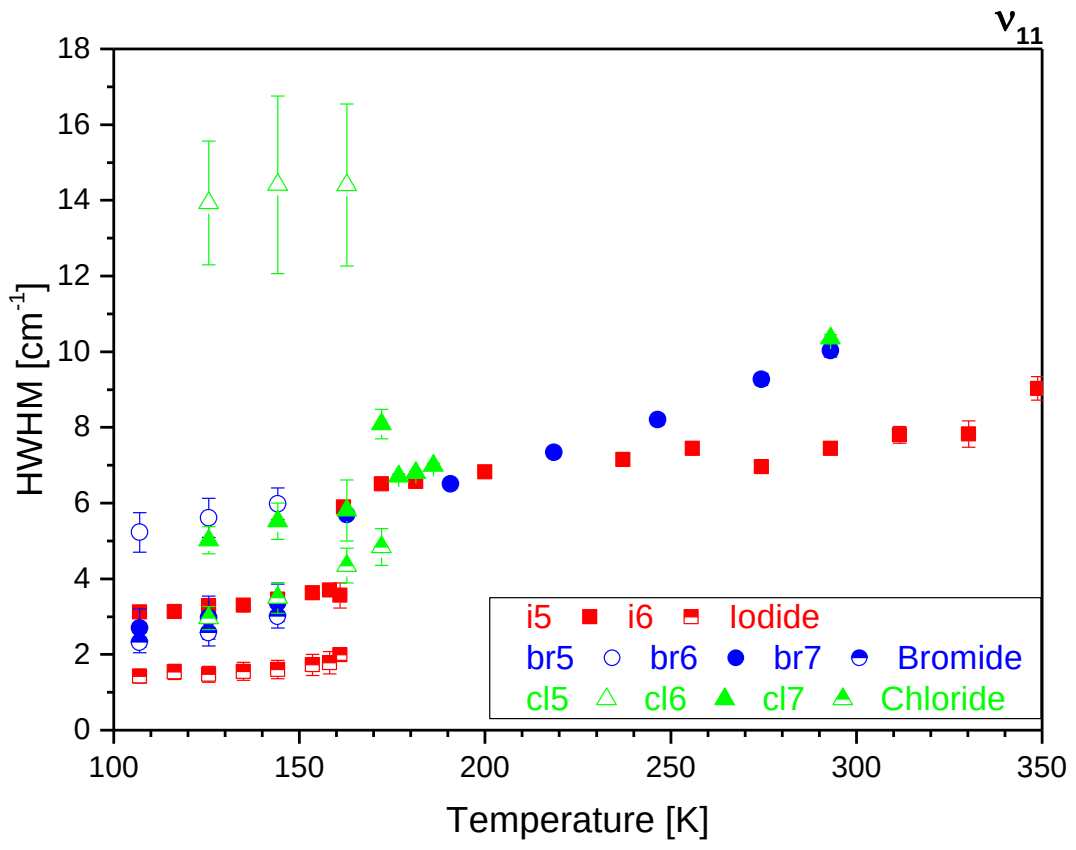

c)

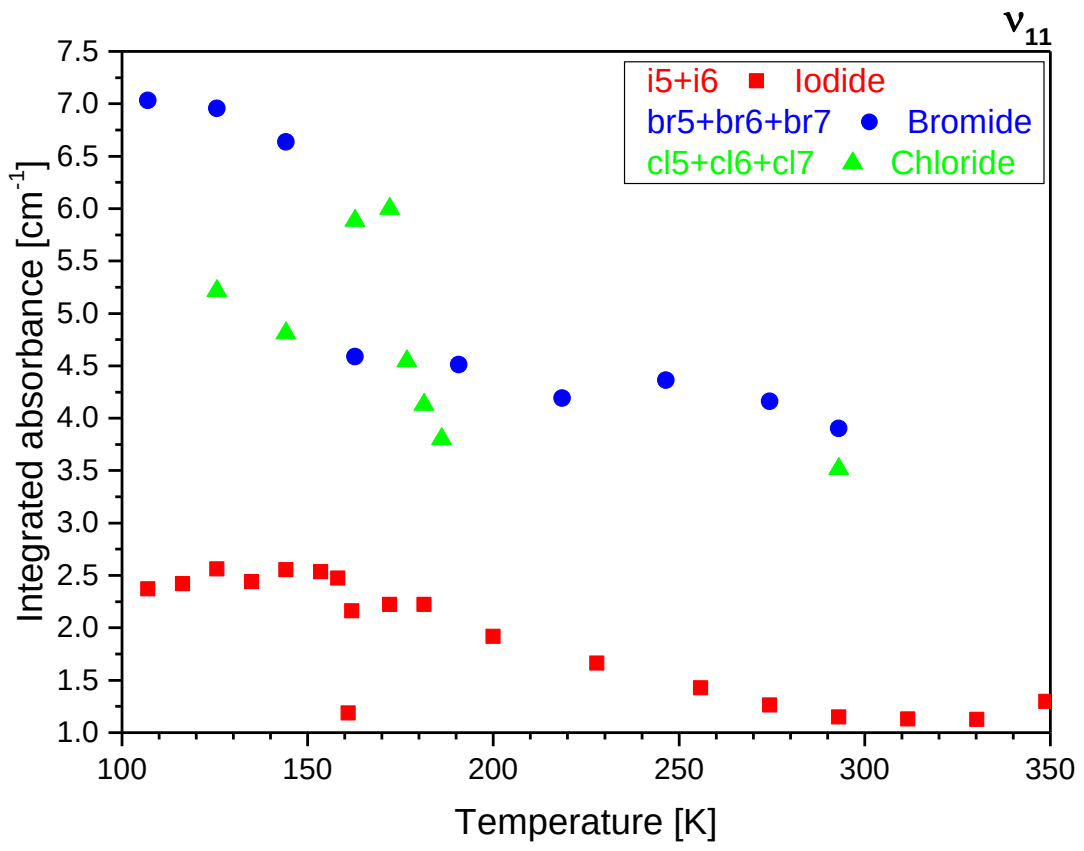

**Figure S8** Temperature-dependence of a) frequencies, b) HWHM and c) integrated absorbance of $NH_3$ rocking mode $\nu_{11}$. In red color: $MAPbI_3$, in blue color: $MAPbBr_3$ and in green color: $MAPbCl_3$. The open and half filed symbols in a) corresponds to the same frequencies in b). The integrated absorbance of degenerated modes are summed up.

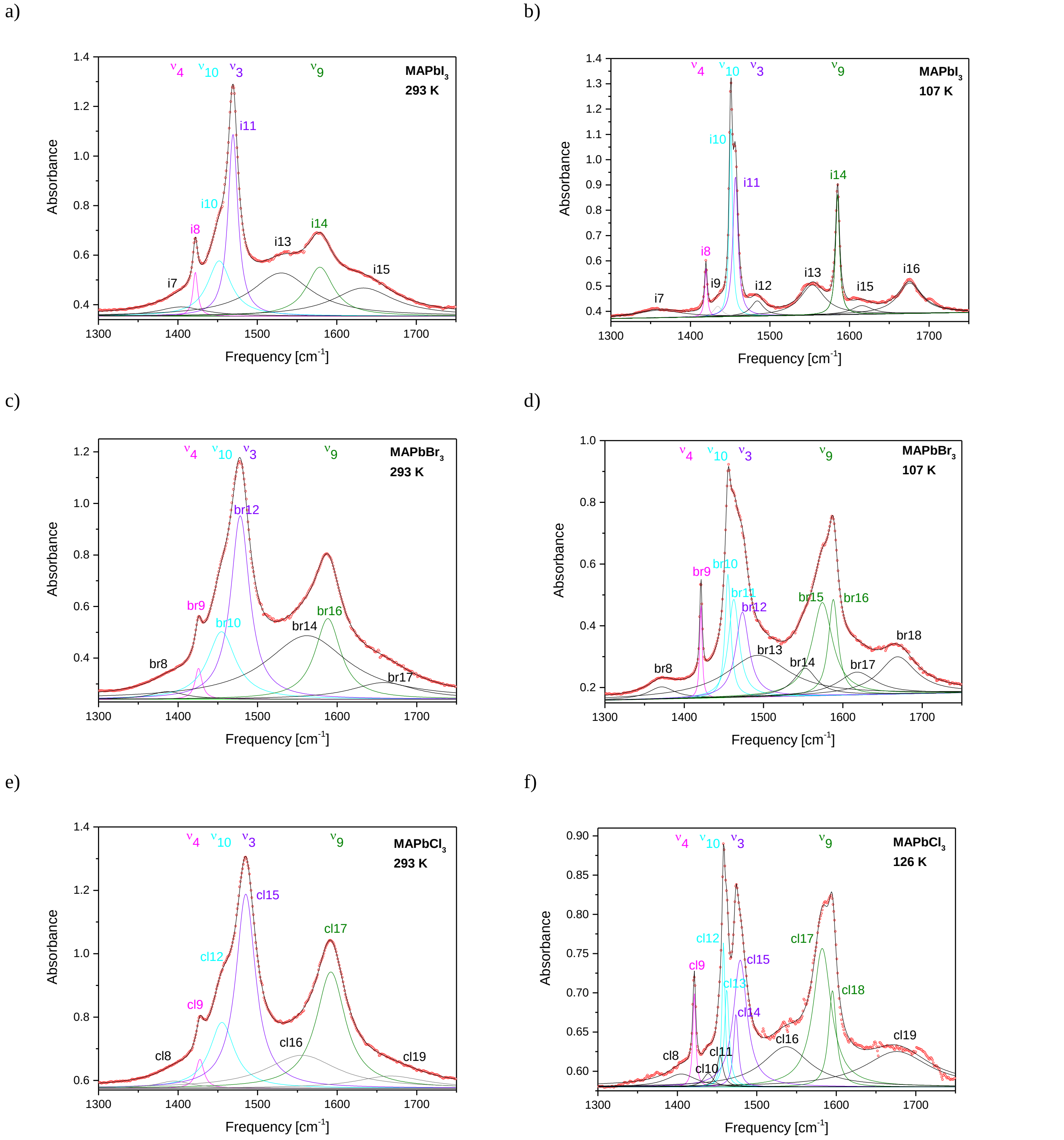

**Figure S9** Infrared spectra of a), b) $MAPbI_3$, c), d) $MAPbBr_3$ and e), f) $MAPbCl_3$ between 1300 $cm^{-1}$ and 1750 $cm^{-1}$ at a), c), e) 293 K (tetragonal/cubic) and at b), d), 107 K and f) 126 K (orthorhombic phase); measurements of small single crystals using a FTIR-microscope spectrometer; fit of the measured spectra (red circles) with a), c), e) seven Lorentzian functions in the tetragonal/cubic phase and b) ten, d) eleven and f) twelve Lorentzian functions in the orthorhombic phase, as well as with a linear background function (green); the solid black line corresponds to summing the Lorentzian functions plotted in colors. Magenta: symmetric $CH_3$ bending mode $\nu_4$; cyan: asymmetric $CH_3$ bending mode $\nu_{10}$; violet: symmetric $NH_3$ bending mode $\nu_3$ and in dark green color: asymmetric $NH_3$ bending mode $\nu_9$. In black color plotted Lorentzian functions correspond to modes that could not clearly be assigned to MA vibrational modes.

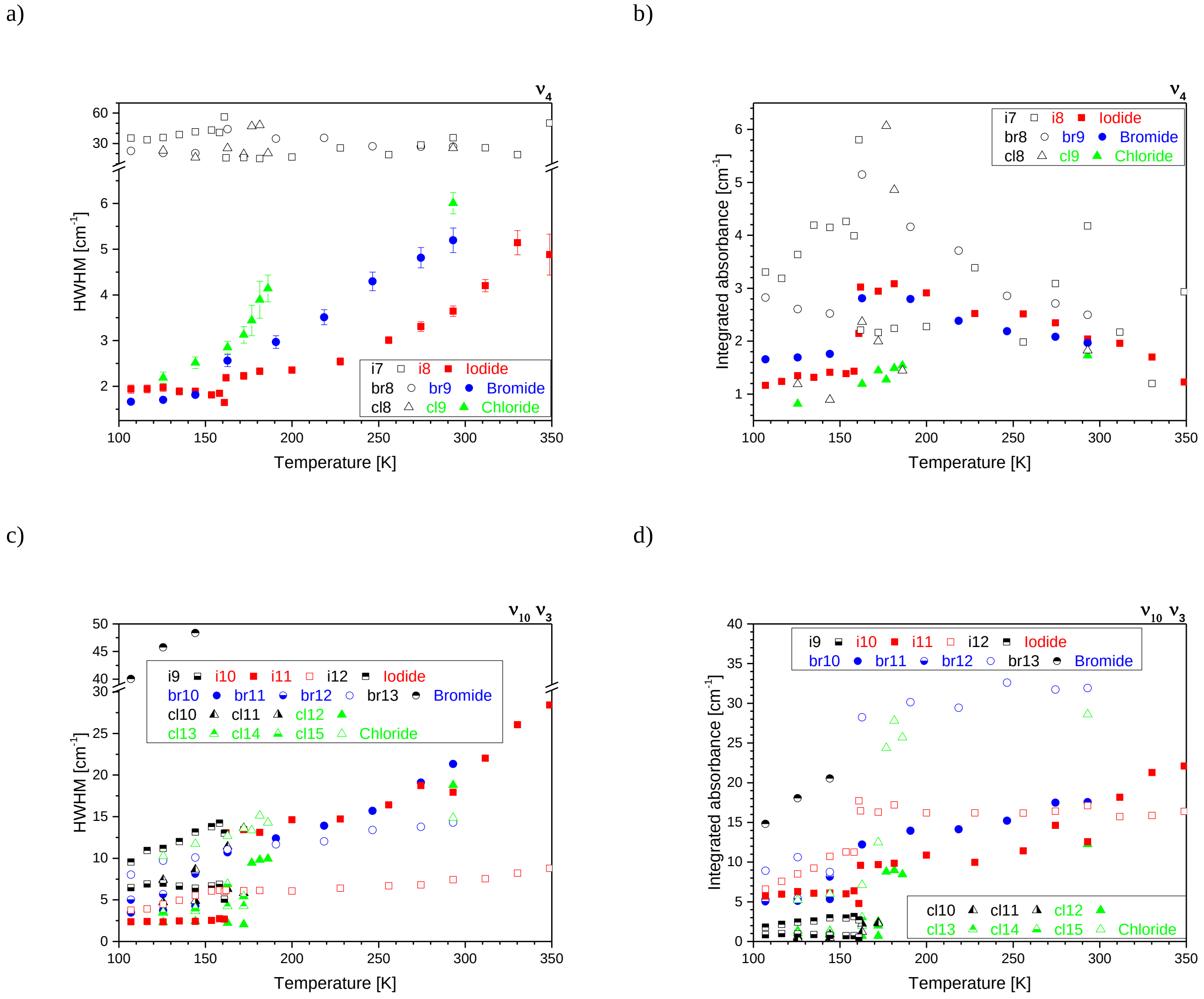

**Figure S10** Temperature-dependence of a), c) HWHM and b), d) integrated absorbance of symmetric $CH_3$ bending mode $\nu_4$ (a, b); asymmetric $CH_3$ bending mode $\nu_{10}$ and symmetric $NH_3$ bending mode $\nu_3$ (c, d). In red color: $MAPbI_3$, in blue color: $MAPbBr_3$ and in green color: $MAPbCl_3$. For most data points no error bars are provided for simplicity.

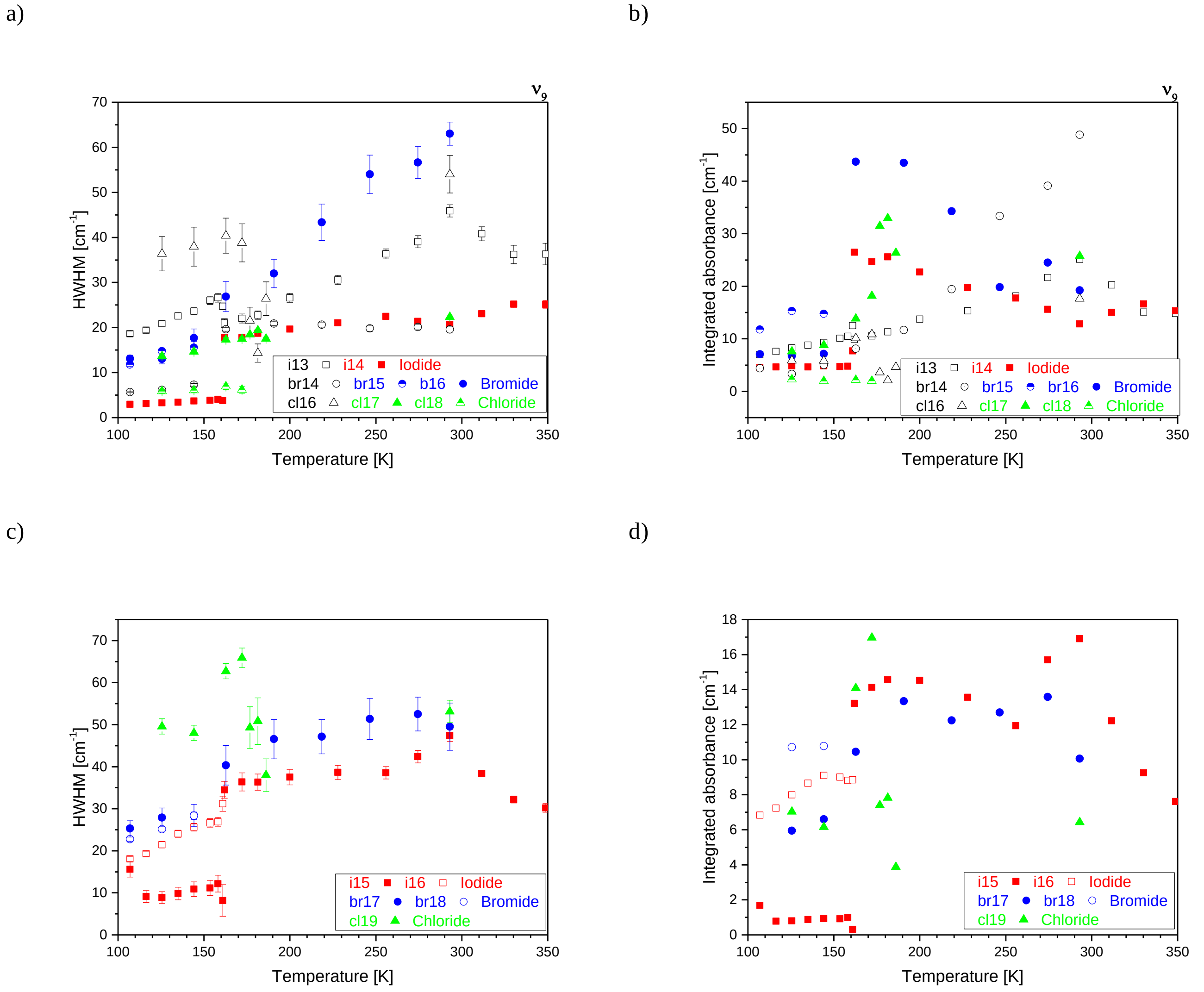

**Figure S11** Temperature-dependence of a), c) HWHM and b), d) integrated absorbance of asymmetric $NH_3$ bending mode $\nu_9$ (a, b), and combinations (c, d). In red color: $MAPbI_3$, in blue color: $MAPbBr_3$ and in green color: $MAPbCl_3$. For most data points no error bars are provided for simplicity.

a)

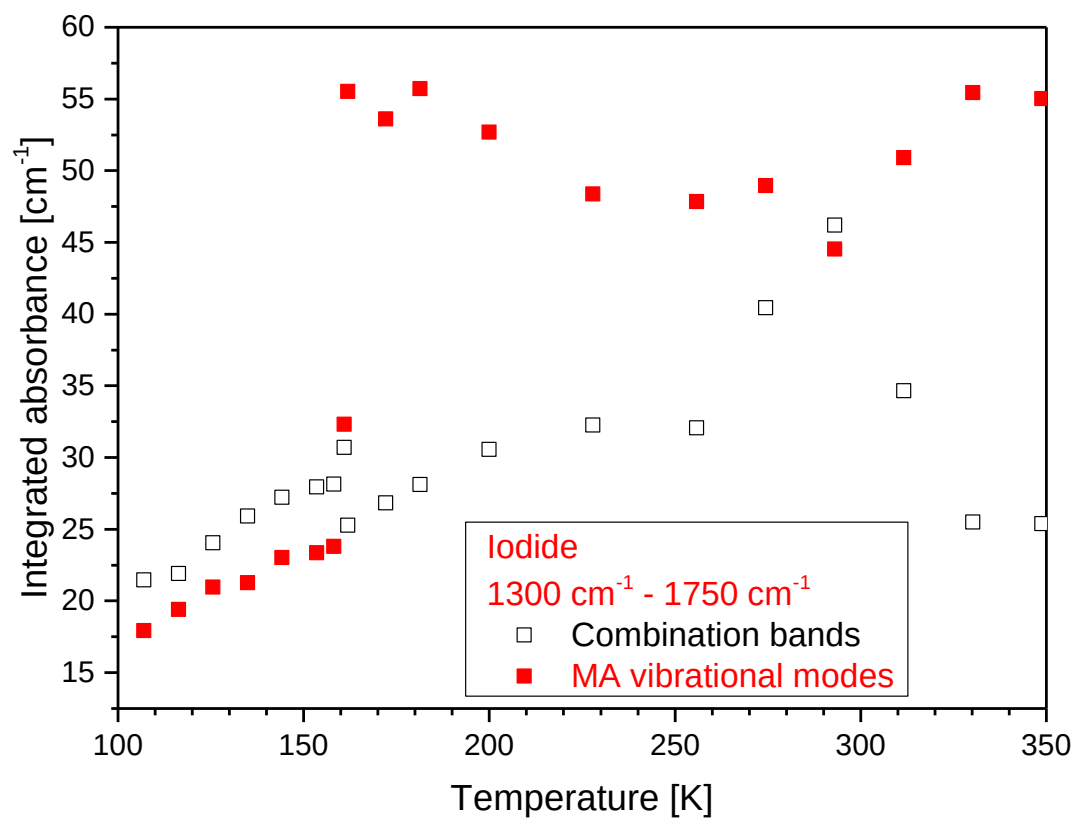

b)

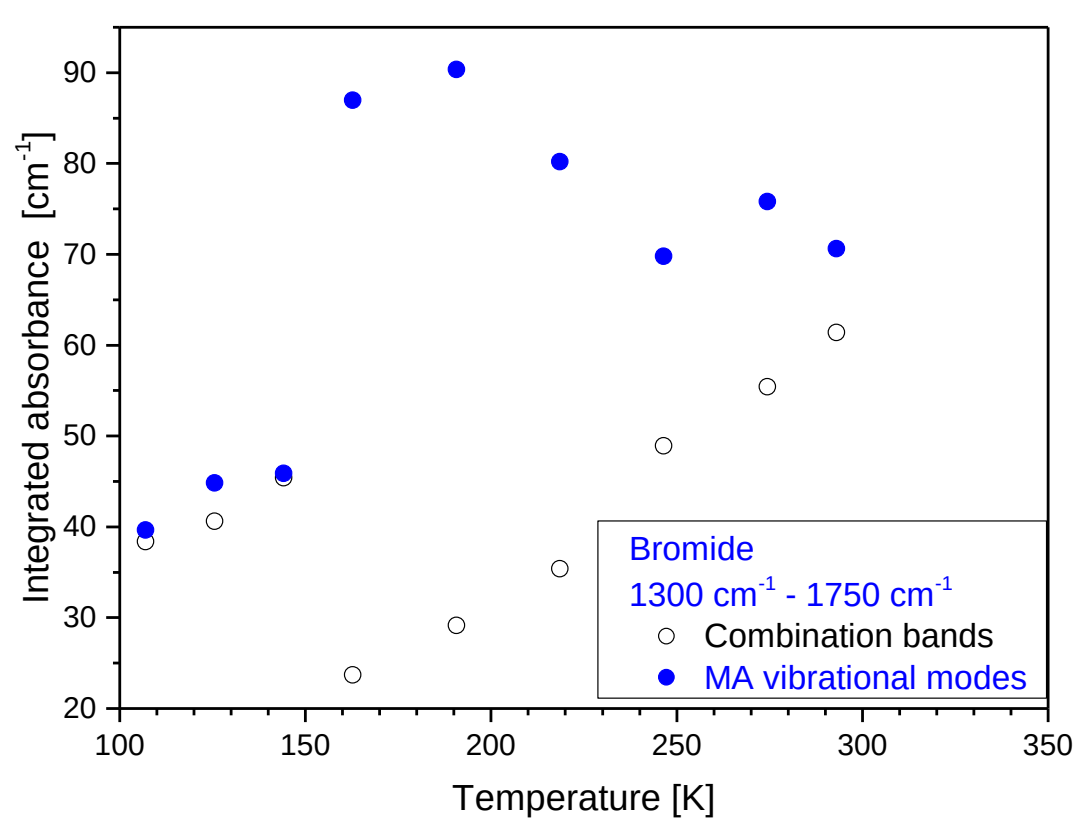

c)

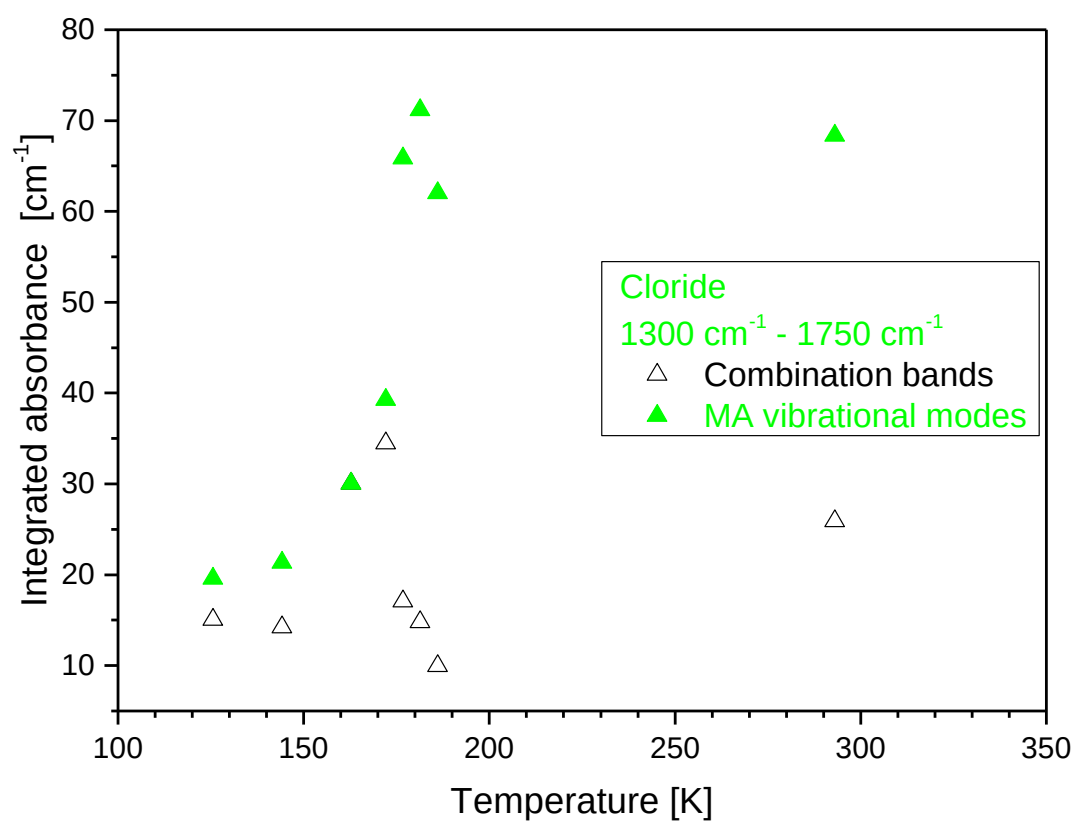

**Figure S12** Summed integrated absorbance of all Lorentzians of a) $MAPbI_3$, b) $MAPbBr_3$ and c) $MAPbCl_3$ between 1300 $cm^{-1}$ and 1750 $cm^{-1}$; in black color summed integrated absorbance of all Lorentzians that was assigned to combinations ($MAPbI_3$ : i7, i9, i12, i13, i15 and i16; $MAPbBr_3$ : br8, b13, br14, br17 and br18; $MAPbCl_3$ : cl8, cl10, cl11, cl16 and cl19).

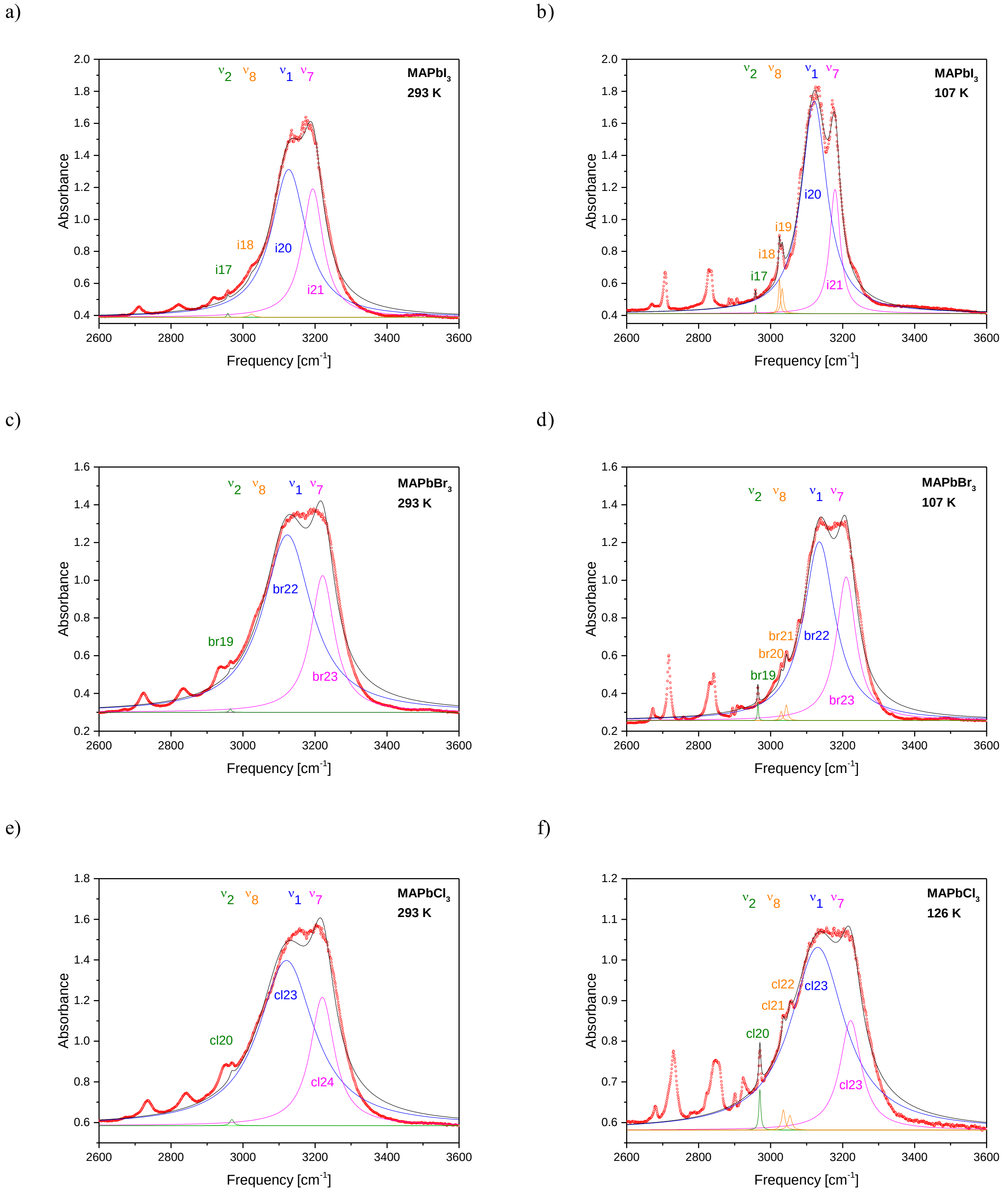

**FIGURE S13** Infrared spectra of a), b) $MAPbI_3$, c), d) $MAPbBr_3$, and e), f) $MAPbCl_3$ between 2600 $cm^{-1}$ and 3600 $cm^{-1}$ at b), d) 107 K and f) 126 K and at a), b), c) 293 K; measurements of small single crystals using a FTIR-microscope spectrometer; fit of the measured spectra (red circles) with two Lorentzian functions for $\nu_1$ and $\nu_7$ and fixed Lorentzian functions for $\nu_2$ and $\nu_8$, as well as with a fixed constant background function (green) that was fitted in the 3600-4000 $cm^{-1}$ frequency range; the solid black line corresponds to the summed-up Lorentzian functions plotted in colors. In green color: C-H stretching vibrational mode $\nu_2$, orange color: C-H stretching vibrational modes $\nu_8$, blue color: N-H stretching vibrational mode $\nu_1$, and in magenta color: N-H stretching vibrational mode $\nu_7$.

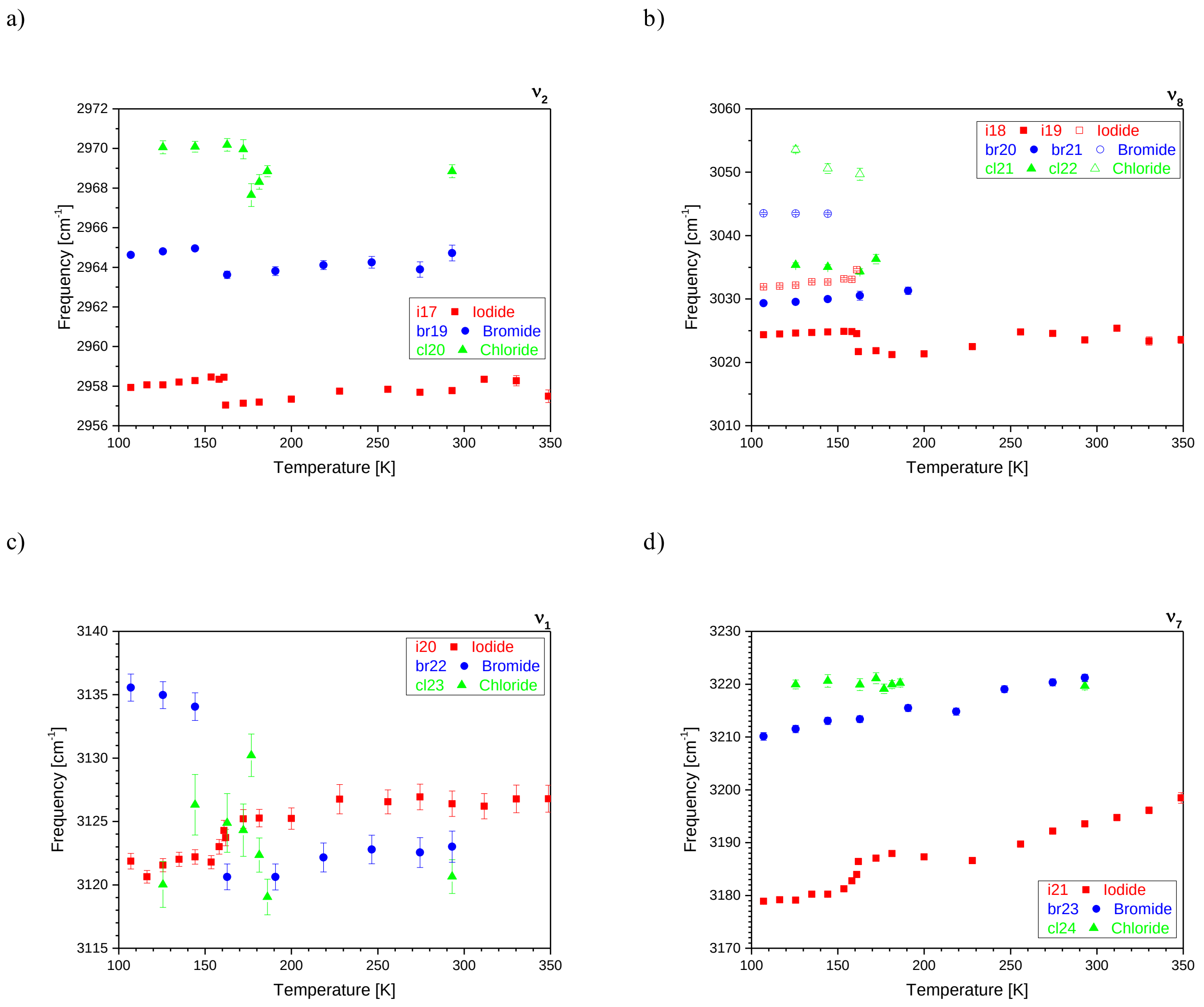

**Figure S14** Temperature-dependence of frequencies of C-H stretching vibrational modes a) $\nu_2$, b) $\nu_8$, and N-H stretching vibrational modes c) $\nu_1$, and d) $\nu_7$. In red color: $MAPbI_3$, in blue color: $MAPbBr_3$ and in green color: $MAPbCl_3$.

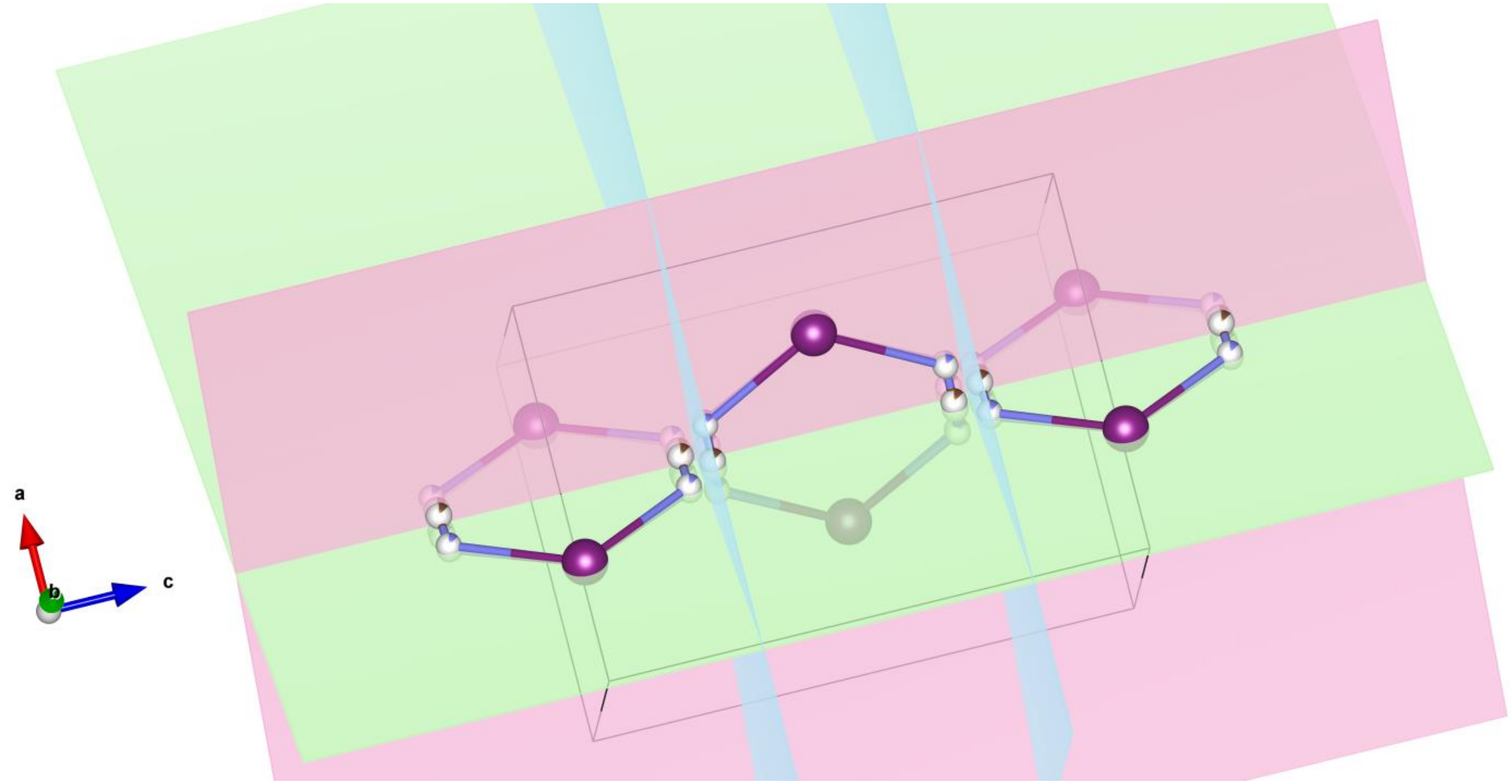

**Figure S15** Symmetry reduced MA molecules in the orthorhombic $MAPbI_3$ crystal structure, drawn in blue color are the *Pnma* mirror planes (0,1/4,0) and (0,3/4,0), d) mirror planes in the tetragonal space group symmetry *I*4/*mcm*, green: (1/4,3/4,0), red: (3/4,3/4,0) and blue: orthorhombic mirror planes not present in the tetragonal structure.

$\nu_{12}$: 77 $B_{2u}$ 82 $B_{3u}$ 84 $B_{1u}$ b)

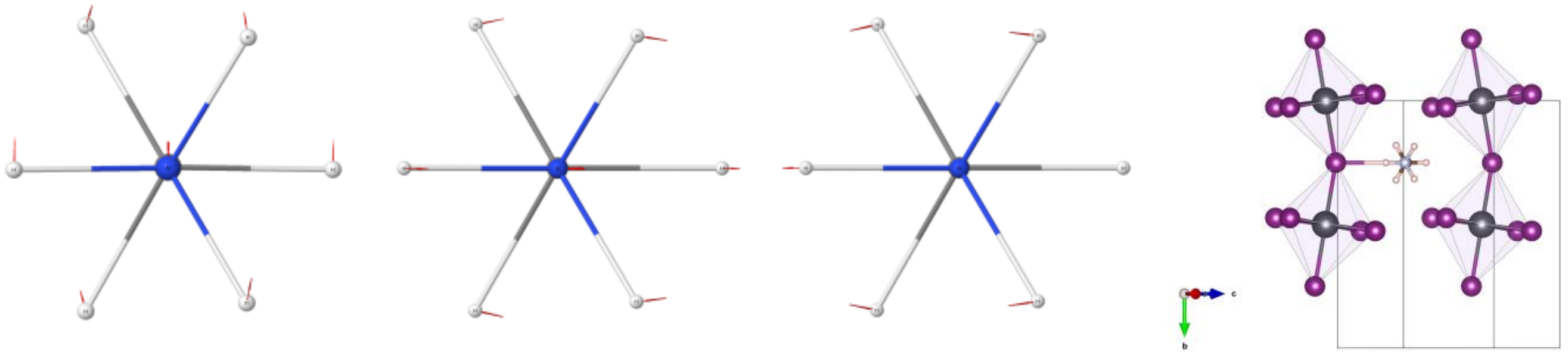

$\nu_5$: 86 $B_{3u}$ 87 $B_{1u}$

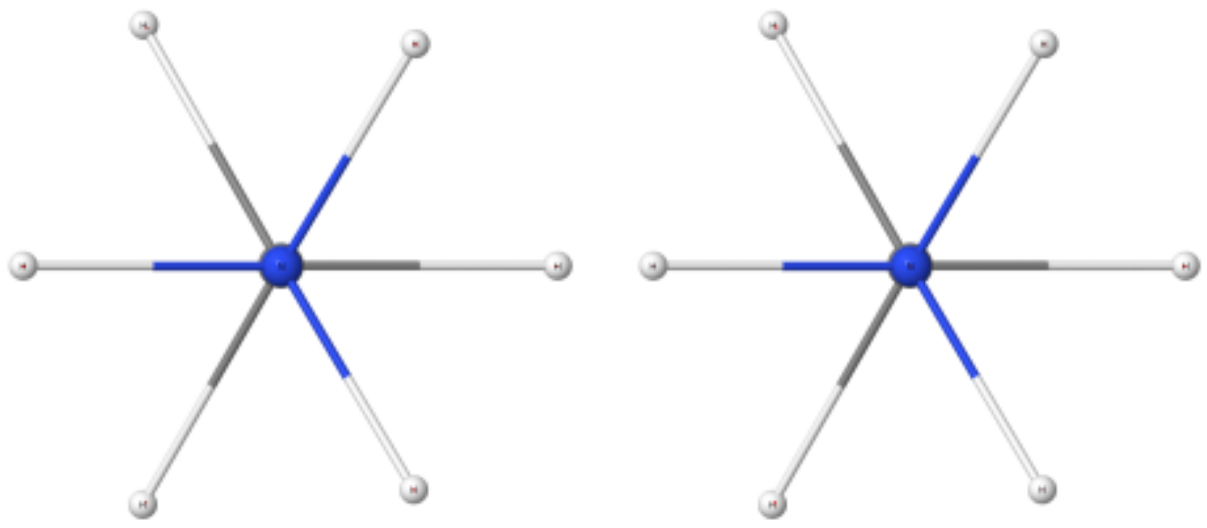

$\nu_{11}$: 91 $B_{2u}$ 93 $B_{1u}$ 95 $B_{3u}$

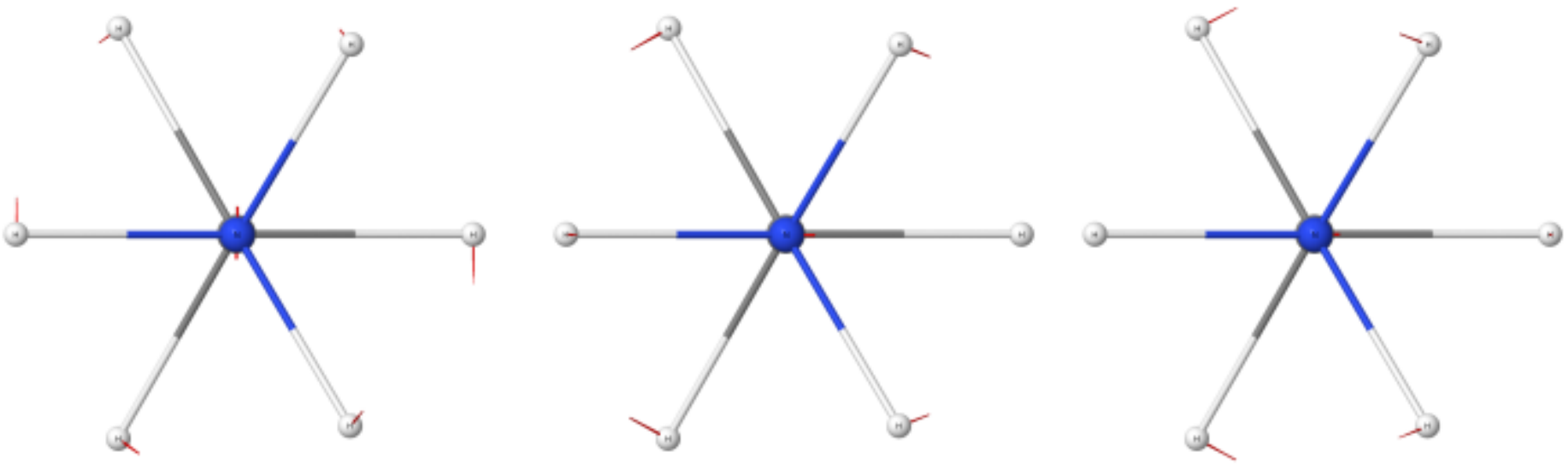

**Figure S16** Selected infrared active vibrational modes of $CH_3NH_3$ (MA) in **$MAPbI_3$** calculated with DFT (DFT mode No. see also Table S3), b) crystallographic orientation of the selected MA molecule, it can be distinguished between out-of-plane movements ($B_{2u}$) and in-plane movements ($B_{3u}$ and $B_{1u}$) relative to the *Pnma* mirror plane (0,1/4,0).

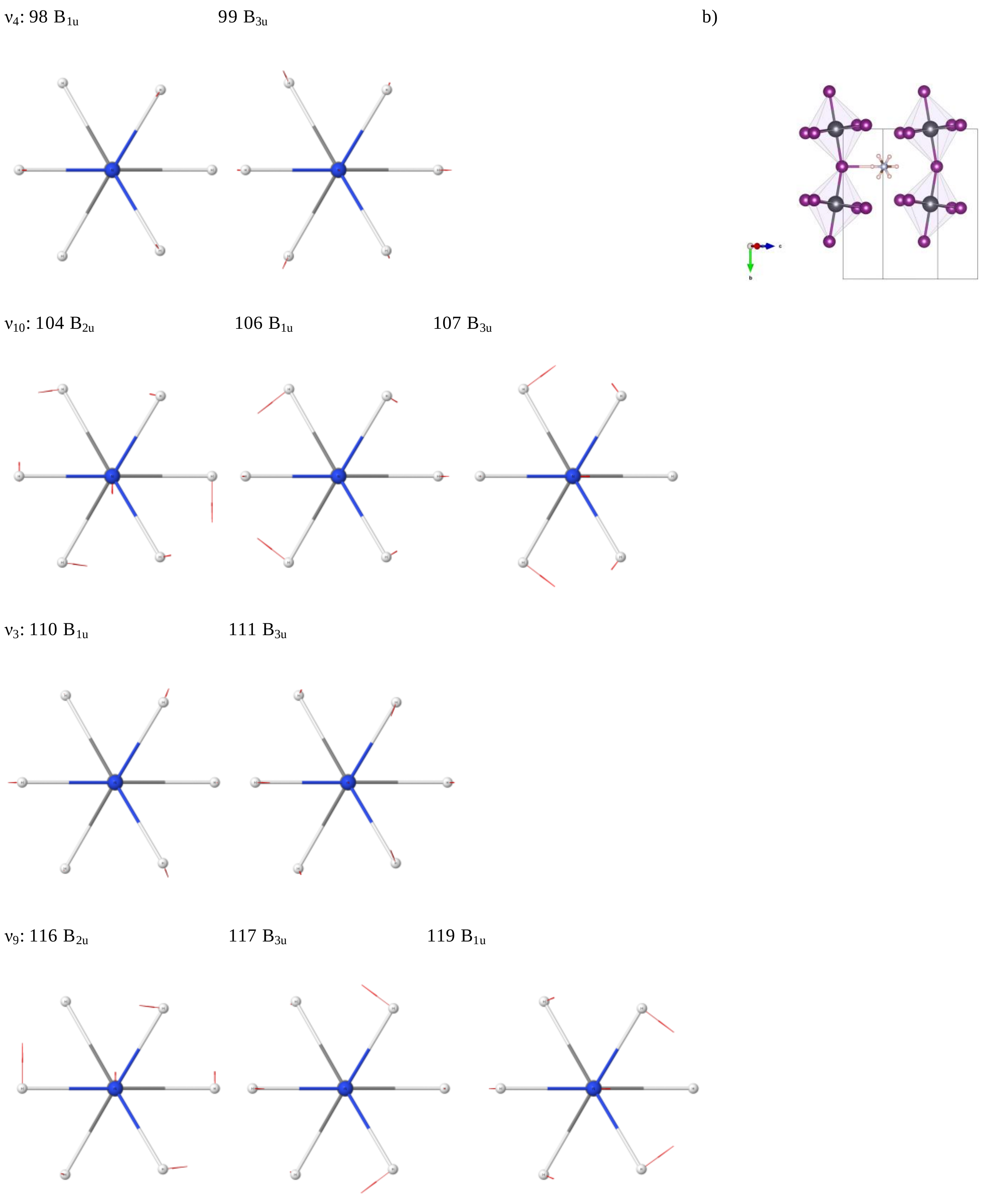

**Figure S17** Selected infrared active vibrational modes of $CH_3NH_3$ (MA) in **$MAPbI_3$** calculated with DFT (DFT mode No. see also Table S3), b) crystallographic orientation of the selected MA molecule, it can be distinguished between out-of-plane movements ($B_{2u}$) and in-plane movements ($B_{3u}$ and $B_{1u}$) relative to the *Pnma* mirror plane (0,1/4,0).

$\nu_2$: 121 $B_{3u}$ 124 $B_{1u}$ b)

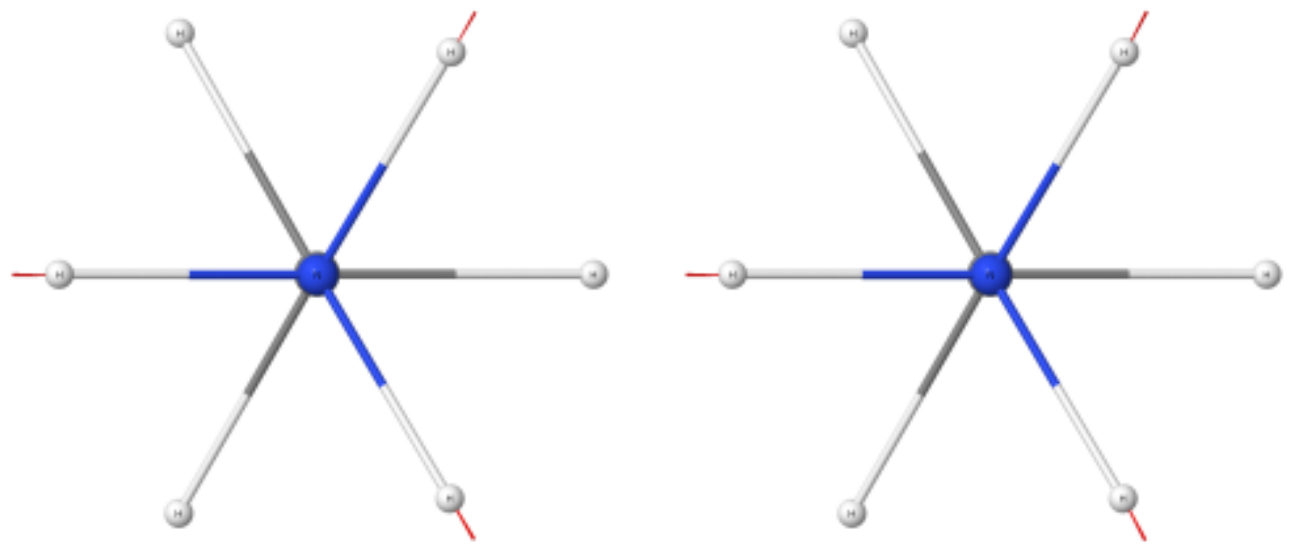

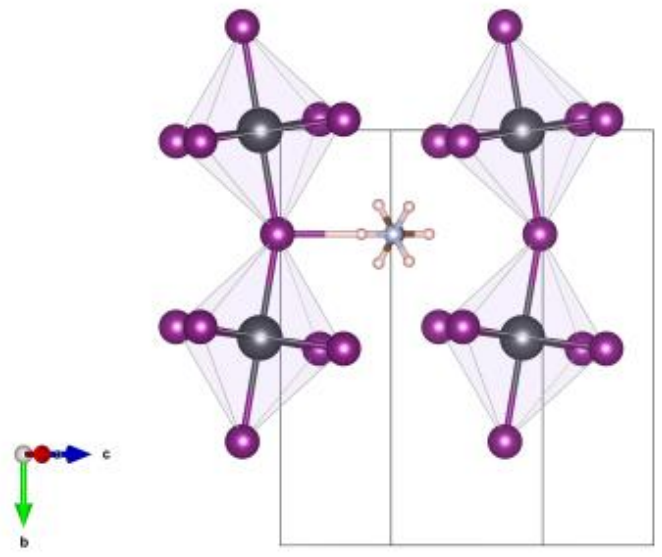

$\nu_8$: 125 $B_{1u}$ 127 $B_{3u}$ 132 $B_{2u}$

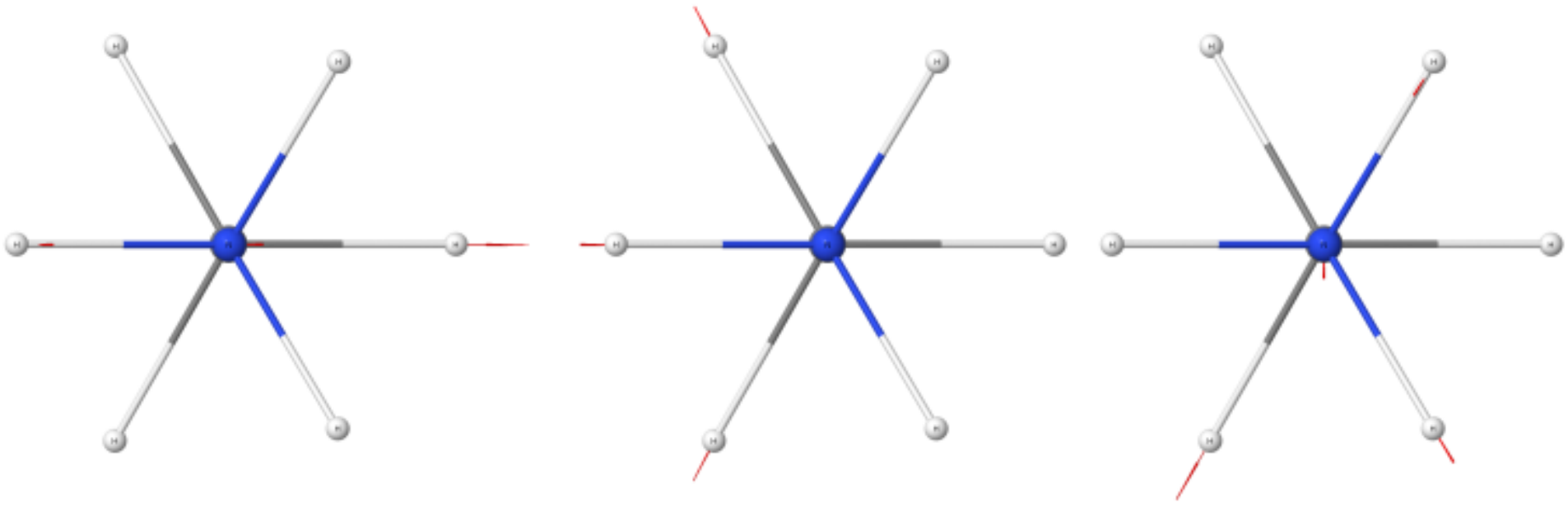

$\nu_1$: 133 $B_{3u}$ 135 $B_{1u}$

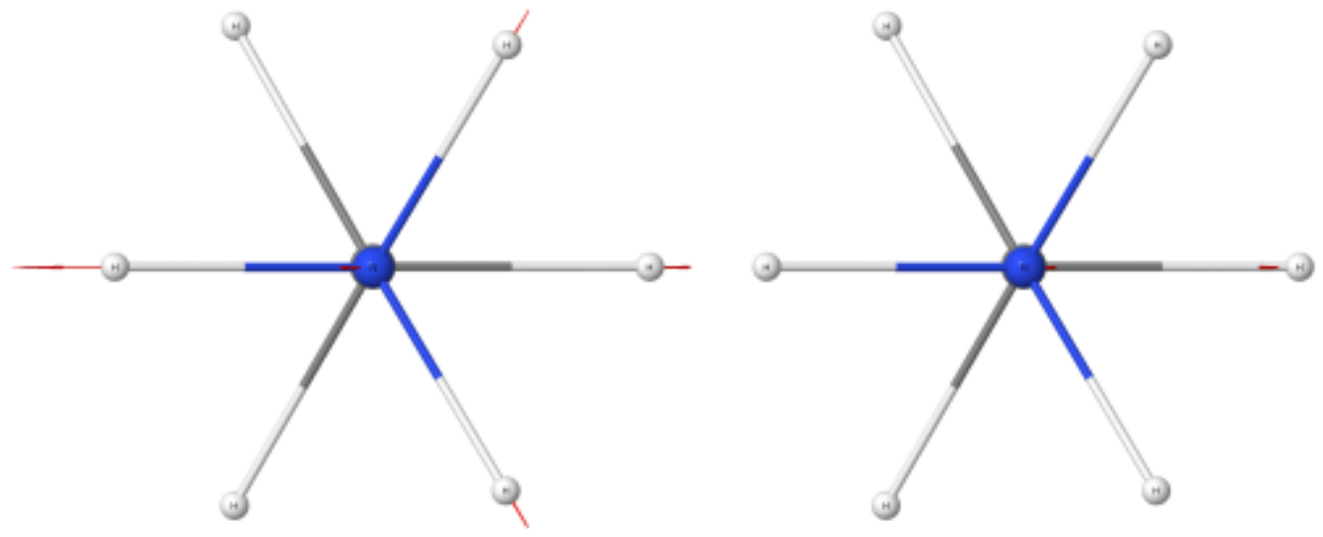

$\nu_7$: 139 $B_{1u}$ 140 $B_{3u}$ 141 $B_{2u}$

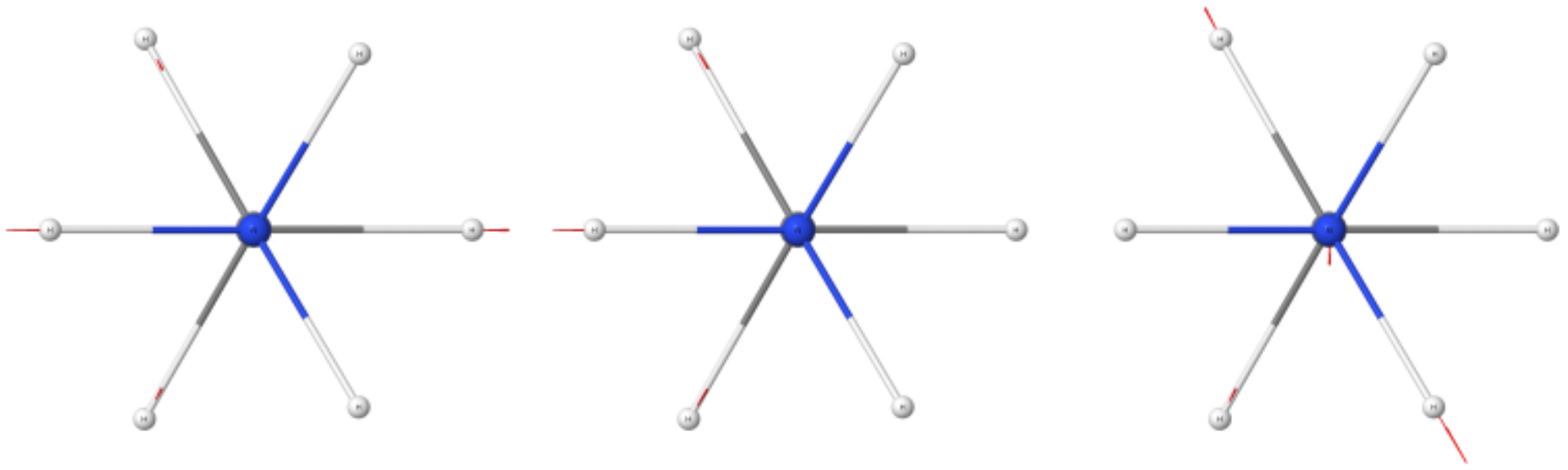

**Figure S18** Selected infrared active vibrational modes of $CH_3NH_3$ (MA) in **$MAPbI_3$** calculated with DFT (DFT mode No. see also Table S3), b) crystallographic orientation of the selected MA molecule, it can be distinguished between out-of-plane movements ($B_{2u}$) and in-plane movements ($B_{3u}$ and $B_{1u}$) relative to the *Pnma* mirror plane (0,1/4,0).

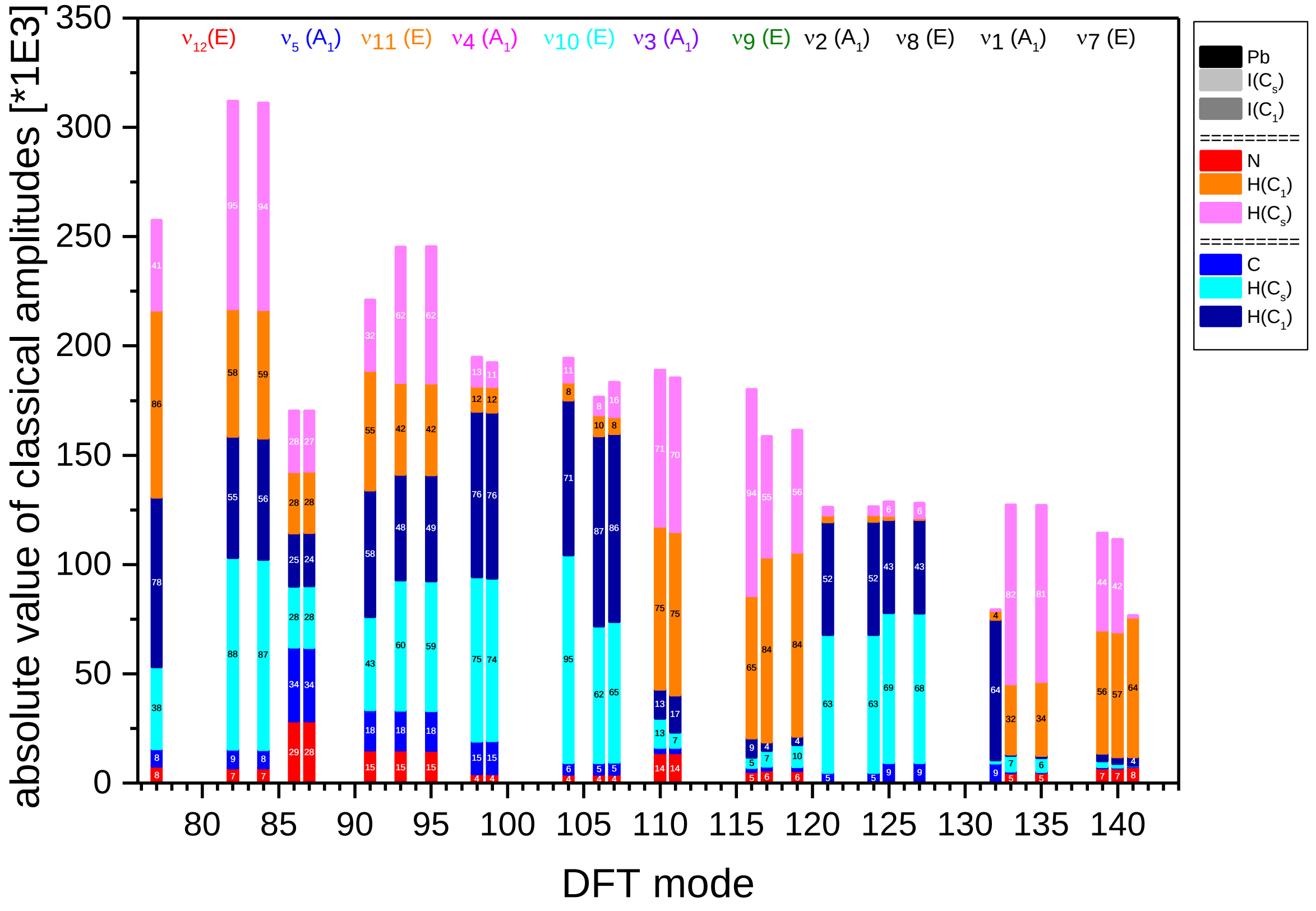

**Figure S19** Orthorhombic $MAPbI_3$; DFT results: absolute values of classical amplitudes (*1000) for selected IR active modes (DFT mode numbers $\nu_{12}$: 77, 82, 84; $\nu_5$: 86, 87; $\nu_{11}$: 91, 93, 95; $\nu_4$: 98, 99; $\nu_{10}$: 104, 106, 107; $\nu_3$: 110, 111; $\nu_9$: 116, 117, 119; $\nu_2$: 121, 124; $\nu_8$: 125, 127, 132; $\nu_1$: 133, 135; $\nu_7$: 139, 140, 141 ) (see also Table S3). Out-of-plane ($B_{2u}$) modes: 77, 91, 104, 116, 132 and 141.